\documentclass[aps,prd,showpacs,showkeys,superscriptaddress,nofootinbib]{revtex4-2}

\usepackage{epsfig,graphicx,amsfonts,amsmath,amssymb,bbm}
\usepackage{bbold}
\RequirePackage{mathrsfs}
\usepackage{color}
\usepackage{epsf}
\usepackage{epstopdf}

\def\eq#1{{Eq.~(\ref{#1})}}
\newcommand{\Le}{\left(}
\newcommand{\Ra}{\right)}

\newcommand{\beq}{\begin{equation}}
	\newcommand{\eeq}{\end{equation}}
\newcommand{\beqar}{\begin{eqnarray}}
	\newcommand{\eeqar}{\end{eqnarray}}
\newcommand{\D}{\partial}

\newcommand{\om}{\omega}

\newcommand{\T}{{\cal T}}
\begin{document}
	\title{Spinor fields, CPTM symmetry and smallness of cosmological constant in framework of extended manifold}
	\author{S. Bondarenko}
	\author{V. De La Hoz-Coronell}
	\affiliation{Ariel University, Ariel 4070000, Israel}

	%\date{\today}
	
	\begin{abstract}
	
	A model of an extended manifold for the Dirac spinor field is considered. Two Lagrangians related by  CPTM (charge-parity-time-mass) symmetry are constructed for a pair of the Dirac spinor fields with each spinor field defined in a separate manifold. An interaction between the matter fields in the manifolds is introduced through gravity. A fermionic effective action of the general system is constructed and a tadpole one-loop spinor diagram and  part of the one-loop vacuum diagrams  with two external gravitational off-shell fields which contribute to the effective action are calculated. It is demonstrated that among different versions of the second spinor Lagrangian there is a special one for which a cancellation of the mentioned diagrams in the total effective action takes place. As a result, the diagrams do not contribute to the cosmological constant, as well there is a zero contribution of the zero point energies of the spinor fields
to the action.
The non-zero leading order value of the cosmological constant for each manifold in the framework 
%arises due an interaction between the manifolds, it 
is proportional to the matter density of each separated manifold or difference of the densities,
depending on the chosen model of interaction of gravitational fields with fermions.
An appearance of the dark matter in the model is shortly discussed as well as further applications of the approach. 
		
	\end{abstract}
	
	\maketitle

\section{Introduction}
\label{S1}

  In the present paper we continue an investigation of the model of an extended manifold introduced in  \cite{Ser1,Ser2,Ser3}. There we discussed a model of manifold which consists two, at least, parts
related by the classical symmetry which reverse charge, parity, time and mass   (CPTM symmetry) preserving a form of the metric $g$ at the case of the zero cosmological constant:
\beqar\label{Add1}
&\,&
q\,\rightarrow\,-\,\tilde{q}\,,r\,\rightarrow\,-\,\tilde{r}\,,
t\,\rightarrow\,-\,\tilde{t}\,,m_{grav}\,\rightarrow\,-\,\tilde{m}_{grav}\,;\,\,\,
\tilde{q}\,,
\,\tilde{r}\,,
\,\tilde{t}\,,
\,\tilde{m}_{grav}\,>\,0\,; \\
&\,&
CPTM(g_{\mu \nu}(x))\,=\,\tilde{g}_{\mu \nu}(\tilde{x})\,=\,g_{\mu \nu}(\tilde{x})\,\label{Add102}.
\eeqar
The general manifold we consider  consists of different separate manifolds with the gravitational masses of different signs for matter fields in each one, see 
details in \cite{Ser1} and discussion in \cite{Chandr} for example.
The motivation of the introduction of the negative mass in the different cosmological models is very clear. In any scenario, see 
\cite{Villata,Chardin} for example, the presence of some kind of the repulsive gravitation forces or an additional gravitational field in our Universe helps with an explanation of the existence of dark energy in the models, see also 
\cite{DamKog,Petit1,Hoss,Kofinas,Fluid,Villata1,Hajd,Najera,Narita} and references therein. In the present construction we, nevertheless, consider a different appearance of the negative mass, the way of two metrics 
introduced is more similarly to the bimetric models perhaps, see \cite{DamKog,Petit1} for some examples. Namely, there are two different bare actions we consider as well, but, firstly, the only interactions between the manifolds we have are through  the weak gravity fields. Secondly, the matter fields of both manifolds are related by the CPTM symmetry as well. These matter fields do no interact each with other directly but only through gravity at least until non-diagonal Green's functions of Schwinger-Keldysh type are not introduced, see about in \cite{Shw,Kel,Kel1}. Additionally, in the case of spinor fields, the whole model's construction depends on the way of an introduction of the gravity field  instead vierbeins in the fermion's Lagrangians. There are different variants of this replacement we have and there is a need in some procedure which will fix the possible form of the B-manifold matter Lagrangian with gravity field included.

  The gravitation properties of the matter of new B-manifold 
are also described by Einstein equations in the presented approach, see \cite{Villata,Villata1} or \cite{Chardin} and \cite{Souriau} for examples of the 
application of the discrete symmetries in the case of the quantum and classical systems. In the proposed model, due the CPTM symmetry, we need to account the possible different directions of the
time's arrows for the different manifolds, therefore there are variants of the theory with closed time paths for the fields of the 
general manifold related by the CPTM symmetry. The interaction between these fields leads to the some quantum effective action, the cosmological constant arise then as a part of this effective action. Once arising, the constant's value can be different for both manifolds related as well by the introduced symmetry. Consequently, we will have for the case of different cosmological constants for A and B manifolds:
\beq\label{Add103}
CPTM(\Lambda)\,=\,\tilde{\Lambda}\,,\,\,\,CPTM(g_{\mu \nu}(x, \Lambda))\,=\,\tilde{g}_{\mu \nu}(\tilde{x},\tilde{\Lambda})\,.
\eeq
This appearance of the cosmological constant satisfies the naturalness criteria of 't Hooft, see \cite{Hooft1}. It's zero value correspondence
to the precise symmetry between the metrics \eq{Add102} whereas it's small non-zero value reduces the symmetry to \eq{Add103} relation.
In any extend, the constant in the formalism is not a constant anymore but it is a functional of the fields in the approach. It, therefore, can acquires different values due it's evolution, the important question we need to clarify is the present smallness of the constant.

 As mentioned above, the model assumes a presence of two copies of any matter fields for two manifolds, the case of a scalar fields was discussed in the \cite{Ser2} . In this article we consider two spinor fields, A-spinor and B-spinor fields correspondingly. Therefore, the first question we need to answer is about the construction of the spinor action for the B-manifold by the use of CPTM symmetry, see
also \cite{Ser1,Ser2}. In our case we base the whole construction on the following propositions. First of all, for the matter fields, we begin from the sum of the spinor's actions, their relative sign is fixed by the possible signs of vierbeins in the the B-manifold metric and reverse sign of the mass term in the B Lagrangian,
see also \cite{Linde} for a different realization of the similar idea. The vierbein's sign in general can be positive or negative\footnote{See in \cite{Volovik,Verg} discussion about the
possible true vacuum symmetry breaking in respect to the  vierbein sign transformation and possibility of an existence of different vierbein fields coupling to the same matter fields. }
 due the squared appearance of vierbeins in the metric's definition, see
as well \cite{CPT, Volovik} and references therein. Secondly, we construct the quantized 
one-particle spinor state, due the M-symmetry present we do not know a priory the commutation properties of the creation and annihilation operators of new state for the B-manifold. 
Therefore, we fix the commutation relations for the operators  by
the request that the corresponding vacuum energy-momentum vector for both fields together will be equal to zero, this provides an absence of the zero energy modes in the model for one-particle quantum states.
There are  two possibilities of the commutation relations for each sign of vierbeins, correspondingly 
it determines a form of the propagators for the each variant of the action for the B-matter field. On this stage we stay with four different theories of B-manifold spinor action, among them there is an action which
similar to the action arises in the Schwinger-Keldysh approach, see \cite{Shw,Kel,Kel1}. So, at the next step we calculate an one-loop effective action which as well provides a large contribution to the cosmological constant value through the corresponding one-loop energy-momentum tensor. In the present manuscript we do not consider the full form of the one-loop spinor action in the external gravitational field, for our purposes it was enough to restrict the calculations only by linear vierbein approximation for the corresponding connection fields. 
The obtained terms of the  effective action are represented by diagrams with one external gravitational field and some diagrams with two external gravitational fields attached to the fermion 
loop\footnote{The possible cancellation of different terms in the effective action is due the fermion's propagators in the loop, the number of the classical gravitational fields attached to the vertex does not matter in this case. }. These diagrams are vacuum and particular tadpoles contributions to the energy-momentum tensor.
By requiring these contributions to be zero for both fields togetherA and B , we end up with only one version of the B-manifold spinor action. It is found that this total spinor effective action is formally equivalent to the
Schwinger-Keldysh action for a fermionic system. We need to underline, that the cancellation between the different terms of the fermion's effective action takes place when the external
gravitational fields attached to the diagrams are the same for the all parts of the action. This is not ordinary assumption in a case of the standard Schwinger-Keldysh formalism, there all
fields, including gravitational, must be different in the case of two time's branches, see Discussion section. Therefore, requiring the cancellation of the terms in the effective action, we need to introduce a gravity coupling in the fermion action different from the usual Schwinger-Keldysh formalism set-up. 

 Having in hand the unique effective spinor's action for the A and B manifolds together, next problem we face is an introduction of the weak gravity field in the actions instead vierbeins fields. In general,
there is no clear prescription how to do it when we talk about an extended manifold. Again, we have different possibilities for the A,B spinor actions with gravity included. In a simplest variant, we can follow to the Schwinger-Keldysh approach and introduce two different weak gravity fields for the different manifolds separately. In this case in the each spinor action arise only corresponding weak gravity field, the interaction between the manifolds can arise only due an interaction similar to the Schwinger-Keldysh interaction mechanism. As mentioned above, this variant does not resolve really the cosmological constant problem, the contributions of the vacuum and tadpole contributions to the energy-momentum tensor can be weaker in this case but do not disappear.
An another variant we discuss is an variant based on the assumption that any matter's energy-momentum tensor is a source of the two weak fields appear as a linear combination, a sum in the simplest case. In the corresponding gravity action the weak field, therefore, can appear or separately for the corresponding separate manifold or as a linear combination of two weak fields in each separate Lagrangian, in that particular case the only one weak field for the both manifolds exists. In this case the vierbein field in each spinor's action is equivalent to the sum of the two weak fields and the cancellation of the corresponding terms in the spinor's one-loop effective action is precise. There are additional one-loop contributions which do are not canceling  when we consider the non-diagonal Green's function appear in the ordinary Schwinger-Keldsyh formalism, but their structure and contributions are quite different from the ordinary QFT loops.
When the weak field expansion for both manifolds consists of bare metric plus two weak fields 
from the both manifolds, i.e. only one weak field mutual for the manifolds,  some more complicated system of interacting fields in comparison to the ordinary gravity action arises. In this article we do not really explore this possibility, it does not change the main results concern the cosmological constant value. For the both variants, nevertheless, the cosmological constant is turned out to be equal to the difference of the matter densities of the manifolds.
 
 An another important observation is about an appearance of "dark" matter in the framework. In any set-up of the problem when we have two manifolds, we have A and B matter fields which do not interact directly or can  interacts through the non-diagonal propagators as in Schwinger-Keldysh formalism. In both cases the structure of the gravitational propagator, aka gravitational potential, is changing due a presence of the interactions with B-manifold and/or with
B-spinors. 
%In the Schwinger-Keldysh formalism the change is due additional corresponding spinor and gravity propagators appear in the theory, there is no B-field present in %A-manifold.  
The modification of the propagator in the general case depends on the considered variant of the theory, the simplest variants are when the B-spinor fields can be created in the A-manifold by the gravity fields, these B-spinor field modifies the propagator similarly to a dark matter influence.
%that leads to an additional change of the gravitational propagator even in comparison to the Schwinger-Keldysh variant.

 Respectively, the article is organized as following. In the Sections \ref{S2}-\ref{S4} we discuss a structure of the bare spinor action for the B-manifold and corresponding spinor propagators.
The Section \ref{S5} is about the standard introduction of the gravity in the spinor's Lagrangian, approximation we made and corresponding spinor's action in an external gravitational field obtained. 
The next Section \ref{S6} dedicated to the results of the calculation of the one-loop of effective spinor action in an external gravity fields, the calculations are present in Appendix \ref{appendix:c}.
There we explore also in which variant of the effective action the cancellation of corresponding  action's terms is happening.
The Section \ref{S7} is about general structure of the resulting action in the gravitational fields and in the Sections \ref{S8}-\ref{S9} we discuss a different possibilities of the vierbeins replacement on the
weak gravity field in the action. In the Section \ref{S10} and Section \ref{S11} the main results of the article and an overview of the formalism are discussed correspondingly.

\section{CPTM symmetry for Dirac equation }
\label{S2}

 Consider the expression for the quantized spinor field:
\beqar\label{Spin1}
&\,&\psi(x)\,= \,\psi_{A}(x)\,=\,\int\,\frac{d^3 p}{(2\pi)^{3/2}}\,\frac{1}{\sqrt{2\mathcal{E}{p}}}\,\sum_{s}\Le
a_{\bold{p}}^{s}\,u^{s}(p)\,e^{-\imath\,p x}\,+\,b_{\bold{p}}^{s\,\dag}\,v^{s}(p)\,e^{\imath\,p x}\Ra\,
\\
&\,&\{\,a_{\bold{p}}^{r}\,a_{\bold{k}}^{s \dag}\,\}\,=\,\{\,b_{\bold{p}}^{r}\,b_{\bold{k}}^{s \dag}\,\}\,=\,\delta^{3}_{p\,k}\,\delta^{r s}
,\label{Spin1001}
\eeqar
with definitions borrowed from \cite{Peskin}. The CPT transformed function has the \eq{Spin14} form
\beq\label{Spin101}
\psi_{\mathcal{CPT}}(x)\,=\,
\int\,\frac{d^3 p}{(2\pi)^{3/2}}\,\frac{1}{\sqrt{2\mathcal{E}{p}}}\,\sum_{s}\Le
b_{\bold{p}}^{s\,\dag}\,v^{s}(p)\,e^{-\imath\,p x}\,+\,a_{\bold{p}}^{s}\,u^{s}(p)\,e^{\imath\,p x}\Ra\,.
\eeq
and the last transformation we need is to define the $\mathcal{M}$ transformation for our function obtaining
the quantum field solution for the Dirac equation in the B manifold.
So, first of all, we put attention now that under the $\mathcal{CPTM}$ transform the  form of the Dirac equation is changed. Whereas the
$\mathcal{CPT}$ transformation preserves the form of the Dirac equation, the additional  $\mathcal{M}$ transformation clearly change the mass sign:
\beq\label{Spin1801}
\mathcal{CPTM}(\mathcal{L})\,=\,\mathcal{L}_{B}\,=\,\overline{\psi}_{B}\,\Le \imath\gamma^{a}(\pm\,E^{\mu}_{a})\D_{\mu}\,+\,m\,\Ra\,\psi_{B}\,.
%=\,
%\overline{\psi}_{CPTM}\,\Le \imath \gamma^{\mu}\D_{\mu}\,+\,m \Ra\,\psi_{CPTM}\,.
\eeq
In the expression an another interesting observation was made amd implemented. Whereas we require the same form of the metric tensor for both manifolds, the sign of the vierbein fields can be changed even at the zero order,
see further Section \ref{S4}. Therefore, we separate the possible forms of the Dirac equation in B manifold denoting  them as for the positive and negative vierbeins and writing 
corresponding Lagrangians as $\mathcal{L}_{1B}$ and $\mathcal{L}_{2B}$. We have then:
\beq\label{Spin1805}
\mathcal{L}_{1B}\,=\,\overline{\psi}_{B}\,\Le \imath\gamma^{\mu}\D_{\mu}\,+\,m\,\Ra\,\psi_{B}\,
%=\,
%\overline{\psi}_{CPTM}\,\Le \imath \gamma^{\mu}\D_{\mu}\,+\,m \Ra\,\psi_{CPTM}\,.
\eeq
and 
\beq\label{Spin1806}
\mathcal{L}_{2B}\,=\,-\,\overline{\psi}_{B}\,\Le \imath\gamma^{\mu}\D_{\mu}\,-\,m\,\Ra\,\psi_{B}\,.
%=\,
%\overline{\psi}_{CPTM}\,\Le \imath \gamma^{\mu}\D_{\mu}\,+\,m \Ra\,\psi_{CPTM}\,.
\eeq
Our next step is to adopt the \eq{Spin101} solution in the form suitable for the Dirac equation for the B manifold.
Consider firstly the \eq{Spin1805} form of the B-manifold Dirac Lagrangian.
For the $A$ manifold we have for the spinor classical plane wave solution
of Klein-Gordon equation 
\beq\label{Spin6}
\Psi(x)\,=\,u(p)\,e^{-\imath\,x p}\,+\,v(p)\,e^{\imath\,x p}\,,
\eeq
where
\beq\label{Spin7}
\Le \gamma^{\mu}\,p_{\mu}\,-\,m\Ra\,u(p)\,=\,0\,\,,\,\,\,\Le \gamma^{\mu}\,p_{\mu}\,+\,m\Ra\,v(p)\,=\,0\,
\eeq
are the plane wave solutions for the Dirac equation and it's conjugated. 
We see, that the difference in \eq{Spin7} solutions is in the change of the mass sign, and, therefore, for the B-manifold we can write:
\beq\label{Spin601}
\Psi_{\mathcal{CPTM}}(x)\,=\,u(p)\,e^{\imath\,x p}\,+\,v(p)\,e^{-\imath\,x p}\,
\eeq
with the same functional form of spinors preserved of course. The
$v(p)$ spinor here is a plane wave solution of the  Dirac equation obtained from \eq{Spin1805} Lagrangian and $u(p)$ is it's conjugated.
From this point of view, the \eq{Spin101} function has a correct expansion in terms of the properly defined plane waves of B-manifold, we write then:
\beq\label{Spin10101}
\psi_{1B}(x)\,=\,
\int\,\frac{d^3 p}{(2\pi)^{3/2}}\,\frac{1}{\sqrt{2\mathcal{E}{p}}}\,\sum_{s}\Le
\Le \hat{\mathcal{M }}b_{\bold{p}}^{s\,\dag}\Ra\,v^{s}(p)\,e^{-\imath\,p x}\,+\,\Le \hat{\mathcal{M }}a_{\bold{p}}^{s}\Ra\,u^{s}(p)\,e^{\imath\,p x}\Ra\,,
\eeq
with $\hat{\mathcal{M }}$ operator which determines the corresponding change of $b_{\bold{p}}^{s}$ and $a_{\bold{p}}^{s}$ operators under the $\mathcal{M}$ transform.

 The Lagrangian \eq{Spin1806} is a replica of the A-manifold Lagrangian, therefore for it we can use the \eq{Spin1} classical filed expression with, of course, redefined operators of creation and annihilation:
\beq\label{Spin1010101}
\psi(x)_{2B}\,=\,\int\,\frac{d^3 p}{(2\pi)^{3/2}}\,\frac{1}{\sqrt{2\mathcal{E}{p}}}\,\sum_{s}\Le
\Le \hat{\mathcal{M }}a_{\bold{p}}^{s}\Ra\,u^{s}(p)\,e^{-\imath\,p x}\,+\,\Le \hat{\mathcal{M }}b_{\bold{p}}^{s\,\dag}\Ra\,v^{s}(p)\,e^{\imath\,p x}\Ra\,.
\eeq
In both cases, in order to define the anti-commutation relations for the new operators, we need to define some additional conditions to satisfy. This problem we consider in the next Section.

\section{Spinor field of B-manifold: first variant of quantization for the Lagrangian with positive vierbein  }
\label{S22}

 Now we will use the obtained results in order to introduce the commutation relations for the new operators of $\psi_{1B}$ field and consequent calculation of it's propagator.
The condition we require is that the energy-momentum tensor for the both manifold will equal to zero at the vacuum state, so we define:
\beq\label{Sec1}
\left\{ 
\begin{array}{c}
\hat{\mathcal{M }} a_{\bold{p}}^{s} \,=\,c_{\bold{p}}^{s\dag}\,\,\leftrightarrow\,a_{\bold{p}}^{s\dag} \,\\
%a_{\bold{p}}^{s}\,\,\leftrightarrow\,c_{\bold{p}}^{s\dag}\,\\
\hat{\mathcal{M }} \,b_{\bold{p}}^{s} \,=\,d_{\bold{p}}^{s\dag}\,\leftrightarrow\,b_{\bold{p}}^{s \dag}\,
\end{array}
\right.\,\rightarrow\,
\left\{ 
\begin{array}{c}
\{\,a_{\bold{p}}^{r}\,a_{\bold{k}}^{s \dag}\,\}\,\rightarrow\,\{\,c_{\bold{p}}^{r}\,c_{\bold{k}}^{s \dag}\,\}\,=\,-\,\delta^{3}_{p\,k}\,\delta^{r s}\,,\,\,\,
c_{\bold{p}}^{r}|0>\,=\,0\,\\
\{\,b_{\bold{p}}^{r}\,b_{\bold{k}}^{s \dag}\,\}\,\rightarrow\,\{\,d_{\bold{p}}^{r}\,d_{\bold{k}}^{s \dag}\,\}\,=\,-\,\delta^{3}_{p\,k}\,\delta^{r s}\,,\,\,\,
d_{\bold{p}}^{r}|0>\,=\,0\,
\end{array}
\right.\,
\eeq
with
\beq\label{Sec2}
\psi_{B}(x)\, = \,
\int\,\frac{d^3 p}{(2\pi)^{3/2}}\,\frac{1}{\sqrt{2\mathcal{E}{p}}}\,\sum_{s}\Le
d_{\bold{p}}^{s}\,v^{s}(p)\,e^{-\imath\,p x}\,+\,c_{\bold{p}}^{s\dag}\,u^{s}(p)\,e^{\imath\,p x}\Ra\,.
\eeq
Correspondingly considering the mutual vacuum\footnote{We do not discuss here the construction of this state, in feneral it can be also considered as superposition of two vacuum states for the A and B manifolds operatros separately. } state for the creation and annihilation operators in both manifolds we have:
\beq\label{Sec3}
\left\{ 
\begin{array}{c}
a_{\bold{p}}^{s}\,|0>\,=\,b_{\bold{p}}^{s}\,|0>\,=\,0\,\\
<0|\,a_{\bold{p}}^{s\dag}\,=\,<0|b_{\bold{p}}^{s\dag}\,=\,0\,
\end{array}
\right.\,\stackrel{CPTM}{\leftrightarrow}\,
\left\{ 
\begin{array}{c}
<0|\,c_{\bold{p}}^{s\dag}\,=\,<0|d_{\bold{p}}^{s\dag}\,=\,0\,\\
c_{\bold{p}}^{s}\,|0>\,=\,d_{\bold{p}}^{s}\,|0>\,=\,0\,.
\end{array}
\right.
\eeq
The energy-momentum vector in turn acquires the following form:
\beqar\label{Sec4}
P^{\mu}\,& = &\,P^{\mu}_{A}\,+\,P^{\mu}_{B}\,=\,\sum_{s}\,\int\,d^3 k\,k^{\mu} 
\Le
a_{\bold{p}}^{s\dag}a_{\bold{p}}^{s}\,-\,
b_{\bold{p}}^{s}b_{\bold{p}}^{s\dag}\,+\,
d_{\bold{p}}^{s\dag}d_{\bold{p}}^{s}\,-\,
c_{\bold{p}}^{s}c_{\bold{p}}^{s\dag}
\Ra\,=\,
\nonumber \\
&=&\,
\sum_{s}\,\int\,d^3 k\,k^{\mu} 
\Le
a_{\bold{p}}^{s\dag}a_{\bold{p}}^{s}\,+\,
b_{\bold{p}}^{s\dag}b_{\bold{p}}^{s}\,+\,
d_{\bold{p}}^{s\dag}d_{\bold{p}}^{s}\,+\,
c_{\bold{p}}^{s\dag}c_{\bold{p}}^{s}
\Ra\,
%=\,P^{\mu}_{A}\,+\,CPTM(P^{\mu}_{A})\,=\,P^{\mu}_{B}\,+\,CPTM(P^{\mu}_{B})\,
\eeqar
that provides with the use of \eq{Sec1}
\beq\label{Sec5}
<0|\,P^{\mu}\,|0>\,=\,0\,.
\eeq
From the \eq{Sec1} formal definition we see that we can define the 
$\mathcal{M}$ transform as following:
\beq\label{Sec5001}
\hat{\mathcal{M }} a_{\bold{p}}^{s} \,=\,\eta a_{\bold{p}}^{s} \,,\,\,\,\hat{\mathcal{M }} a_{\bold{p}}^{s\dag} \,=\,\eta^{*} a_{\bold{p}}^{s\dag}\,,\,\,\,
\eta^{2}\,=\,\eta^{2*}\,=1\,,\,\,\,\eta\,\eta^{*}\,=\,-1\,.
\eeq
It provides in turn
\beq\label{Sec5002}
\{\,a_{\bold{p}}^{r}\,a_{\bold{k}}^{s \dag}\,\}\,\rightarrow\,\{\,c_{\bold{p}}^{r}\,c_{\bold{k}}^{s \dag}\,\}\,=\,\eta\,\eta^{*}\,
\{\,a_{\bold{p}}^{r}\,a_{\bold{k}}^{s \dag}\,\}\,=\,-\,\delta^{3}_{p\,k}\,\delta^{r s}\,
\eeq
as defined in \eq{Sec1}.

 The next we define it is a Feynman propagator for the B-manifold. We begin from 
the Feynman propagator of $A$ manifold which is defined as usual:
\beqar\label{Sec8}
S_{F}(x-y)\, & = & \,-\imath\,<0|\mathbf{T} \,\Le \psi(x) \overline{\psi}(y)\Ra |0>\,=\,
\nonumber \\
&=&\,
-\imath\,\Le \theta(x^{0}-y^{0})\,<0|\psi(x) \overline{\psi}(y) |0>\,-\,\theta(y^{0}-x^{0})\,<0|\overline{\psi}(y) \psi(x) |0>\Ra\,
\eeqar
where the Wightman functions for the fermion fields of the $A$ manifold can be defined through
\beqar\label{Sec6}
D_{1}(x-y)\,& = &\,<0|\psi(x) \overline{\psi}(y) |0>\,=\,\Le \imath \hat{\D}_{x}\,+\,m\Ra\,\int \frac{d^3 p}{(2\pi)^{3}}\,\frac{1}{2\mathcal{E}{p}}\,e^{-\imath p (x-y)} \,=\,\nonumber \\
&=&
\Le \imath \hat{\D}_{x}\,+\,m\Ra\,D(x-y) \,;\\
D_{2}(x-y)\,& = &\,<0|\overline{\psi}(y) \psi(x) |0>\,=\,-\,\Le \imath \hat{\D}_{x}\,+\,m\Ra\,\int \frac{d^3 p}{(2\pi)^{3}}\,\frac{1}{2\mathcal{E}{p}}\,e^{\imath p (x-y)}\,=\, \nonumber \\
&=&
-\,\Le \imath \hat{\D}_{x}\,+\,m\Ra\,\,D(y-x) \,
\label{Sec7}
\eeqar
with $D(x-y)$  are  Wightman function of a scalar field.
Consequently, using corresponding definition of the propagator for $B$ manifold with \eq{Sec2} spinor fields and \eq{Sec1} anti-commutation relations, we obtain
for the Feynman Green's function of the second manifold:
\beq\label{Sec11}
\tilde{S}_{F}(x-y)\,=\,CPTM(S_{F}(x-y))\,=\,-\,\Le \imath\hat{\D_{x}}\,-\,m \Ra\,\int\,\frac{d^{4}k}{(2\pi)^{4}}\,\frac{e^{-\imath\,k\,(x-y)}}{k^2\,-\,m^2\,+\,\imath\varepsilon }\,
\eeq
similarly to scalar field results obtained in \cite{Ser2,Ser3}, an additional minus here is due the \eq{Sec1} commutation relations.

\section{Spinor field of B-manifold: second variant of quantization for the Lagrangian with positive vierbein   }
\label{S3}

 We again consider \eq{Sec2} function
\beq\label{Sect1}
\psi_{B}(x)\, = \,
\int\,\frac{d^3 p}{(2\pi)^{3/2}}\,\frac{1}{\sqrt{2\mathcal{E}{p}}}\,\sum_{s}\Le
d_{\bold{p}}^{s}\,v^{s}(p)\,e^{-\imath\,p x}\,+\,c_{\bold{p}}^{s\dag}\,u^{s}(p)\,e^{\imath\,p x}\Ra\,.
\eeq
but change the properties of the operators, we define this time: 
\beq\label{Sect2}
\left\{ 
\begin{array}{c}
\hat{\mathcal{M}}\,a_{\bold{p}}^{s} \,=\,c_{\bold{p}}^{s\dag}\,\,\leftrightarrow\,a_{\bold{p}}^{s\dag} \,\\
%a_{\bold{p}}^{s}\,\,\leftrightarrow\,c_{\bold{p}}^{s\dag}\,\\
\hat{\mathcal{M}}\,b_{\bold{p}}^{s} \,=\,d_{\bold{p}}^{s\dag}\,\leftrightarrow\,b_{\bold{p}}^{s \dag}\,
\end{array}
\right.\,\rightarrow\,
\left\{ 
\begin{array}{c}
\{\,a_{\bold{p}}^{r}\,a_{\bold{k}}^{s \dag}\,\}\,\rightarrow\,\{\,c_{\bold{p}}^{r}\,c_{\bold{k}}^{s \dag}\,\}\,=\,\delta^{3}_{p\,k}\,\delta^{r s}\,,\,\,\,
<0| c_{\bold{p}}^{r}\,=\,0\,\\
\{\,b_{\bold{p}}^{r}\,b_{\bold{k}}^{s \dag}\,\}\,\rightarrow\,\{\,d_{\bold{p}}^{r}\,d_{\bold{k}}^{s \dag}\,\}\,=\,\delta^{3}_{p\,k}\,\delta^{r s}\,,\,\,\,
<0| d_{\bold{p}}^{r}\,=\,0\,.
\end{array}
\right.\,
\eeq
Similarly to done before, we have for the
energy-momentum vector in this case the following expression:
\beqar\label{Sect3}
P^{\mu}\,& = &\,P^{\mu}_{A}\,+\,P^{\mu}_{B}\,=\,\sum_{s}\,\int\,d^3 k\,k^{\mu} 
\Le
a_{\bold{p}}^{s\dag}a_{\bold{p}}^{s}\,-\,
b_{\bold{p}}^{s}b_{\bold{p}}^{s\dag}\,+\,
d_{\bold{p}}^{s\dag}d_{\bold{p}}^{s}\,-\,
c_{\bold{p}}^{s}c_{\bold{p}}^{s\dag}
\Ra\,=\,
\nonumber \\
&=&\,
\sum_{s}\,\int\,d^3 k\,k^{\mu} 
\Le
a_{\bold{p}}^{s\dag}a_{\bold{p}}^{s}\,+\,
b_{\bold{p}}^{s\dag}b_{\bold{p}}^{s}\,-\,
d_{\bold{p}}^{s}d_{\bold{p}}^{s\dag}\,-\,
c_{\bold{p}}^{s}c_{\bold{p}}^{s\dag}
\Ra\,
%=\,P^{\mu}_{A}\,+\,CPTM(P^{\mu}_{A})\,=\,P^{\mu}_{B}\,+\,CPTM(P^{\mu}_{B})\,
\eeqar
that provides 
\beq\label{Sect7}
<0|\,P^{\mu}\,|0>\,=\,0\,
\eeq
as expected.
Now, using again  \eq{Sec8} definition of the propagator together with \eq{Sec1} definition of the spinor field, we obtain in turn
\beq\label{Sect8}
\tilde{S}_{F}(x-y)\,=\,\Le \imath\hat{\D_{x}}\,-\,m \Ra\,\int\,\frac{d^{4}k}{(2\pi)^{4}}\,\frac{e^{-\imath\,k\,(x-y)}}{k^2\,-\,m^2\,-\,\imath\varepsilon }\,
\eeq
as a second possibility for the form of the Feynman propagator in the $B$ manifold.

\section{Spinor field of B-manifold: quantization for the Lagrangian with negative vierbein }
\label{S4}

 Now we consider the B-manifold fermion as a precise replica of the A-manifol quantum field taking
\beq\label{TS1}
\psi(x)_{2B}\,=\,\int\,\frac{d^3 p}{(2\pi)^{3/2}}\,\frac{1}{\sqrt{2\mathcal{E}{p}}}\,\sum_{s}\Le
\Le \hat{\mathcal{M }}a_{\bold{p}}^{s}\Ra\,u^{s}(p)\,e^{-\imath\,p x}\,+\,\Le \hat{\mathcal{M }}b_{\bold{p}}^{s\,\dag}\Ra\,v^{s}(p)\,e^{\imath\,p x}\Ra\,.
\eeq
For the simplest possible variant 
\beq\label{TS2}
\left\{ 
\begin{array}{c}
\hat{\mathcal{M}}\,a_{\bold{p}}^{s} \,=\,c_{\bold{p}}^{s}\,\,\leftrightarrow\,a_{\bold{p}}^{s} \,\\
%a_{\bold{p}}^{s}\,\,\leftrightarrow\,c_{\bold{p}}^{s\dag}\,\\
\hat{\mathcal{M}}\,b_{\bold{p}}^{s} \,=\,d_{\bold{p}}^{s}\,\leftrightarrow\,b_{\bold{p}}^{s}\,
\end{array}
\right.\,\rightarrow\,
\left\{ 
\begin{array}{c}
\{\,a_{\bold{p}}^{r}\,a_{\bold{k}}^{s \dag}\,\}\,\rightarrow\,\{\,c_{\bold{p}}^{r}\,c_{\bold{k}}^{s \dag}\,\}\,=\,\delta^{3}_{p\,k}\,\delta^{r s}\,,\,\,\,
c_{\bold{p}}^{r}|0>\,=\,0\,\\
\{\,b_{\bold{p}}^{r}\,b_{\bold{k}}^{s \dag}\,\}\,\rightarrow\,\{\,d_{\bold{p}}^{r}\,d_{\bold{k}}^{s \dag}\,\}\,=\,\delta^{3}_{p\,k}\,\delta^{r s}\,,\,\,\,
    d_{\bold{p}}^{r}|0>\,=\,0\,
\end{array}
\right.\,
\eeq
we, of course will arrive to the 
\beqar\label{Sec201}
P^{\mu}\,& = &\,P^{\mu}_{A}\,+\,P^{\mu}_{B}\,=\,\sum_{s}\,\int\,d^3 k\,k^{\mu} 
\Le
a_{\bold{p}}^{s\dag}a_{\bold{p}}^{s}\,-\,
b_{\bold{p}}^{s}b_{\bold{p}}^{s\dag}\,-\,
c_{\bold{p}}^{s\dag}c_{\bold{p}}^{s}\,+\,
d_{\bold{p}}^{s}d_{\bold{p}}^{s\dag}\,
\Ra\,=\,
\nonumber \\
&=&\,
\sum_{s}\,\int\,d^3 k\,k^{\mu} 
\Le
a_{\bold{p}}^{s\dag}a_{\bold{p}}^{s}\,+\,
b_{\bold{p}}^{s\dag}b_{\bold{p}}^{s}\,-\,
c_{\bold{p}}^{s\dag}c_{\bold{p}}^{s}\,-\,
d_{\bold{p}}^{s\dag}d_{\bold{p}}^{s}
\Ra\,
%=\,P^{\mu}_{A}\,+\,CPTM(P^{\mu}_{A})\,=\,P^{\mu}_{B}\,+\,CPTM(P^{\mu}_{B})\,
\eeqar
here the additional minus in front of the Lagrangian of B-manifold fermion, see \eq{Spin1806}, is accounted.
Correspondingly we have
\beq\label{TS3}
<0|\,P^{\mu}\,|0>\,=\,0\,
\eeq
as it must be for the case. 
The propagator here is the usual, Feynman, one:
\beq\label{TS4}
\tilde{S}_{F}(x-y)\,=\,\Le \imath\hat{\D_{x}}\,+\,m \Ra\,\int\,\frac{d^{4}k}{(2\pi)^{4}}\,\frac{e^{-\imath\,k\,(x-y)}}{k^2\,-\,m^2\,+\,\imath\varepsilon }\,
\eeq
see Appendix \ref{appendix:b}.

 There is an additional possibilities to define the B-manifold quantization which is similar to the considered previously, this time we define:
\beq\label{TS5}
\left\{ 
\begin{array}{c}
\hat{\mathcal{M}}\,a_{\bold{p}}^{s} \,=\,c_{\bold{p}}^{s}\,\,\leftrightarrow\,a_{\bold{p}}^{s} \,\\
%a_{\bold{p}}^{s}\,\,\leftrightarrow\,c_{\bold{p}}^{s\dag}\,\\
\hat{\mathcal{M}}\,b_{\bold{p}}^{s} \,=\,d_{\bold{p}}^{s}\,\leftrightarrow\,b_{\bold{p}}^{s}\,
\end{array}
\right.\,\rightarrow\,
\left\{ 
\begin{array}{c}
\{\,a_{\bold{p}}^{r}\,a_{\bold{k}}^{s \dag}\,\}\,\rightarrow\,\{\,c_{\bold{p}}^{r}\,c_{\bold{k}}^{s \dag}\,\}\,=\,-\,\delta^{3}_{p\,k}\,\delta^{r s}\,,\,\,\,
<0| c_{\bold{p}}^{r}\,=\,0\,\\
\{\,b_{\bold{p}}^{r}\,b_{\bold{k}}^{s \dag}\,\}\,\rightarrow\,\{\,d_{\bold{p}}^{r}\,d_{\bold{k}}^{s \dag}\,\}\,=\,-\,\delta^{3}_{p\,k}\,\delta^{r s}\,,\,\,\,
   <0|d_{\bold{p}}^{r}\,=\,0\,
\end{array}
\right.\,
\eeq
The energy-momentum vector in this case acquires the following form:
\beqar\label{TS6}
P^{\mu}\,& = &\,P^{\mu}_{A}\,+\,P^{\mu}_{B}\,=\,\sum_{s}\,\int\,d^3 k\,k^{\mu} 
\Le
a_{\bold{p}}^{s\dag}a_{\bold{p}}^{s}\,-\,
b_{\bold{p}}^{s}b_{\bold{p}}^{s\dag}\,-\,
c_{\bold{p}}^{s\dag}c_{\bold{p}}^{s}\,+\,
d_{\bold{p}}^{s}d_{\bold{p}}^{s\dag}\,
\Ra\,=\,
\nonumber \\
&=&\,
\sum_{s}\,\int\,d^3 k\,k^{\mu} 
\Le
a_{\bold{p}}^{s\dag}a_{\bold{p}}^{s}\,+\,
b_{\bold{p}}^{s\dag}b_{\bold{p}}^{s}\,+\,
c_{\bold{p}}^{s}c_{\bold{p}}^{s\dag}\,+\,
d_{\bold{p}}^{s}d_{\bold{p}}^{s\dag}
\Ra\,
%=\,P^{\mu}_{A}\,+\,CPTM(P^{\mu}_{A})\,=\,P^{\mu}_{B}\,+\,CPTM(P^{\mu}_{B})\,
\eeqar
that also provides
\beq\label{TS7}
<0|\,P^{\mu}\,|0>\,=\,0\,.
\eeq
Correspondingly, for the quantum field
\beq\label{TS8}
\psi(x)_{2B}\,=\,\int\,\frac{d^3 p}{(2\pi)^{3/2}}\,\frac{1}{\sqrt{2\mathcal{E}{p}}}\,\sum_{s}\Le
c_{\bold{p}}^{s}\,u^{s}(p)\,e^{-\imath\,p x}\,+\,d_{\bold{p}}^{s\,\dag}\,v^{s}(p)\,e^{\imath\,p x}\Ra\,.
\eeq
and \eq{TS5} definition of the operators properties, 
the propagator is the same as in \eq{Sect8} with only $m\,\rightarrow\,-m$ replace, we have:
\beq\label{TS9}
\tilde{S}_{F}(x-y)\,=\,-\,\Le \imath\hat{\D_{x}}\,+\,m \Ra\,\int\,\frac{d^{4}k}{(2\pi)^{4}}\,\frac{e^{-\imath\,k\,(x-y)}}{k^2\,-\,m^2\,-\,\imath\varepsilon }\,.
\eeq
The additional minus here is due the initial minus we accounted in \eq{Spin1806},  see the minus also in anti-commutator of \eq{TS5}.
This is  Dyson propagator as it defined by \eq{A5}. We see, that this particular choice of the operators relations determines an analog of the Keldysh approach 
in the non-equilibrium condensed matter theory.

\section{Actions for the spinors in an external gravitational field}
\label{S5}

 In this article we do not discuss Einstein-Cartan gravity Lagrangian and action, see \cite{Krasnov} for the examples of the Einstein-Cartan gravity formulation. The only assumption we made is that the gravity Lagrangian describes a torsionless relativity theory equivalent to the Einsten general relativity. The torsionless connections is defined then through the vierbein fields:
\beq\label{FG1}
\omega_{c a b}\,=\,\Le C_{c a b } - C_{a b c }+ C_{b c a} \Ra
\eeq
where
\beq\label{FG2}
C_{a b c }\,=\,E^{\mu}_{a} E^{\nu}_{b}\,\Le \D_{\mu} e_{\nu c}\,-\,\D_{\nu} e_{\mu c}\Ra\,=\,E^{\mu}_{a} E^{\nu}_{b}\,\D_{[\mu} e_{\nu] c}\,.
\eeq
The vierbein fields here are defined as usual, we have:
\beq\label{FG3}
e_{\mu}^{a}\,E^{\mu}_{b}\,=\,\delta^{a}_{b}\,;\,\,\, g_{\mu \nu}\,=\,\eta_{a b}\,e_{\mu}^{a}  e_{\nu}^{b}\,,  
\eeq
with $\eta_{ab}$ as some classical solution for the metric in the case of a non-flat space-time, further for sake of simplicity we will take $\eta_{ab}$ as the Minkowski metric.
Some truncation scheme is requested in this description when a perturbative solution for the vierbein is considered. Namely, in general in such kind of problem we define the 
vierbein field as some perturbative series:
\beq\label{FG4}
e_{\mu}^{a}\,=\,\sum_{k=0}^{\infty}\,e_{k\,\mu}^{a}\,,\,\,\,\eta_{a b}\,e_{0\,\mu}^{a}\,e_{0\,\mu}^{b}\,=\,\eta_{\mu \nu}\,.
\eeq
The $e_{1,\mu}^{a}$ vierbein in the series can be found solving equations of motion of a coresponding part of the full gravity Lagrangian quadratic with respect to $e_{1}$ without or with the
energy-momentum tensor in the r.h.s. of the equations, i.e. from the corresponding wave equation with or without source. 
%In first case the $e_{1}$ vierbein will describe the linearized free solution of the Einstein-Cartan
%gravity in any formulation, it will not be depend on the fermion fields\footnote{We always can use a freedom to choose an appropriate r.h.s. of the gravity classical equations. In general, the Dirac 
%action includes the first order degree $e_{1}$ vierbein term, therefore the corresponding energy-momentum tensor could be accounted or in the equations of motion for $e_{1}$ vierbein or further in respect %to the perturbative scheme, in our case we can correspondingly consider the  $e_{1}$ order vierbein as a graviton or as a gravitational wave from known source plus graviton. }. The next order vierbein,% $e_{2}$, will account the $e_{1}^{3}$ terms of the coresponding order Lagrangian and a corresponding fermion action contribution included in the right hand side of the vierbein's equations of motion. 
%Namely, in this perturbative scheme, the dependence of the vierbein as a function of the spinor fields appears the first time in this perturbative order, i.e. at this order we will have %$e_{2}\,=\,f(e_{0},\,e_{1},\,\psi)$ and so further. 
%Consequently, it will produce additional induced vertices in the fermion action and step by step implementation of this scheme of course will produce an infinite tower of the induced vertices in the %Dirac's Lagrangian. 
For our aims it is enough to account only two first orders in \eq{FG4}, we will take
\beq\label{FG5}
e_{\mu}^{a}\,=\,e_{0\,\mu}^{a}\,+\,e_{1\,\mu}^{a}\,
\eeq
with the $e_{1\,\mu}^{a}$ vierbein as Einstein-Cartan gravity counterpart of the usual graviton field. For the B-manifold, correspondingly, we require that the metric tensor will remain the same
%\footnote{The same metric will remain also when we have $e_{B\,\mu}^{a}\,=\,-\,e_{\mu}^{a}$ idenity. In this case we will have $-\delta^{a}_{\mu}$ delta already in the zero order in respect to gravitation %Dirac equation that redefines $\tilde{S}_{F}$ to $\,-S_{F}$ or $S_\,{D}$ defined in Appendix\ref{App1}, first of all, and with additional minus from this zero order vierbein we will arrive to the same %form% of free Dirac equation. Secondly, this redefinition of the B-vierbein will change the sign of the
% additional parts of the Dirac equation responsible for the interactions with the gravitational fields. So we will get some replica of the A-manifold Dirac equation with the
%interactions part with changed sign. } 
and define
% It means the 
%same flat geometry for both manifolds and
%undefined sign for the B-manifold vierbein fields:
\beq\label{FG51}
\,e_{B1\,\mu}^{a}\,=\,\pm\,\Le e_{0\,\mu}^{a}\,+\,e_{1\,\mu}^{a}\Ra\,,\,\,\,E_{B1\,a}^{\mu}\,=\,\pm\,\Le E_{0\,a}^{\mu}\,-\,e_{1\,a}^{\mu}\Ra\,.
\eeq
As discussed above, there are two possible vierbein's signs are possible here in general.

 The main purposes of the approach is an investigation of the model of two manifolds which interact by gravity. 
%For our aims it is enough to account the two orders of 
%a vierbein's expansion. 
Similarly to done in \cite{Ser2,Ser3}, firstly we assume that the $e_{1\,\mu}^{a}$ is the same  for the both Dirac-Einstein-Cartan systems in A and B manifolds, till the sign  of the vierbein.
%Also, we do not consider also the Schwinger-Keldysh like formulation of the interacting fermion systems, see for examples \cite{Kel1}, this problem we will discuss %in an additional publication.
Therefore, we have for the action of the interacting system of fermions:
\beqar\label{FG6}
S\,& = &\,S_{A}\,+\,S_{B}\,=\,\int d^4 x_{A}\,e_{A}\,\overline{\psi}_{A}\Le\,\imath\,E^{\mu}_{c}\,\gamma^{c}\,D_{\mu}\,-\,m\,\Ra\,\psi_{A}\,\pm\,
\nonumber \\
&\pm&
\int d^4 x_{B}\,e_{B}\,\overline{\psi}_{B}\,\Le\,\imath\,E^{\mu}_{c}\,\gamma^{c}\,D_{\mu}^{\pm}\,\pm\,m\,\Ra\,\psi_{B}\,,
\eeqar
with
\beq\label{FG7}
e_{A,B}\,=\,\sqrt{-g(e_{A,B})}\,
\eeq
and
\beq\label{FG8}
D_{\mu}^{\pm}\,=\,\D_{\mu}\,+\,\frac{1}{8}\,\omega_{\mu a b}^{\pm}\,[\gamma^{a} \gamma^{b}]\,,
\eeq
the subscripts $A$ and $B$ denote the corresponding manifolds, further we will use these indexes only in the need of special clarification. The subscripts $\pm$, in turn, denotes
the terms with positive or negative sign of B-manifold vierbeins accounted.
% Assuming the same initial geometries for the both manifolds we will obtain the same covariant derivatives for the both parts of the action. Now, taking into account the additional minus sign in both %\eq{Sec11} and \eq{Sect8}, we can write the same action as  
%\beq\label{FG9}
%S\,= \,S_{A}\,+\,S_{B}\,=\,\int d^4 x_{A}\,e_{A}\,\overline{\psi}_{A}\,G_{F}^{-1}(e,E)\,\psi_{A}\,-\,
%\int d^4 x_{B}\,e_{B}\,\overline{\psi}_{B}\,\tilde{G}_{F}^{-1}(e,E)\psi_{B}\,,
%\eeq
%with $\tilde{G}_{F}\,\rightarrow\,-\,\tilde{G}_{F}$ replace performed in the second term of the action. 
Next we write the positive linearized vierbein fields for A manifold as
\beq\label{FG10}
e_{\mu}^{a}\,=\,\delta_{\mu}^{a}\,+\,e_{1\,\mu}^{a}\,,\,\,\,E_{a}^{\mu}\,=\,\delta^{\mu}_{a}\,-\,\eta^{\mu \nu}\,\eta_{a b}\,e_{1\,\nu}^{b}
\eeq
with $\eta^{\mu \nu}$ as Minkowski metric. The spinor field we write in the form of a fluctuation around classical solution taken as background field:
\beq\label{FG11}
\psi_{A,B}\,\rightarrow\,\psi_{A,B\,cl}\,+\,\chi_{A,B}
\eeq
with $\psi_{A,B\,cl}$ in the r.h.s. as solution of the classical Dirac equation without gravity present for $A$ or $B$ manifold. Further we will omit the $\psi_{A,B\,cl}$ notations
writing these fields simply as $\psi_{A,B}$.
%for the case of $B$ these fields are given by \eq{Sec2} or by \eq{Sect1}.
%We note, that whereas the spinors are considered as quantum fields, for the gravity we stay in the paradigm of the classical solution, there is no quantum gravity in this set-up. 
Preserving in the connections fields only terms linear with respect to the vierbeins, we obtain for the \eq{FG1} expression:
\beq\label{FG12}
\omega_{c a b}\,=\,\D_{c}\,\Le e_{1\,a b} \,-\,e_{1\,b a} \Ra\,-\,\D_{a}\,\Le e_{1\,c b} \,+\,e_{1\,b c} \Ra\,+\,\D_{b}\,\Le e_{1\,c a} \,+\,e_{1\,a c } \Ra\,.
\eeq
Further in the calculations the only symmetrical vierbein combination is considered:
\beq\label{FG13}
s_{c b} \,=\, \frac{1}{2}\,\Le\,e_{1\,c b} \,+\,e_{1\,b c}\Ra\,,\,\,\,s_{c b} \,=\,s_{b c}
\eeq
that corresponds to the some chosen form of an external gravitational field of course, see also \cite{FLoop} for example.
Now we note, that considering $PT$ transformation of the \eq{FG1}-\eq{FG2} we obtain that for the case of the positive vierbein's transformation when  $\Le e,E\Ra_{A}\,\rightarrow\,\Le e,E\Ra_{B}$, the connection changes sign:
due the   $\D\,\rightarrow\,-\,\D$ PT transform:
\beq\label{FG1201}
\omega_{c a b}^{+}\,\rightarrow\,-\,\omega_{c a b}\,,\,\,\,D_{\mu}^{+}\,=\,\D_{\mu}\,-\,\frac{1}{8}\,\omega_{\mu a b}\,[\gamma^{a} \gamma^{b}]\,
\eeq
with $\omega$ provided by \eq{FG12} of course.
%But due the pairing of the connection's indexes with $[\gamma^{a} \gamma^{b}]$, see further, this change of sign does not affect on the overall sign in the expressions.
Whereas we have the negative vierbein's transformation, $\Le e,E\Ra_{A}\,\rightarrow\,-\,\Le e,E\Ra_{B}$, the connection does not change the sign
\beq\label{FG120101}
\omega_{c a b}^{-}\,\rightarrow\,\omega_{c a b}\,,\,\,\,D_{\mu}^{-}\,=\,\D_{\mu}\,+\,\frac{1}{8}\,\omega_{\mu a b}\,[\gamma^{a} \gamma^{b}]\,.
\eeq
These transformation properties we account in the corresponding Lagrangians.
%We note, that at the case of vierbein's sigh change the sign of the graviton field $s$ must change sign as well. Nevertheless, for sake of simplicity, we do not account the sign 
%of the graviton in general expressions further writing the correct sign only when we precisely consider negative sign vierbeins. 
Thereafter we define:
\beq\label{FG14}
\omega_{c a b}\,=\,\D_{b}\,\Le e_{1\,c a} \,+\,e_{1\,a c } \Ra\,-\,\D_{a}\,\Le e_{1\,c b}+\,e_{1\,b c} \Ra \,=\,2\Le \D_{b} s_{c a}\,-\,\D_{a}s_{c b}\Ra\,.
\eeq
For the \eq{FG8} covariant derivative part we will have correspondingly:
\beq\label{FG15}
\imath\,\frac{1}{8}\,\omega_{\mu a b}\,\gamma^{c}\,[\gamma^{a} \gamma^{b}]\,=\,\frac{\imath}{2}\,\Le \D_{b} s_{c a}\,-\,\D_{a}s_{c b}\Ra\,\gamma^{c}\,\gamma^{a}\, \gamma^{b}
\eeq
Due the simplest perturbative scheme we consider, further we will take:
\beq\label{FG151}
\sqrt{-g}\,\approx\,1\,+\,\frac{1}{2}\,\eta^{\mu \nu}\,h_{\mu \nu}\,=\,1\,+\,\eta^{a b}\,s_{a b}\,=\,1\,+\,s\,.
\eeq
Collecting all terms in the background expansion of the fields and accounting classical equation of motion, we have for the whole action:
\beq\label{FG16}
S\,= \,S_{A}\,+\,S_{B}^{\pm}\,=\,S_{0 A}\,+\,S_{1 A}\,+\,S_{0 B}^{\pm}\,+\,S_{1B}^{\pm}\,,
\eeq
with $S_{B}^{+}$ and $S_{B}^{-}$ as Lagrangians defined for the B-manifold positive and negative vierbeins correspondingly, with all corresponding changes of signs accounted.
The \eq{FG16} terms, therefore, have the following from:
%that to the first perturbative order with respect to the external vierbeines provides:
\beqar\label{FG17}
S_{A}\,& = & \,\int d^4 x_{A}\,\overline{\chi}_{A}\Le\,\imath\,\gamma^{\mu}\,\D_{\mu}\,-\,m\,\Ra\,\chi_{A}\,+\,
%\int d^4 x_{A}\,e_{A}\,\overline{\chi}_{A}\Le\,\imath\,\gamma^{\mu}\,\D_{\mu}\,-\,m\,\Ra\,\psi_{A}\,+\,
%\nonumber \\
%&+&
%\int d^4 x_{A}\,e_{A}\,\overline{\psi}_{A}\Le\,\imath\,\gamma^{\mu}\,\D_{\mu}\,-\,m\,\Ra\,\chi_{A}\,+\,
\nonumber \\
&+&\,
\int d^4 x_{A}\,\overline{\psi}_{A}\Le\,\imath\,\Le s\delta^{\,\mu}_{c}\,-\,\eta^{\mu \nu} s_{c \nu}\Ra\,\gamma^{c}\,\D_{\mu}\,+\,
\frac{\imath}{2}\,\Le \D_{b} s_{\,c a}\,-\,\D_{a}s_{\,c b}\Ra\,\gamma^{c}\,\gamma^{a}\, \gamma^{b} \,-\,m\,s\Ra\,\psi_{A}\,+\,
\nonumber \\
&+&
\int d^4 x_{A}\,\overline{\chi}_{A}\Le\,\imath\,\Le s\delta^{\,\mu}_{c}\,-\,\eta^{\mu \nu} s_{c \nu}\Ra\,\gamma^{c}\,\D_{\mu}\,+\,
\frac{\imath}{2}\,\Le \D_{b} s_{\,c a}\,-\,\D_{a}s_{\,c b}\Ra\,\gamma^{c}\,\gamma^{a}\, \gamma^{b}\,-\,m\,s\Ra\,\psi_{A}\,+\,
\nonumber \\
&+&
\int d^4 x_{A}\,\overline{\psi}_{A}\Le\,\imath\,\Le s\delta^{\,\mu}_{c}\,-\,\eta^{\mu \nu} s_{c \nu}\Ra\,\gamma^{c}\,\D_{\mu}\,+\,
\frac{\imath}{2}\,\Le \D_{b} s_{\,c a}\,-\,\D_{a}s_{\,c b}\Ra\,\gamma^{c}\,\gamma^{a}\, \gamma^{b}\,-\,m\,s\Ra\,\chi_{A}\,+\,
\nonumber \\
&+&
\int d^4 x_{A}\,\overline{\chi}_{A}\Le\,\imath\,\Le s\delta^{\,\mu}_{c}\,-\,\eta^{\mu \nu} s_{c \nu}\Ra\,\gamma^{c}\,\D_{\mu}\,+\,
\frac{\imath}{2}\,\Le \D_{b} s_{\,c a}\,-\,\D_{a}s_{\,c b}\Ra\,\gamma^{c}\,\gamma^{a}\, \gamma^{b}\,-\,m\,s\Ra\,\chi_{A}\,.
\eeqar
Correspondingly for the case of the positive B-manifold vierbein we have:
\beqar\label{FG1801}
S_{B}^{+}\,& = & \,\int d^4 x_{B}\,\overline{\chi}_{B}\Le\,\imath\,\gamma^{\mu}\,\D_{\mu}\,+\,m\,\Ra\,\chi_{B}\,+\,
%\int d^4 x_{B}\,e_{B}\,\overline{\chi}_{B}\Le\,\imath\,\gamma^{\mu}\,\D_{\mu}\,-\,m\,\Ra\,\psi_{B}\,+\,
%\nonumber \\
%&+&
%\int d^4 x_{B}\,e_{B}\,\overline{\psi}_{B}\Le\,\imath\,\gamma^{\mu}\,\D_{\mu}\,-\,m\,\Ra\,\chi_{B}\,+\,
\nonumber \\
&+&\,
\int d^4 x_{B}\,\overline{\psi}_{B}\Le\,\imath\,\Le s\delta^{\,\mu}_{c}\,-\,\eta^{\mu \nu} s_{\,c \nu}\Ra\,\gamma^{c}\,\D_{\mu}\,-\,
\frac{\imath}{2}\,\Le \D_{b} s_{\,c a}\,-\,\D_{a}s_{\,c b}\Ra\,\gamma^{c}\,\gamma^{a}\, \gamma^{b} \,+\,m\,s\Ra\,\psi_{B}\,+\,
\nonumber \\
&+&
\int d^4 x_{B}\,\overline{\chi}_{B}\Le\,\imath\,\Le s\delta^{\,\mu}_{c}\,-\,\eta^{\mu \nu} s_{\,c \nu}\Ra\,\gamma^{c}\,\D_{\mu}\,-\,
\frac{\imath}{2}\,\Le \D_{b} e_{B1\,c a}\,-\,\D_{a}e_{1\,c b}\Ra\,\gamma^{c}\,\gamma^{a}\, \gamma^{b}\,+\,m\,s\Ra\,\psi_{B}\,+\,
\nonumber \\
&+&
\int d^4 x_{B}\,\overline{\psi}_{B}\Le\,\imath\,\Le s\delta^{\,\mu}_{c}\,-\,\eta^{\mu \nu} s_{\,c \nu}\Ra\,\gamma^{c}\,\D_{\mu}\,-\,
\frac{\imath}{2}\,\Le \D_{b} e_{B1\,c a}\,-\,\D_{a}e_{1\,c b}\Ra\,\gamma^{c}\,\gamma^{a}\, \gamma^{b}\,+\,m\,s\Ra\,\chi_{B}\,+\,
\nonumber \\
&+&
\int d^4 x_{B}\,\overline{\chi}_{B}\Le\,\imath\,\Le s\delta^{\,\mu}_{c}\,-\,\eta^{\mu \nu} s_{\,c \nu}\Ra\,\gamma^{c}\,\D_{\mu}\,-\,
\frac{\imath}{2}\,\Le \D_{b} e_{B1\,c a}\,-\,\D_{a}e_{1\,c b}\Ra\,\gamma^{c}\,\gamma^{a}\,\gamma^{b}\,+\,m\,s \,\Ra\,\chi_{B}\,,
\eeqar
whereas for the negative B-manifold vierbein we obtain
\beqar\label{FG18}
S_{B}^{-}\,& = & \,-\,\int d^4 x_{B}\,\overline{\chi}_{B}\Le\,\imath\,\gamma^{\mu}\,\D_{\mu}\,-\,m\,\Ra\,\chi_{B}\,-\,
%\int d^4 x_{B}\,e_{B}\,\overline{\chi}_{B}\Le\,\imath\,\gamma^{\mu}\,\D_{\mu}\,-\,m\,\Ra\,\psi_{B}\,+\,
%\nonumber \\
%&+&
%\int d^4 x_{B}\,e_{B}\,\overline{\psi}_{B}\Le\,\imath\,\gamma^{\mu}\,\D_{\mu}\,-\,m\,\Ra\,\chi_{B}\,+\,
\nonumber \\
&-&\,
\int d^4 x_{B}\,\overline{\psi}_{B}\Le\,\imath\,\Le s\delta^{\,\mu}_{c}\,-\,\eta^{\mu \nu} s_{\,c \nu}\Ra\,\gamma^{c}\,\D_{\mu}\,+\,
\frac{\imath}{2}\,\Le \D_{b} s_{\,c a}\,-\,\D_{a}s_{\,c b}\Ra\,\gamma^{c}\,\gamma^{a}\, \gamma^{b} \,-\,m\,s\Ra\,\psi_{B}\,-\,
\nonumber \\
&-&
\int d^4 x_{B}\,\overline{\chi}_{B}\Le\,\imath\,\Le s\delta^{\,\mu}_{c}\,-\,\eta^{\mu \nu} s_{\,c \nu}\Ra\,\gamma^{c}\,\D_{\mu}\,+\,
\frac{\imath}{2}\,\Le \D_{b} e_{B1\,c a}\,-\,\D_{a}e_{1\,c b}\Ra\,\gamma^{c}\,\gamma^{a}\, \gamma^{b}\,-\,m\,s\Ra\,\psi_{B}\,-\,
\nonumber \\
&-&
\int d^4 x_{B}\,\overline{\psi}_{B}\Le\,\imath\,\Le s\delta^{\,\mu}_{c}\,-\,\eta^{\mu \nu} s_{\,c \nu}\Ra\,\gamma^{c}\,\D_{\mu}\,+\,
\frac{\imath}{2}\,\Le \D_{b} e_{B1\,c a}\,-\,\D_{a}e_{1\,c b}\Ra\,\gamma^{c}\,\gamma^{a}\, \gamma^{b}\,-\,m\,s\Ra\,\chi_{B}\,-\,
\nonumber \\
&-&
\int d^4 x_{B}\,\overline{\chi}_{B}\Le\,\imath\,\Le s\delta^{\,\mu}_{c}\,-\,\eta^{\mu \nu} s_{\,c \nu}\Ra\,\gamma^{c}\,\D_{\mu}\,+\,
\frac{\imath}{2}\,\Le \D_{b} e_{B1\,c a}\,-\,\D_{a}e_{1\,c b}\Ra\,\gamma^{c}\,\gamma^{a}\,\gamma^{b}\,-\,m\,s \,\Ra\,\chi_{B}\,=\,
\nonumber \\
&=&\,-\,S_{A}\,,
\eeqar
%here  we accounted that $m\,\sqrt{-g}\,\approx\,m\,\Le 1\,-\,s\Ra\,$ for the case of the negative vierbein's sign, we see that overall sign here is defined by expansion of teh determinant 
%and not by the change of the vierbein's sign in thta action in \eq{Spin1806}.
In general formulation, the each action has the following form therefore:
%\beqar
\beq\label{FG19}
S\,= \,\int d^4 x\,\overline{\chi}\,S_{0}\,\chi\,+\,\int d^4 x\,\overline{\chi}\,M_{1}\,\chi\,+\,
\int d^4 x\,\overline{\chi}\,J(x)\,+\,
\int d^4 x\,\overline{J}(x)\,\chi\,+\,
%\nonumber \\
%&+&
\int d^4 x\,\overline{\psi}\,M_{1}(x)\,\psi\,,
%\int d^4 x_{B}\,e_{B}\,\overline{\psi}_{B}\Le\,\imath\,E^{\mu}_{B1\,c}\,\gamma^{c}\,\D_{\mu}\,+\,
%\frac{\imath}{2}\,\Le \D_{b} e_{B1\,c a}\,-\,\D_{a}e_{1\,c b}\Ra\,\gamma^{c}\,\gamma^{a}\, \gamma^{b} \,\Ra\,\psi_{B}\Ra\,
\eeq
%\eeqar
with corresponding redefinition of the terms in the Lagrangian. 

 Now we have to account the all variants of the propagators for the B-manifold. We begin from the general expression for the whole action in the case of
the positive vierbeins, we have:
\beqar\label{FG20}
S_{B1,2}^{+}\,& = & \,\int d^4 x_{B}\,\overline{\chi}_{B}\Le\,\imath\,\gamma^{\mu}\,\D_{\mu}\,+\,m\,\Ra\,\chi_{B}\,+\,
\int d^4 x_{B}\,\overline{\chi}_{B}\,J_{1B}^{+}(x)\,+\,
\int d^4 x_{B}\,\overline{J}_{1B}^{+}(x)\,\chi_{B}\,+\,
\nonumber \\
&+&\,
\int d^4 x_{B}\,\overline{\chi}_{B}\,M_{1B}^{+}(x)\,\chi_{B}\,+\,
\int d^4 x_{B}\,\overline{\psi}_{B}\,M_{1B}^{+}(x)\,\psi_{B}\,,
%\int d^4 x_{B}\,e_{B}\,\overline{\psi}_{B}\Le\,\imath\,E^{\mu}_{B1\,c}\,\gamma^{c}\,\D_{\mu}\,+\,
%\frac{\imath}{2}\,\Le \D_{b} e_{B1\,c a}\,-\,\D_{a}e_{1\,c b}\Ra\,\gamma^{c}\,\gamma^{a}\, \gamma^{b} \,\Ra\,\psi_{B}\Ra\,
\eeqar
the notations here and further are shortened notations from the \eq{FG18}.
In the cases of the \eq{Sec11} and \eq{Sect8} propagators, we have 
\beqar\label{FG2001}
&\,&S_{F1}(x-y)\,= \,-\,\Le \imath\hat{\D_{x}}\,-\,m \Ra\,\int\,\frac{d^{4}k}{(2\pi)^{4}}\,\frac{e^{-\imath\,k\,(x-y)}}{k^2\,-\,m^2\,+\,\imath\varepsilon }\,,\nonumber \\
&\,& \Le\,\imath\,\hat{\D}_{\mu x}\,+\,m\,\Ra\,S_{F1}(x-y)\,=\,-\,\delta^{4}(x-y)\,
\eeqar
and 
\beqar\label{FG2002}
&\,& S_{F2}(x-y)\,=\,\Le \imath\hat{\D_{x}}\,-\,m \Ra\,\int\,\frac{d^{4}k}{(2\pi)^{4}}\,\frac{e^{-\imath\,k\,(x-y)}}{k^2\,-\,m^2\,-\,\imath\varepsilon }\,,
\nonumber \\
&\,&\Le\,\imath\,\hat{\D}_{\mu x}\,+\,m\,\Ra\,S_{F2}(x-y)\,=\,\delta^{4}(x-y)\,.
\eeqar
for the the $S_{B1,2}$ actions correspondingly.

 For the case of the negative vierbein's sign, we have correspondingly the following action:
\beqar\label{FG204}
S_{B3,4}^{-}\,& = & \,-\,\int d^4 x_{B}\,\overline{\chi}_{B}\Le\,\imath\,\gamma^{\mu}\,\D_{\mu}\,-\,m\,\Ra\,\chi_{B}\,-\,
\int d^4 x_{B}\,\overline{\chi}_{B}\,J_{1B}^{-}(x)\,-\,
\int d^4 x_{B}\,\overline{J}_{1B}^{-}(x)\,\chi_{B}\,-\,
\nonumber \\
&-&\,
\int d^4 x_{B}\,\overline{\chi}_{B}\,M_{1B}^{-}(x)\,\chi_{B}\,-\,
\int d^4 x_{B}\,\overline{\psi}_{B}\,M_{1B}^{-}(x)\,\psi_{B}\,,
%\int d^4 x_{B}\,e_{B}\,\overline{\psi}_{B}\Le\,\imath\,E^{\mu}_{B1\,c}\,\gamma^{c}\,\D_{\mu}\,+\,
%\frac{\imath}{2}\,\Le \D_{b} e_{B1\,c a}\,-\,\D_{a}e_{1\,c b}\Ra\,\gamma^{c}\,\gamma^{a}\, \gamma^{b} \,\Ra\,\psi_{B}\Ra\,
\eeqar
see \eq{Spin1806} definition. The first variant of the propagator is given by \eq{A5}
\beqar\label{FG2041}
&\,& S_{F3}(x-y)\, = \,
\Le \imath\hat{\D_{x}}\,+\,m \Ra\,\int\,\frac{d^{4}k}{(2\pi)^{4}}\,\frac{e^{-\imath\,k\,(x-y)}}{k^2\,-\,m^2\,+\,\imath\varepsilon }\,,\nonumber \\
&\,& \Le\,\imath\,\hat{\D}_{x}\,-\,m\,\Ra\,S_{F3}(x,y)\,=\,\delta^{4}(x-y)\,
\eeqar
whereas the second is
\beqar\label{FG2042}
&\,& S_{F4}(x-y)\, = \,-\,
\Le \imath\hat{\D_{x}}\,+\,m \Ra\,\int\,\frac{d^{4}k}{(2\pi)^{4}}\,\frac{e^{-\imath\,k\,(x-y)}}{k^2\,-\,m^2\,-\,\imath\varepsilon }\,,\nonumber \\
&\,& \Le\,\imath\,\hat{\D}_{x}\,-\,m\,\Ra\,S_{F4}(x,y)\,=\,-\,\delta^{4}(x-y)\,
\eeqar
for the two \eq{FG204} cases.

% With the given $M_{1}$ (or $M_{1B}$) effective vertex, which form is clear from the 
%\eq{FG17}-\eq{FG18} expressions, the general expressions for the full Green's function $G_{F}$ can be written as following in the form of a perturbative series: 
%\beqar\label{FG21}
%&\,& G_{F}\,= \,\left[\imath\,\gamma^{\mu}\,\D_{\mu}\,\pm\,m\,+\,M_{1}\,\right]^{-1}\,=\,\left[S_{F}^{-1}\,+\,M_{1}\,\right]^{-1}\,; 
%\\
%&\,& G_{F}(x,y)\,= \,S_{F}(x,y)\,\mp\,\int \,d^4 z\,S_{F}(x,z)\,M_{1}(z)\,G_{F}(z,y)\,;
%\\
%&\,& \Le \imath\,\gamma^{\mu}\,\D_{x\mu}\,\pm\,m\,\Ra\,S_{F}(x,y)\,=\,\pm\,\delta^4(x-y)\,
%\eeqar
%and
%\beqar\label{FG21}
%&\,&\tilde{G}_{F}\, =  \,\left[\imath\,\gamma^{\mu}\,\D_{\mu}\,+\,m\,-\,M_{1B}\,\right]^{-1}\,=\,\left[\tilde{S}_{F}^{-1}\,-\,M_{1B}\,\right]^{-1}\,; 
%\\
%&\,&\tilde{G}_{F}(x,y)\, =  \,\tilde{S}_{F}(x,y)\,+\,\int \,d^4 z\,\tilde{S}_{F}(x,z)\,M_{1B}(z)\,\tilde{G}_{F}(z,y)\,;
%\\
%&\,& \Le \imath\,\gamma^{\mu}\,\D_{x\mu}\,+\,m\,\Ra\,\tilde{S}_{F}(x,y)\,=\,\delta^4(x-y)\,.
%\eeqar
%Now we can calculate the one loop spinor contribution to the action, the calculaions are presented in the next section.

\section{One loop effective action for spinors: simplest perturbative analysis}
\label{S6}

 In this Section we calculate one-loop effective action for the A manifold and compare it with the corresponding results for the actions of B manifold. 
The calculation we perform are restricted by the two simplest diagrams, we calculate a one-loop tadpole diagram and self-energy diagram with one and two external gravitons legs attached correspondingly.
This restriction does not change the main consequences of the calculations. The cancellation of the one-loop terms in the effective action is due the loop's internal momenta integral and not affected 
%by the number of external vertexes
%inserted in the loop but not 
by the number of the external gravity fields attached to the particular vertex. As mentioned above, of course the cancellation is possible only if we assume that the
gravity fields attached to the loop is the same for A and B manifolds, see discussion further. Therefore,
the simplest diagrams we considered are enough for the demonstration of the main idea of the proposed framework, more complete perturbative analysis of the full action will be done somewhere else.
For the \eq{FG17} action, after the integration with respect to $\chi$, we will obtain the following expression for the effective action with one loop precision:
\beq\label{OL1}
Z[\overline{\psi}_{A},\psi_{A}]\,=\,Z_{0}^{-1}\,e^{\imath\,\Gamma_{A}(\overline{\psi}_{A},\psi_{A})}
\eeq
with
\beqar\label{OL2}
\Gamma_{A}(\overline{\psi}_{A},\psi_{A})\,& = &\,\int d^4 x\,\overline{\psi}_{A}\,M_{1}(x)\,\psi_{A}\,-\,\int d^4 x\,d^{4}y\,\overline{\psi}_{A}(x)\,M_{1}(x)\,S_{F}(x,y)\,M_{1}(y)\,\psi_{A}(y)\,+\, 
\nonumber \\
&+&\,
Tr\, \ln\Le S_{0}\,+\,M_{1} \Ra\,.
\eeqar
In this definition we omit the gravitational field, it is considered as some background field and it's contribution in the generating functional we will account further.
For our purposes it is enough to consider the diagrams with only two external gravitons included\footnote{We did not consider the vertex of interactions of two vierbeins (gravitons) with a fermion-antifermion pair, therefore the two external gravitons contribution we account is not full one of course. }. An ordinary expansion of the last term in the \eq{OL2} provides to the required precision:
\beqar\label{OL3}
Tr\, \ln\Le S_{0}\,+\,M_{1} \Ra\,& = &\,Tr\,\ln (S_{0})\,+\,Tr\,\ln \Le I\,+\,S_{F}\,M_{1}\Ra\,\approx\,Tr\,\ln (S_{0})\,+\,Tr\,\Le S_{F}\,M_{1}\Ra\,-\,
\nonumber \\
&-&
\frac{1}{2}\,
Tr\,\Le S_{F}\,M_{1}\,S_{F}\,M_{1}\Ra\,.
\eeqar 
We fix the $Z_{0}$ normalization constant in the \eq{OL1} by condition\footnote{In general we can define the constant as $Z_{0}\,=\,Z[\phi_{cl}\,=\,0]$ with $\phi_{cl}$ as all possible classical background fields in the problem. In this case the $Z_{0}$ defines the contribution of the bubble diagrams in the action, i.e. diagrams without external legs. This definition 
of the bubble diagrams must be not confused with the bubble 
diagrams arise in the equation of motion. There it appears in the form of the momentum-energy tensor, i.e. after the variation of the action's diagrams with one external leg with respect to some $\phi_{cl}$,
see \eq{OL1001} definition further.}
\beq\label{OL4}
Z_{0}^{-1}\,e^{\imath\,Tr\,\ln (S_{0})\,}\,=\,1\,
\eeq
leaving us with two contributions we need to calculate. Also, further we will not write explicitly the trace in the expressions if it will not lead to the confusion.

 For the second term in the \eq{OL3} we have formally:
\beq\label{OL401}
\Gamma_{1}\,= \,\int\,d^{4}x\,S_{F}(x,x)\,M_{1}(x)\,.
\eeq
This expression can be understood as 
\beq\label{OL5}
\, \Gamma_{1}\,= \,\imath\,S_{FR}^{E}(0)\,\int\,d^{4}x\,
\Big(
-\imath\,\Le \delta^{\,\mu}_{c}\,\D_{\mu} s\,-\,\eta^{\mu \nu} \D_{\mu} s_{c \nu}\Ra\,\gamma^{c}\,+\,
%\right.
%\nonumber \\
%&+&
%\left.
\frac{\imath}{2}\,\Le \D_{b} s_{\,c a}\,-\,\D_{a}s_{\,c b}\Ra\,\gamma^{c}\,\gamma^{a}\, \gamma^{b}\,-\,m\,s\,
\Big)\,
\eeq
%=\,
%\nonumber \\
%&=&\,
%\int\,d^{4}x\,\lim_{y\to x}\Le\Big(
%\imath \gamma^{c}\Le \delta^{\,\mu}_{c} s \D_{\mu}-\eta^{\mu \nu}  s_{c \nu}\D_{\mu}\Ra -
%\frac{\imath}{2} \gamma^{c} [\gamma^{a}, \gamma^{b}] s_{\,c a}\D_{b} - m\,s\,\Big)_{x}\Le S_{F\,sing}(x,y)+S_{F\,fin}(x,y)\Ra\Ra\,,
%\eeqar
%The classical vierbein we have is a simple plane wave:
%\beq\label{OL6}
%s_{a b}(x)\,=\,\Pi_{a b}\,e^{i\,p_{f}\,x}\,,\,\,\,\eta^{a b}\,\Pi_{a b}\,=\,0\,,\,\,\,p^{a}_{f} \Pi_{a b}\,=\,0\,.
%\eeq
%here $S_{F\,sing}$ and $S_{F\,fin}$ are singular and final (analytical) parts of $S_{F}$ correspondingly. Having in mind the dimensional regularization of the $S_{F\,sing}$,
%we firstly calculate the $y\to x$ limit of the expressions, isolating the singularities of this limit, and take the limit of four dimensional space in integrals only after that.
with $S_{FR}^{E}(0)$ as Feynman propagator with some regularization applied in the momentum space after the rotation in an Euclidean space.
The only non-zero term  we have in this order is the following one:
\beq\label{OL7}
\Gamma_{1}(p_{f})\,=\,-\,4\,\imath\,m^{2}\,G_{FR}^{E}(0)\,\int d^{4}x\, s(x)\,
%=\, 
%2\,\imath\,p_{f b}\,\Pi_{c a}\,\gamma^{c}\,\,S^{a b}\,\int d^{4} x\,e^{i\,p_{f}\,x}\,S_{F}(x,x)\,.
\eeq
where $G_{FR}^{E}(0)$ is a regularized Feynman propagator of the scalar field.
%\beq\label{OL8}
%I_{1}(p_{f})\,=\,
%-\,\frac{1}{2}\,(2\pi)^{4}\,p_{f b}\,\Pi_{c a}\,Tr\left[ \gamma^{c}\,[\gamma^{a},\gamma^{b}]\,S_{FR}(x,x)\right]\,\delta^{4}(p_{f})\,=\,0\,,
%2\,\imath\,(2\pi)^{4}\,p_{f b}\,\Pi_{c a}\,\gamma^{c}\,\,S^{a b}\,\delta^{4}(p_{f})\,\,S_{FR}(x,x)\,=\,\hat{V}(p_{f})\,\,S_{FR}(x,x)\,
%\eeq
%with
%\beq\label{OL8001}
%\hat{V}(p_{f})\,=\,2\,\imath\,p_{f b}\,\Pi_{c a}\,\gamma^{c}\,\,S^{a b}\,.
%\eeq
%The corresponding contribution to S-matrix, in turn, has the following form:
%\beq\label{OL9}
%\int\,d^{4}p_{f}\,I_{1R}(p_{f})\,=\,(2\pi)^{4}\,S_{FR}(x,x)\,\int\,d^{4}p_{f}\,\hat{V}(p_{f})\,\delta^{4}(p_{f})\,=\,0\,,
%2\,\imath\,(2\pi)^{4}\,p_{f b}\,\Pi_{c a}\,\gamma^{c}\,\,S^{a b}\,S_{FR}(x,x)\,\int\,d^{4}p_{f}\,\hat{p}_{f}\,\delta^{4}(p_{f})\,=\,0\,,
%\eeq
%i.e. it does not contribute to the corresponding S-matrix element. 

 The third term in \eq{OL3} has the following form:
%\beqar
\beq\label{OL11}
\Gamma_{2}\,= \,-\frac{1}{2} \,\int\,d^{4}x\,\int\,d^{4}y\,M_{1}(x)\,S_{F}(x,y)\,M_{1}(y)\,S_{F}(y,x)\,
\eeq
%=\,
%\nonumber \\
%&=&\,
%-\frac{1}{2}\,(2\pi)^{4}\,\delta^{4}(p_{f}+q_{f})\,
%\int\,\frac{d^{4} p}{(2\pi)^{4}}\,\hat{V}(p_{f})\,\hat{S}_{F}(p)\,\hat{V}(q_{f})\,\hat{S}_{F}(p-p_{f})\,=\,
%\nonumber \\
%&=&\,
%2\,(2\pi)^{4}\,\delta^{4}(p_{f}+q_{f})\,\Le p_{f b} \Pi_{c a}\,\Ra \,\Le q_{f b_{1}} \Pi_{c_{1} a_{1}}\,\Ra
%\int\,\frac{d^{4} p}{(2\pi)^{4}}\,\gamma^{c}\,S^{a b}\,\hat{S}_{F}(p)\,\gamma^{c_{1}}\,\,S^{a_{1} b_{1}}\hat{S}_{F}(p-p_{f})\,
%\eeqar
which we write formally as
\beqar\label{OL12}
\Gamma_{2}\,&=& \,-\frac{1}{2} \,\int\,d^{4}x\,\int\,d^{4}y\,
\Big(
\imath\,\gamma^{c}\,\Le \delta^{\,\mu}_{c} s \D_{\mu} -\eta^{\mu \nu} s_{c \nu}\D_{\mu} \Ra - 
\frac{\imath}{2}\,\gamma^{c}\,[\gamma^{a}, \gamma^{b}] s_{\,c a}\D_{b} - m\,s\,\Big)_{x}\,S_{F}(x,y)\,
\nonumber \\
&\,&
\Big(
\imath\,\gamma^{c_{1}}\,\Le \delta^{\,\mu_{1}}_{c_{1}} s \D_{\mu_{1}} -\eta^{\mu_{1} \nu_{1}} s_{c_{1} \nu_{1}}\D_{\mu_{1}} \Ra - 
\frac{\imath}{2}\,\gamma^{c_{1}}\,[\gamma^{a_{1}}, \gamma^{b_{1}}] s_{\,c_{1} a_{1}}\D_{b_{1}} - m\,s\,\Big)_{y}
S_{F}(y,x)\,
\eeqar
where we performed integration by parts and derivative with respect to $x$ and $y$ act on the both Green's function.
In terms of regularized integrals we obtain the following answer for this expression:
\beqar
\Gamma_{2}\,&=& \,-\,
\frac{3\imath}{8}\,
\int \frac{d^4 p_{i}}{(2\pi)^{4}}\,\int d^4 p_{f}\, \tilde{s}(p_{i})\,\tilde{s}(p_{f})\, p_{f}^{2}\,\delta^{4}(p_{i}+p_{f})\,
\int_{0}^{1}\,dx\,\int \frac{d^{4} l_{E}}{(2\pi)^{4}}
\frac{l_{E}^{2}-5 p_{f}^{2} x\Le 1-x \Ra/2 +2 m^2  }{\Le l^{2}_{E}\,+\,\Delta\Ra^{2}}\,-\,
\nonumber \\
&-&\,
\,\frac{3\imath}{16}\,
\int  \frac{d^4 p_{i}}{(2\pi)^{4}}\,\tilde{s}_{c \nu}(p_{i})\,\int d^4 p_{f} \tilde{s}^{c \nu}(p_{f})p_{f}^{2}\,\delta^{4}(p_{i}+p_{f})\,
\int_{0}^{1}\,dx\,\int \frac{d^{4} l_{E}}{(2\pi)^{4}}\,
\frac{l_{E}^{2} + 2  p_{f}^{2} x\Le 1-x \Ra +2 m^2  }{\Le l^{2}_{E}\,+\,\Delta\Ra^{2}}\,+\,
\nonumber \\
&+&\,
2\,\imath\, m^2\,\int \frac{d^4 p_{i}}{(2\pi)^{4}} \tilde{s}(p_{i})\int d^4 p_{f} \tilde{s}(p_{f})\delta^{4}(p_{i}+p_{f})\,
\int_{0}^{1}\,dx\,\int \frac{d^{4} l_{E}}{(2\pi)^{4}}\,
\frac{l_{E}^{2} + p_{f}^{2} x\Le 1-x \Ra - m^2  }{\Le l^{2}_{E}\,+\,\Delta\Ra^{2}}\,.
\label{OL13}
\eeqar
see the calculations in Appendix ~\ref{appendix:c}. 
An expression for the corresponding energy-momentum tensor for the arbitrary weak field 
we can obtain from the usual definition of the tensor in the flat space-time:
\beq\label{OL1001}
\frac{1}{2}\,T^{\rho \sigma}\,= \,\frac{\delta \,\Gamma}{\delta h_{\rho \sigma}(z)}\,-\,
\D_{\mu}\,\frac{\delta\, \Gamma}{\delta \,\Le \D_{\mu}h_{\rho \sigma}(z)\Ra}\,.
\eeq
For example, the tensor to the first order approximation has the following form:
%\beqar
\beq\label{OL10}
T^{\rho \sigma}_{0}\, = \,2\,\frac{\delta \Gamma_{1}}{\delta h_{\rho \sigma}(z)}\,=\,\frac{\delta \Gamma_{1}}{\delta s_{\rho \sigma}(z)}\,=\,
-\,4\,\int\,d^{4}x\,\delta^{4}(x-z)\,m^{2}\,\eta^{\rho \sigma}\,G_{F}(x,x)\,=\,-\,4\,\imath\,m^{2}\,G_{FR}^{E}(z,z)\,\eta^{\rho \sigma}\,
%\frac{1}{2}\,\eta^{\rho \sigma}\,I_{1}\,+\,\frac{1}{2}\,\frac{\delta I_{1}}{\delta s_{\rho \sigma}(z)}\,-\,\frac{1}{2}\,
%\frac{\delta I_{1}}{\delta \Le\D_{\mu} h_{\rho \sigma}(z)\Ra}\,\Le \D_{\mu} h\Ra\,\rightarrow\,\frac{1}{2}\,\frac{\delta I_{1}}{\delta s_{\rho \sigma}(z)}\,=\,
%\nonumber \\
%&=&
%\int\,d^{4}x\,\delta^{4}(x-z)\,\Big(
%\imath\,\Le \eta^{\rho \sigma}\,\gamma^{\mu}\,\D_{\mu} \,-\,\eta^{\mu \sigma}\,\gamma^{\rho}\, \D_{\mu} \Ra\,-\,
%\right.
%\nonumber \\
%&+&
%\left.
%\frac{\imath}{2}\,\gamma^{\rho}\,[\gamma^{\sigma}, \gamma^{b}]\,\D_{b}\,-\,m\,\eta^{\rho \sigma}\,\Big)\,S_{F}(x,x)\,=\,
%\nonumber \\
%&=&\,
%-\,m\,S_{FR}(z,z)\,\eta^{\rho \sigma}\,.
%\eeqar
\eeq 
with $G_{FR}(z,z)$ as a regularized propagator of scalar field.
We see, that this term has a form of the cosmological term contribution in the equations of motion, as expected.

\subsection{First variant of the effective action for the B manifold }

 The first variant of the effective action for the B-manifold is determined by the it's bare action $S_{B1}$ and propagator $S_{F1}$, \eq{FG20} and \eq{FG2001} correspondingly.
For this action the following expansion holds:
\beqar\label{AcB11}
Tr\, \ln\Le S_{0B1}^{+}\,+\,M_{1B} \Ra\,& = &\,Tr\,\ln (S_{0B1}^{+})\,+\,Tr\,\ln \Le I\,-\,S_{F1}\,M_{1B}\Ra\,\approx\,Tr\,\ln (S_{0}^{+})\,-\,Tr\,\Le S_{F1}\,M_{1B}\Ra\,-\,
\nonumber \\
&-&
\frac{1}{2}\,
Tr\,\Le S_{F1}\,M_{1B}\,S_{F1}\,M_{1B}\Ra\,.
\eeqar 
Correspondingly, we have for the tadpole contribution:
\beq\label{AcB12}
\Gamma_{1\,B1}\,= \,-\,\int\,d^{4}x\,S_{F1}(x,x)\,M_{1B}(x)\,=\,-\,4\,m^{2}\,G_{FR}(0)\,\int d^{4}x\, s(x)\,=\,\Gamma_{1\,A}\,,
\eeq
i.e. the answer is the same as obtained for the \eq{OL7} tadpole contribution in the effective action of A-manifold. For the the one-loop contribution expression,
first of all, we put attention that the 
differences between the 
third terms in \eq{AcB11} and \eq{OL3}  is in a sign of the $m$ in the corresponding propagators and in  $M_{1}$ terms. 
The second difference between the contributions, is in a sign of the connection field term in the Lagrangians.
Because the \eq{OL13} answer is invariant with respect to the $m\,\rightarrow\,-m$
replace we conclude, therefore, that for this variant of the B-manifold action
\beq\label{AcB13}
\Gamma_{B1}\,+\,\Gamma_{A}\,\neq\,0
\eeq
to the given precision and correspondingly there is no any cancellation of contributions in the momentum-energy tensor.

\subsection{Second variant of the effective action for the B manifold }

 In the case of \eq{FG2002} propagator we face the following picture. First of all, we have:
\beqar\label{AcB21}
Tr\, \ln\Le S_{0B1}^{+}\,+\,M_{1B} \Ra\,& = &\,Tr\,\ln (S_{0B1}^{+})\,+\,Tr\,\ln \Le I\,+\,S_{F2}\,M_{1B}\Ra\,\approx\,Tr\,\ln (S_{0B1}^{+})\,+\,Tr\,\Le S_{F2}\,M_{1B}\Ra\,-\,
\nonumber \\
&-&
\frac{1}{2}\,
Tr\,\Le S_{F2}\,M_{1B}\,S_{F2}\,M_{1B}\Ra\,.
\eeqar 
For the tadpole contribution we write again:
\beq\label{AcB22}
\Gamma_{1\,B2}\,= \,-\,\int\,d^{4}x\,S_{F2}(x,x)\,M_{1B}(x)\,=\,4\,\imath\,m^{2}\,G_{FR}^{E}(0)\,\int d^{4}x\, s(x)\,=\,\Gamma_{1\,A}\,.
\eeq
Here the rotation in an Euclidean space is given by $G_{FR}\,\rightarrow\,-\,\imath\,G_{FR}^{E}$ expression, there is a Dyson propagator we use.
For the one-loop contribution the situation is a bit more complicated. For this term the symmetry in respect to $m\,\rightarrow\,-m$ exists as well, but now again we have the Dyson propagator 
in the expressions instead the Feynman one. Performing the $l_{0}\,\rightarrow\,-\,\imath\,l_{0E}$ continuation into the Euclidean we 
obtain some part of the terms of  \eq{OL13} expression with $\imath\,\rightarrow\,-\,\imath\,$ replace done. Nevertheless, due the different sign of the connection field in teh "long" derivative
, see \eq{FG1801} action, not all the terms in the effective action change sign. Therefore, in general we have
\beq\label{AcB23}
\Gamma_{2\,B2}\,\neq \,-\,\Gamma_{2\,A}\,
\eeq
and in this variant of the B-manifold we stay with the contributions to the effective action arise from the one-loop diagrams with two external off-shell gravity fields attached, i.e.
with some diagrams which formally are of the graviton's mass like type.

\subsection{Third variant of the effective action for the B manifold }

In this particular case we have:
\beqar\label{AcB31}
Tr\, \ln\Le S_{0B3}^{-}\,+\,M_{1B}^{-}) \Ra\,& = &\,Tr\,\ln (S_{0B3}^{-})\,+\,Tr\,\ln \Le I\,-\,S_{F3}\,M_{1B}^{-}\Ra\,\approx\,Tr\,\ln (S_{0B3})\,-\,Tr\,\Le S_{F3}\,M_{1B}^{-}\Ra\,-\,
\nonumber \\
&-&
\frac{1}{2}\,
Tr\,\Le S_{F3}\,M_{1B}^{-}\,S_{F3}\,M_{1B}^{-}\Ra\,.
\eeqar
In this particular case the masses in answer for the tadpole contribution comes as
\beq\label{AcB32}
\Gamma_{1\,B3}\,= \,-\,4\,\imath\,m^{2}\,G_{FR}^{E}(0)\,\int d^{4}x\, s(x)\,=\,\Gamma_{1\,A}\,
%\,\int\,d^{4}x\,S_{F1}(x,x)\,M_{1B}(x)\,=\,4\,m^{2}\,G_{FR}(0)\,\int d^{4}x\, s(x)\,=\,-\,\Gamma_{1\,A}\,,
\eeq
i.e. it is the same as contribution in the $\Gamma_{1\,A}\,$, they are doubling and nit canceling. The same is correct for the $\Gamma_{2\,B2}$ and $\Gamma_{2\,A}$
contributions in the general effective action. Correspondingly have a doubling of the contribution into the momentum-energy tensor.

\subsection{Fourth variant of the effective action for the B manifold }

 The one-loop contributions we consider in this case is the following one:
\beqar\label{AcB41}
Tr\, \ln\Le S_{0B4}^{-}\,+\,M_{1B}^{-} \Ra\,& = &\,Tr\,\ln (S_{0B4}^{-})\,+\,Tr\,\ln \Le I\,+\,S_{F4}\,M_{1B}^{-}\Ra\,\approx\,Tr\,\ln (S_{0B4})\,+\,Tr\,\Le S_{F4}\,M_{1B}^{-}\Ra\,-\,
\nonumber \\
&-&
\frac{1}{2}\,
Tr\,\Le S_{F4}\,M_{1B}^{-}\,S_{F4}\,M_{1B}^{-}\Ra\,.
\eeqar
In this case we have as well
\beq\label{AcB42}
\Gamma_{1\,B4}\,= \,-\,\int\,d^{4}x\,S_{F4}(x,x)\,M_{1B}(x)\,=\,4\,m^{2}\,\imath\,G_{FR}^{E}(0)\,\int d^{4}x\, s(x)\,=\,-\,\Gamma_{1\,A}\,,
\eeq
the change of the sigh here is due the use of Dyson propagator of course. So,
again, a cancellation of tadpole contributions from $S_{A}$ and $S_{B4}^{-}$ actions is present here. 
%but, differently from the previous $S_{B3}$ action, here we have the Dyson propagators in action.
%As mentioned above, it means the
More important, that here the identity
%$\imath\,\rightarrow\,-\,\imath\,$ replace performed in \eq{OL15} expression that in turn leads to the
\beq\label{AcB43}
\Gamma_{2\,B4}\,= \,-\,\Gamma_{2\,A}\,
\eeq
takes place, i.e. these terms are canceling in the total action as well. We obtained, that for this particular variant of the B-manifold action we have a total cancellation of the considered diagrams into the final  effective action. Correspondingly, this cancellation occurs also for the the momentum-energy tensor and for the contributions into the cosmological constant value. 
Only this variant of the effective action, therefore, is considered further and we will denote it as 
$\Gamma_{B}$.

\section{Effective action with gravity included}
\label{S7}

 Now we ready to discuss the form of an appearance of gravity field in the model. Firstly consider the matter's fields action we obtained:
\beqar\label{CosC1}
&\,&\Gamma(\overline{\psi},\psi,h)\,= \,\Gamma_{A}(\overline{\psi}_{A},\psi_{A},s)\,+\,\Gamma_{B}(\overline{\psi}_{B},\psi_{B},s)\,= \,
\nonumber\\
&=&
\int d^4 x_{A}\,\overline{\psi}_{A}\,M_{1}(x_{A})\,\psi_{A}\,-\,\int d^4 x_{A}\,d^{4}y_{A}\,\overline{\psi}_{A}(x_{A})\,M_{1}(x_{A})\,S_{F}(x_{A},y_{A})\,M_{1}(y_{A})\,\psi_{A}(y_{A})\,-\,
\nonumber \\
&-&\,
\int d^4 x_{B}\,\overline{\psi}_{B}\,M_{1}(x_{B})\,\psi_{B}\,-\,\int d^4 x_{B}\,d^{4}y_{B}\,\overline{\psi}_{B}(x_{B})\,M_{1}(x_{B})\,S_{D}(x_{B},y_{B})\,M_{1}(y_{B})\,\psi_{B}(y_{B})\,,
\eeqar 
here the $S_{F4}$ propagator  is denoted as the Dyson's one $S_{D}$. 
%we also do not denote the coordinates of different manifolds differently till it will not lead to an any confusion. 
The action describes a classical fermion fields in two manifolds which interact with a background gravitational field. 
Writing precisely the form of $M_{1}$ effective vertex, we have:
\beqar\label{CosC101}
M_{1}(x)\,& = &\,
\Le\,\imath\,\Le s\delta^{\,\mu}_{c}\,-\,\eta^{\mu \nu} s_{c \nu}\Ra\,\gamma^{c}\,\D_{\mu}\,+\,
\frac{\imath}{2}\,\Le \D_{b} s_{\,c a}\,-\,\D_{a}s_{\,c b}\Ra\,\gamma^{c}\,\gamma^{a}\, \gamma^{b} \,-\,m\,s\Ra\,=\,
\nonumber \\
&=&
\Le\,\imath\,\Le s\,\gamma^{\mu}\,\D_{\mu}\,-\,\eta^{\mu \nu}\,\gamma^{c}\,s_{c \nu}\,\D_{\mu}\,+\,\gamma^{a}\,\D_{a} s\,-\, \gamma^{b}\,\D_{a} s^{a}_{b}\Ra\,
\,-\,m\,s\Ra\,,
\eeqar
here the \eq{CC2} identity was used. Taking into account that
\beq\label{CosC102}
\Le \imath\,\gamma^{\mu}\,\D_{\mu}\,-\,m\Ra\,\psi_{A,B}\,=\,0\,,\,\,\,\D_{a}\,\Le \overline{\psi}_{A,B}\,\gamma^{a}\,\psi_{A,B}\Ra\,=\,0\,,
\eeq
we stay with
\beq\label{CosC103}
M_{1}(x)\, = \,-\,\imath\,\Le \gamma^{c}\,s_{c}^{ \mu}\,\D_{\mu}\,+\, \gamma^{b}\,\D_{a} s^{a}_{b}\Ra\,.
\eeq
Further, in the gravitational field action, we make an usual redefinition of the weak gravity field which in turn leads to
\beq\label{CosC104}
s_{a b}\,=\,\overline{s}_{a b}\,-\,\frac{1}{2}\,\eta_{a b}\,\overline{s}\,,\,\,\,\D_{a}\,\overline{s}^{a}_{b}\,=\,0\,
\eeq
redefinition and gauge condition for the $s$ field. In terms of redefined $s$ field we obtain for the \eq{CosC103} expression:
\beq\label{CosC105}
M_{1}(x)\, = \,-\,\imath\,\gamma^{c}\,\overline{s}_{c}^{ \mu}\,\D_{\mu}\,+\, \frac{1}{2}\,m\,\overline{s}\,=\,
\overline{s}_{\mu \nu}\,\Le J_{1}^{\mu \nu}\,+\,\frac{1}{2}\,m\,\eta^{\mu \nu} \Ra\,=\,\overline{s}_{\mu \nu}\,J^{\mu \nu}\,.
%=\,\overline{s}_{\mu \nu}\,\Le T^{\mu \nu}\,+\,\eta^{\mu \nu}\,\Lambda \Ra\,,
%\overline{s}_{\mu \nu}\,J^{\mu \nu}\,,
\eeq
here again the identities \eq{CosC102} exact for the classical fermion fields were used. 

  Our next step is an introduction of the gravity on the stage.
In the framework we introduce separately the gravity field limits for the A and B manifolds defining
\beq\label{CosC2}
S_{gr}\,=\,-\,m_{p}^{2}\,\int_{-\infty}^{\infty} dt_{A}\int\,d^{3}x_{A}\,\sqrt{-g(x_{A})}\,R(x_{A})\,-\,
\,m_{p}^{2}\,\int_{-\infty}^{\infty} dt_{B}\int\,d^{3}x_{B}\,\sqrt{-g_{B}(x_{B})}\,R(x_{B})\,
\eeq
with
\beq\label{CosC3}
m_{p}^2\,=\,\frac{1}{16\pi G}\,=\,\frac{2}{\kappa^2}\,,\,\,\,g_{B}(x_{B})\,=\,CPTM\Le g_{A}(x_{A})\Ra\,,
\eeq
i.e. the B-manifold metric is defined through the metric of A-manifold by corresponding CPTM transform. We also assume that there no any cosmological constant terms in the actions, it will arise dynamically through the interactions between manifolds.
In the case of a standard weak field approximation we have:
\beqar
&\,& g^{A}_{\mu \nu}(x)\,= \,g^{A\,0}_{\mu \nu}(x)\,+\,h_{\mu \nu}^{A}(x)\,; \label{CosC4}\\
&\,& g^{B}_{\mu \nu}(y)\,= \,g^{B\,0}_{\mu \nu}(y)\,+\,h_{\mu \nu}^{B}(y)\, ; \label{CosC5}\,.
\eeqar 
Further we will use the usual redefined $h$ field and gauge condition introduced for it 
\beq\label{CosC6}
h_{\mu \nu}\,=\,\overline{h}_{\mu \nu}\,-\,\frac{1}{2}\,\eta_{\mu \nu}\,\overline{h}\,,\,\,\,\D_{\mu}\,\overline{h}^{\mu}_{\nu}\,=\,0\,,
\eeq
see the \eq{CosC104} in the correspondence. Now, the important question we need to answer on, is about the gravity fields appearance in the \eq{CosC1} expression. Further we will discuss two different possibilities for the inclusion of the gravity in the \eq{CosC1}. Before that, firstly, we will consider a mostly general expression for the
action with gravitational fields and will write the corresponding Green's function and classical solutions adapting them later to the particular cases. In order to avoid a detailed notations in the long expressions, further we will write only the superscripts $A$ and $B$ for the fields without Lorentzian indexes sometimes in the expressions where it will not lead to a confusion.. 
An another important remarks is about the coordinates notations for different manifolds. In general we have no way to separate the 
fields by their dependence on the $x_{A}$ or $x_{B}$ coordinates. Therefore, we will preserve the $A$ and $B$ indexes only as signs of the different fields and will not distinguish the coordinates of both manifolds in the expressions. The mostly general action we have to the given precision is the following 
one\footnote{We adjusted the numerical coefficients and notations here for the simplification of the further calculations.}:
\beqar\label{CC18}
\Gamma\,&=&\,
\frac{m^{2}_{p}}{4}\,\int d^4 x\,\overline{h}^{\mu \nu}_{A}\,G^{-1}_{A}\,\overline{h}_{\mu \nu}^{A}\,+\,
\frac{m^{2}_{p}}{4}\,\int d^4 x\,\overline{h}^{\mu \nu}_{B}\,G^{-1}_{B}\,\overline{h}_{\mu \nu}^{B}\,+\,
\int d^4 x\,\overline{h}^{\mu \nu}_{A}\,T_{\mu \nu\,A}\,+\,
\nonumber \\
&+&
\int d^4 x\,\overline{h}^{\mu \nu}_{B}\,T_{\mu \nu\,B}\,+\,
\eta_{\mu \nu}\,\int d^4 x\,\overline{h}^{\mu \nu}_{A}\,\Lambda_{0\,A}\,+\,
\eta_{\mu \nu}\,\int d^4 x\,\overline{h}^{\mu \nu}_{B}\,\Lambda_{0\,B}\,-\,
\nonumber \\
&-&
\frac{1}{2}\,\int d^4 x\,\int d^4 y\,\overline{h}^{\mu \nu}_{A}(x)\,M^{1\,A A}_{\mu \nu\,\rho \sigma}\,\,\overline{h}^{\rho \sigma}_{A}(y)\,-\,
%\nonumber \\
%&-&
\frac{1}{2}\,\int d^4 x\,\int d^4 y\,\overline{h}^{\mu \nu}_{B}(x)\,M^{1\,B B}_{\mu \nu\,\rho \sigma}\,\overline{h}^{\rho \sigma}_{B}(y)\,-\,
\nonumber \\
&-&
\frac{1}{2}\,\int d^4 x\,\int d^4 y\,\overline{h}^{\mu \nu}_{A}(x)\, M^{1\,A B}_{\mu \nu\,\rho \sigma}\,\overline{h}^{\rho \sigma}_{B}(y)\,-\,
%\nonumber \\
%&-&
\frac{1}{2}\,\int d^4 x\,\int d^4 y\,\overline{h}^{\mu \nu}_{B}(x)\, M^{1\,B A}_{\mu \nu\,\rho \sigma}\,\overline{h}^{\rho \sigma}_{A}(y)\,=\,
\nonumber \\
&=&\,
\Gamma_{0\,gr}\,+\,\Gamma_{int}\,.
\eeqar
The notations used are simple, the $M^{1\,I J}_{\mu \nu\,\rho \sigma}$ here are  vertices of effective interaction between $I=A,B$ and $J=A,B$ gravity fields, the interaction can take place between the fields
in the same or in the different manifolds, all other quantities in the expression are clear. The cosmological constant term in the expression must be understood as appearing from the corresponding terms in the matter's action and not as introduced by "hands". 
The Green's functions equation has the following form in turn:
\beqar\label{CC25}
&\,&
\frac{m_{p}^{2}}{2}\,\left (
\begin{array}{cc} 
G_{A}^{-1}-\frac{2}{m_{p}^{2}}\,M_{1}^{\,A A} & -\frac{2}{m_{p}^{2}}\,M_{1}^{\,A B}\\
-\frac{2}{m_{p}^{2}}\,M_{1}^{\,B A} & G_{B}^{-1}-\frac{2}{m_{p}^{2}}\, M_{1}^{\,B B} 
\end{array}\right)
%\cdot
%\\
%&\,&\cdot
\left (
\begin{array}{cc} 
G_{A A}& G_{AB}\\
G_{BA} & G_{BB}
\end{array}\right)\,
=\,
\left (
\begin{array}{cc} 
\delta_{A A} & 0\\
0 & \delta_{B B}
\end{array}\right)\,.
\nonumber
\eeqar
The $S_{AB}$ and $S_{BA}$ propagators here are analog of the 
Wightman propagators $S_{>}$ and $S_{<}$ which connect the points of the different time's contours in the Schwinger-Keldysh technique, see for examples and definitions
\cite{Kel1}. In order not to write overweighted expressions, we presents here the Green's functions for the A-manifold only, expressions for the second pair of the Green's functions can be obtained by replace of A and B indexes there:
\beqar
&\,&
G_{A }^{-1}\,G_{0\,A A}\,=\,\frac{2}{m_{p}^{2}}\,\delta_{AA}\,,\,\,\,G_{B }^{-1}\,G_{0\,B B}\,=\,\frac{2}{m_{p}^{2}}\,\delta_{B B}\,;
\label{CC25001}\\
&\,&
G_{B }^{-1}\,G_{0\,B A}\,=\,0\,,\,\,\,G_{A }^{-1}\,G_{0\,A B}\,=\,0\,;
\label{CC25003}\\
&\,&
G_{A A}\,=\,G_{0\,A A}\,+\,\int\,G_{0\,AA}\,M^{1\,AA}\,G_{A A}\,+\,\int\,G_{0\,AA}\,M^{1\,A B}\,G_{B A}\,+\,
\nonumber \\
&\,&+\,
\int\,G_{0\,AB}\,M^{1\,BA}\,G_{A A}\,+\,\int\,G_{0\,AB}\,M^{1\,B B}\,G_{B A}\,;
\label{CC25002}
\\
&\,&
G_{A B}\,=\,G_{0\,A B}\,+\,\int\,G_{0\,A A}\,M^{1\,A B}\,G_{B B}\,+\,\int\,G_{0\,A A}\,M^{1\,A A}\,G_{A B}\,+\,
\nonumber \\
&\,&+\,
\int\,G_{0\,AB}\,M^{1\,BA}\,G_{A B}\,+\,\int\,G_{0\,AB}\,M^{1\,B B}\,G_{B B}\,;
\label{CC25004}
\eeqar
the repeating indexes mean also a corresponding integration in the expression, of course.
The most general expression for the gravitational field we have is the following therefore:
\beq\label{CC25005}
h_{I}\,=\,h_{0\,I}\,-\,\frac{2}{m_{p}^{2}}\,\int\,G_{I J}\,\frac{\delta\, \Gamma_{int}}{\delta h_{J}}\,+\,\xi_{I}\,;\,\,\,G_{I}^{-1}h_{0\,I}\,=\,0\,;\,\,\, I,J=A,B\,, 
\eeq
with $\xi_{I}$ as a fluctuation around a classical solution represented by the first two terms in the expression.

\section{Appearance of cosmological constant in the model: first variant}
\label{S8}

  The terms of the $M^{1\,A B}$ and $M^{1\,B A}$ type arise in the corresponding effective action in the Schwinger-Keldysh approach. We did not consider 
the full analog of the Schwinger-Keldysh action in our framework , it will be done in an separate publication. Nevertheless, in general the formal difference between the present and 
Schwinger-Keldysh frameworks\footnote{See also Discussion section}
it is an absence in the 
\eq{CosC1} effective action of additional vertices of the following form\footnote{The signs of the terms and numerical coefficients in the front of them are not important for the further discussion, 
we simply write them similarly to the other vertices and introduce in the action by "hand".}
\beqar\label{CC25006}
&\,&
M_{1\,A B}^{\mu \nu\,\rho \sigma}(x,y)\,=
\,\frac{1}{2}\,\overline{\psi}_{A}(x)\,J_{x}^{\mu \nu}\,S_{>}(x,y)\,
J_{y}^{\rho \sigma}\,\psi_{B}(y)\,;
\\
\label{CC25007}\,
&\,&
M_{1\,B A}^{\mu \nu\,\rho \sigma}(x,y)\,=
\,\frac{1}{2}\,\overline{\psi}_{B}(x)\,J_{x}^{\mu \nu}\,S_{<}(x,y)\,
J_{y}^{\rho \sigma}\,\psi_{A}(y)\,
\eeqar
attached to the $s$ graviton field similarly to written in \eq{CosC1}, see also \eq{CosC105}. Formally speaking, these vertices generate an additional one loop contributions to the fermionic effective action
through the $S_{>}(x,y)\,S_{<}(y,x)$ type of expressions. We did not discuss and explore these types of contributions postponing that for a separate publication, so further by one loop contributions we will mean only ordinary contributions discussed above.  

 Now, the main question we ask is about a connection between the $s$ and $h_{A,B}$ fields.
In the case of the Schwinger-Keldysh like formulation of the approach we assume that there are two different fields 
\beq\label{CC2500601}
s\,=\,\frac{h_{A}}{2}\,,\,\,\,\,s\,=\,\frac{h_{B}}{2}
\eeq
for the A and B manifolds separately exist. We note immediately, therefore, that then the cancellation between the effective action parts does not take place, see further.
Making the substitutions in \eq{CosC1} action, we will obtain an precise analog of \eq{CC18}:
\beqar\label{CC25008}
\Gamma\,&=&\,
\frac{m^{2}_{p}}{4}\,\int d^4 x\,\overline{h}^{\mu \nu}_{A}\,G^{-1}_{A}\,\overline{h}_{\mu \nu}^{A}\,+\,
\frac{m^{2}_{p}}{4}\,\int d^4 x\,\overline{h}^{\mu \nu}_{B}\,G^{-1}_{B}\,\overline{h}_{\mu \nu}^{B}\,+\,
\int d^4 x\,\overline{h}^{\mu \nu}_{A}\,T_{\mu \nu\,A}\,-\,
\nonumber \\
&-&\,
\int d^4 x\,\overline{h}^{\mu \nu}_{B}\,T_{\mu \nu\,B}\,+\,
\eta_{\mu \nu}\,\int d^4 x\,\overline{h}^{\mu \nu}_{A}\,\Lambda_{0\,A}\,-\,
\eta_{\mu \nu}\,\int d^4 x\,\overline{h}^{\mu \nu}_{B}\,\Lambda_{0\,B}\,-\,
\nonumber \\
&-&
\frac{1}{2}\,\int d^4 x\,\int d^4 y\,\overline{h}^{\mu \nu}_{A}(x)\,M^{1\,A A}_{\mu \nu\,\rho \sigma}\,\,\overline{h}^{\rho \sigma}_{A}(y)\,-\,
%\nonumber \\
%&-&
\frac{1}{2}\,\int d^4 x\,\int d^4 y\,\overline{h}^{\mu \nu}_{B}(x)\,M^{1\,B B}_{\mu \nu\,\rho \sigma}\,\overline{h}^{\rho \sigma}_{B}(y)\,-\,
\nonumber \\
&-&
\frac{1}{2}\,\int d^4 x\,\int d^4 y\,\overline{h}^{\mu \nu}_{A}(x)\, M^{1\,A B}_{\mu \nu\,\rho \sigma}\,\overline{h}^{\rho \sigma}_{B}(y)\,-\,
%\nonumber \\
%&-&
\frac{1}{2}\,\int d^4 x\,\int d^4 y\,\overline{h}^{\mu \nu}_{B}(x)\, M^{1\,B A}_{\mu \nu\,\rho \sigma}\,\overline{h}^{\rho \sigma}_{A}(y)\,
\eeqar
with
\beqar\label{CC25009}
&\,&
M_{1\,A A}^{\mu \nu\,\rho \sigma}(x,y)\,=
\,\frac{1}{2}\,\overline{\psi}_{A}(x)\,J_{x}^{\mu \nu}\,S_{F}(x,y)\,
J_{y}^{\rho \sigma}\,\psi_{A}(y)\,;
\\
\label{CC250010}\,
&\,&
M_{1\,B B}^{\mu \nu\,\rho \sigma}(x,y)\,=
\,\frac{1}{2}\,\overline{\psi}_{B}(x)\,J_{x}^{\mu \nu}\,S_{D}(x,y)\,
J_{y}^{\rho \sigma}\,\psi_{B}(y)\,;
\\
&\,&
T^{\mu \nu}_{A,B}\,=\,\frac{1}{2}\,\overline{\psi}_{A,B}\,J^{\mu \nu}_{1}\,\psi_{A,B}\,,\,\,\,\Lambda_{0\,A,B}\,=\,\frac{1}{4}\,m\,\overline{\psi}_{A,B}\,\psi_{A,B}\,;
\eeqar
see \eq{CosC105} definition again.
Now we can determine what is the cosmological constant in the framework, to the leading order we define it as
\beqar\nonumber
&\,&\Lambda_{A}\,= \,\frac{1}{m_{p}^2}\,\left.\frac{\delta \,\Gamma}{\delta\,\overline{h}_{A}}\,\right|_{\overline{h}_{A}\,=\,0}\,=\,
\\
&=&\,
\frac{m}{4\,m_{p}^{2}}\,\overline{\psi}_{A}\,\psi_{A}\,-\,
\frac{m^2}{16\,m_{p}^{2}}\,\int d^4 y\,\Le \overline{\psi}_{A}(x)\,S_{>}(x,y)\,\psi_{B}(y)+\overline{\psi}_{B}(y)\,S_{<}(y,x)\,\psi_{A}(x)\Ra\,\overline{h}_{B}(y)\,
\label{CC250011}.
%\,\,-\,
%\frac{m^2}{16}\,\int d^4 y\,\overline{h}_{B}(y)\,\,.
\eeqar
and correspondingly
\beqar\nonumber
&\,& \Lambda_{B}\,=\,\frac{1}{m_{p}^2}\,\left.\frac{\delta \,\Gamma}{\delta\,\overline{h}_{B}}\,\right|_{\overline{h}_{B}\,=\,0}\,=\,
\\
&=&
-\frac{m}{4\,m_{p}^{2}}\,\overline{\psi}_{B}\,\psi_{B}-
\frac{m^2}{16\,m_{p}^{2}}\,\int d^4 y\,\Le \overline{\psi}_{A}(y)\,S_{>}(y,x)\,\psi_{B}(x)+\overline{\psi}_{B}(x)\,S_{<}(x,y)\,\psi_{A}(y)\Ra\,\overline{h}_{A}(y).
\label{CC250012}
\eeqar
We see that to leading order the constant is proportional to the following values:
\beqar\label{CC250015}
\Lambda_{0 A}\,&=&\,\frac{1}{4\,m_{p}^{2}}\,m\,\overline{\psi}_{A}\,\psi_{A}\,; \\
%=\,\frac{\rho_{A}}{4\,m_{p}^{2}}\,
\Lambda_{0 B}\,&=&\,-\frac{1}{4\,m_{p}^{2}}\,m\,\overline{\psi}_{B}\,\psi_{B}\,\label{CC250016}; 
%=\,-\,\frac{\rho_{B}}{4\,m_{p}^{2}}\,
\eeqar
with plus or minus sign, the next-order corrections are due the interactions between the fermion fields from different manifolds.

 Now we can comment on the above results, not in depth, but clarifying interesting points. First of all we note that
\beq\label{NewC1}
m\,\overline{\psi}_{A,B}\,\psi_{A,B}\,=\,\Le T^{\,\mu}_{0\,\mu}\Ra_{A,B},
\eeq 
i.e. the cosmological constant to that order is proportional to the leading order of trace of energy-momentum tensor. This result is expected, namely, we do not have Einstein cosmological constant in the r.h.s. of the equations and the trace appears in the in equations in the same position as the constant does:
\beq\label{NewC7}
G_{\mu \nu}\,=\,\frac{2}{m_{p}^{2}}\,\Le T_{\mu \nu}\,+\,\eta_{\mu \nu}\,T^{\rho}_{\rho}\Ra\,,
\eeq
i.e. the constant  occurs in the effective action after the following  appearance of the on-shell energy-momentum tensor in the effective action:
\beq\label{NewC2}
\T_{\mu \nu}\,=\,T_{\mu \nu}\,+\,\eta_{\mu \nu}\,T^{\rho}_{\rho}
\eeq
see \eq{CosC105} above. 
Therefore, it is quite natural to assume that the cosmological constant term is emerging in the \eq{CC25008} action as
\beq\label{NewC3}
\Gamma_{\Lambda}\,=\,
 \eta_{\mu \nu}\,\int d^4 x\,\overline{h}^{\mu \nu}_{A}\,T^{\rho}_{A\,\rho}\,-\,
\eta_{\mu \nu}\,\int d^4 x\,\overline{h}^{\mu \nu}_{B}\,T^{\rho}_{B\,\rho}\,.
\eeq
Having in mind that we in general can write the trace of the tensor as 
\beq\label{NewC4}
T_{\mu}^{\mu}\,=\,T_{cl\,\mu}^{\mu}\,+\,T_{a\,\mu}^{\mu}\,,
%=\,\Le \epsilon\,-\,3p \Ra\,+\,T_{a\,\mu}^{\mu}\,
\eeq
with $T_{cl}$ as a classical part of the trace and $T_{a}$ as trace of the conformal anomaly of the tensor,
the question we face is about a smallness of the constant in the form it appears in \eq{NewC4}. One from the possible solutions in this case is a 
definition of the the constant as 
\beq\label{NewC5}
\Lambda_{0 A}=\frac{1}{4\,m_{p}^{2}}\,\Le T_{cl\,\mu}^{\mu} \,+\,T_{a\,\mu}^{\mu}\,\Ra\,
\eeq
whereas it's smallness is provided by the $T_{cl}\,=\,0$ condition that leaves
\beq\label{NewC501}
\Lambda_{0 A}=\frac{1}{4\,m_{p}^{2}}\,T_{a\,\mu}^{\mu}\,.
\eeq
The second comment we have is about the issue of the constant's sign. In our notations the positive value of the constant (if it positive after all of course) appears at the r.h.s. of the Einstein equations
so the constant is positive as well in the usual definition of the term (and vice verse of course). Therefore, the conclusion is that the
\eq{NewC5} and a direct application of the Schwinger-Keldsyh formalism do provide the correct answer until the \eq{NewC5} value is positive but in the approach the problem of 
large vacuum contributions into the constant is not fully resolved, see above and further.

 In an analog of the  Schwinger-Keldysh formulation of the problem there is no cancellation between the different contributions into the effective spinor's action.
The spinor's action terms under consideration are depend on $s$ field, see \eq{OL7} or/and \eq{OL12}, and when the fields on both manifolds are different, 
see \eq{CC2500601}, these terms are not canceled.
Indeed, we have for the classical fields:
\beqar\label{CC25001501}
s_{A}\,&=&\,\frac{h_{A}}{2}\,=\,\frac{1}{2}\,h_{0\,A}\,-\,\int\,G_{A A}\,\Lambda_{0\,A}\,+\,\cdots \\
s_{B}\,&=&\,\frac{h_{B}}{2}\,=\,\frac{1}{2}\,h_{0\,B}\,-\,\int\,G_{B B}\,\Lambda_{0\,B}\,+\,\cdots\,,\label{CC25001502}
\eeqar
and even assuming that
\beq\label{CC25001503}
h_{0\,A}\,=\,h_{0\,B}\,
\eeq
we have that the cancellation of the contributions occurs only to leading order. In general there is some perturbative equation
for the cosmological constant due the not-full cancellation of the one-loop contribution to the constant. We can achieve
the full cancellation of the mentioned terms assuming that $h_{A}\,=\,h_{B}\,=\,h$. In this case the CPTM symmetry can be, perhaps, understood in the sense of "clone"model of 't Hooft,  see \cite{Hooft}. We have only one gravity field overall in this case, with corresponding change in \eq{CC25008} action we correspondingly will obtain to leading order: 
\beq\label{CC25001504}
\Lambda_{A}\,= \,\frac{1}{m_{p}^2}\,\left.\frac{\delta \,\Gamma}{\delta\,\overline{h}}\,\right|_{\overline{h}\,=\,0}\,=\,
\frac{m}{4\,m_{p}^{2}}\,\Le \overline{\psi}_{A}\,\psi_{A}\,-\,\overline{\psi}_{B}\,\psi_{B}\Ra\,
\eeq
or, generalizing, we can assume that
\beq\label{CC25001505}
\Lambda_{A}\,= \,\frac{1}{m_{p}^2}\,\left.\frac{\delta \,\Gamma}{\delta\,\overline{h}}\,\right|_{\overline{h}\,=\,0}\,=\,
\frac{1}{4\,m_{p}^{2}}\,\Le T^{\mu}_{A\,\mu}\,-\,T^{\mu}_{B\,\mu}\Ra\,
%=\,-\,\frac{1}{4\,m_{p}^{2}}\,\Le \T^{\mu}_{B\,\mu}\,-\,\T^{\mu}_{A\,\mu}\Ra\,
\eeq
The correct value of the constant is provided in this case by
\beq\label{CC25001506}
T^{\mu}_{B\,\mu}\,<\,T^{\mu}_{A\,\mu}
\eeq
condition.
Therefore, the particular version of the formalism can be defined as a "dark" matter model, more similar to the \cite{Villata,Chardin} construction than to the bimetric models 
\cite{DamKog}. The clone B-fields are the "dark" matter in this formulation. The smallness of the constant, in this case, can be achieved by the smallness of the difference of the tensors's
traces in the \eq{CC25001505}, see also some discussion in the next Section. 

%But how this assumption is compatible with the a Schwinger-Keldysh formalism applied only to the matter fields and not to the gravity fields is not clear; 
%in general the problem requires an additional research which we do not perform in this article. 

 We return now to the Green's functions, \eq{CC25002} expression, and consider a change of the function's structure due the B-manifold presence in comparison to the usual gravity
Green's function.
We have:
\beq\label{CC250013}
G_{AA}\,=\,G_{A A}^{1}\,+\,G_{A A}^{2}\,
\eeq
with
\beq\label{CC2500131}
G_{AA}^{1}\,=\,G_{0\,A A}\,+\,\int\,G_{0\,AA}\,M^{1\,AA}\,G_{A A}\,
\eeq
as bare A-manifold graviton's Green's function
and
\beq\label{CC250014}
G_{AA}^{2}\,=\,\int\,G_{0\,AA}\,M^{1\,A B}\,G_{B A}\,+\,
\int\,G_{0\,AB}\,M^{1\,BA}\,G_{A A}\,+\,\int\,G_{0\,AB}\,M^{1\,B B}\,G_{B A}\,+\,\int\,G_{0\,AA}\,G_{0\,A B}\,G_{0\,B A}\,+\,\cdots
\eeq
as corrections to the $G_{AA}^{1}$ propagator arise in the formalism due the B-manifold presence.
Therefore, the  first part of the full propagator is an usual propagator of a graviton, i.e. it represents an usual gravity potential, whereas the second term is a correction
to the potential due an interactions of matter and gravity between A and  B-manifolds. In that extend, the B-manifold plays a role of the dark matter in the approach of the extended manifold.
Subsequently, when we consider only one gravity field, see discussion above, then the corrections to the usual propagator will acquire the following form:
\beq\label{CC2500150001}
G_{AA}^{2}\,=\,\int\,G_{0\,AA}\,M^{1\,A B}\,M^{1\,B A}\,G_{A A}\,+\,\cdots\,.
\eeq
These corrections are due the Schwinger-Keldysh like mechanism of the interactions applied to the different fermions fields only, see \eq{CC25006}-\eq{CC25007} definitions. In the case when we have no any connection between A and B manifolds, i.e. in the case of an absence of the Schwinger-Keldysh type of interactions for all fields, the gravitational propagators are not changing but the one-loop
cancellation of the effective action terms occurs as well in the case of the same gravitational field of course.

\section{Appearance of cosmological constant in the model: second variant}
\label{S9}

 As an another possibility for the introduction of $h$ fields in the action with $s$ vierbein present, we consider the following 
definition of both A,B vierbein fields in the effective fermion's action: 
\beq\label{CosC7}
\overline{s}_{\mu \nu}\,=\,\frac{1}{2}\,\overline{h}_{\mu \nu}\,=\,\frac{1}{4}\,\Le \overline{h}_{\mu \nu}^{A}\,+\,\overline{h}_{\mu \nu}^{B} \Ra\,.
\eeq
The field is the same for the A,B manifolds therefore the precise cancellation of the one loop $\Gamma_{2}$ terms exists in this case.
The assumption made here is that any matter is a source for the both $h_{A,B}$ fields\footnote{An another possibility is that there are perhaps not all but some matter 
fields which interacts with both gravity fields, i.e. which are the same for both A and B-manifolds, see also discussion about Weyl fermions in \cite{Volovik}.} and 
%if the fields are the same we have the same weak gravitational field for the both manifolds.
this is a main difference between the \eq{CosC7} and \eq{CC2500601} expressions. Namely, in the previous section we considered the interaction between the $h_{A}$ and $h_{B}$ fields  only through the Schwinger-Keldysh
type of interaction, each field was present only on it's own manifold.
Now, instead, we discuss the interaction which are initially non-linear because of the momentum-energy tensor which interacts with both fields. 
Consequently, we also can consider different definitions of the full metric through the weak fields for the separated manifolds, it can be defined similarly to done in \eq{CosC4}-\eq{CosC5} or as 
\beqar
&\,& g^{A}_{\mu \nu}(x)\,= \,g^{A\,0}_{\mu \nu}(x)\,+\,\frac{1}{2}\,\Le h_{\mu \nu}^{A}(x)\,+\,h_{\mu \nu}^{B}(x)\Ra\,; \label{CosC8}\\
&\,& g^{B}_{\mu \nu}(y)\,= \,g^{B\,0}_{\mu \nu}(y)\,+\,\frac{1}{2}\,\Le h_{\mu \nu}^{A}(y)\,+\,h_{\mu \nu}^{B}(y)\Ra\, \label{CosC9}\,,
\eeqar 
further we consider the both cases.

 So, for the \eq{CosC4}-\eq{CosC5} weak fields definition, the mostly general Lagrangian we have is the following one:
\beqar\label{CosC10}
\Gamma\,&=&\,
\frac{m^{2}_{p}}{4}\,\int d^4 x\,\overline{h}^{\mu \nu}_{A}\,G^{-1}_{AA} \,\overline{h}_{\mu \nu}^{A}\,+\,
\frac{m^{2}_{p}}{4}\,\int d^4 x\,\overline{h}^{\mu \nu}_{B}\,G^{-1}_{BB} \,\overline{h}_{\mu \nu}^{B}\,+\,
%\frac{m^{2}_{p}}{4}\,\int d^4 x\,\overline{h}^{\mu \nu}_{A}\,G^{-1}_{BB}\,\overline{h}_{\mu \nu}^{A}\,+\,
%\nonumber \\
%&+&
%\frac{m^{2}_{p}}{4}\,\int d^4 x\,\overline{h}^{\mu \nu}_{B}\,G^{-1}_{AA}\,\overline{h}_{\mu \nu}^{B}\,
%\frac{m^{2}_{p}}{2}\,\int d^4 x\,\overline{h}^{\mu \nu}_{A}\,\Le G^{-1}_{AA} + G^{-1}_{BB}\Ra\,\overline{h}_{\mu \nu}^{B}\,+\,
%\frac{m^{2}_{p}}{2}\,\int d^4 x\,\overline{h}^{\mu \nu}_{A}\,G^{-1}_{BB}\,\overline{h}_{\mu \nu}^{B}\,+\,
\int d^4 x\,\overline{h}^{\mu \nu}_{A}\,T_{\mu \nu\,A}\,+\,
\nonumber \\
&+&
\,\int d^4 x\,\overline{h}^{\mu \nu}_{B}\,T_{\mu \nu\,B}\,+\,
\,\eta_{\mu \nu}\,\int d^4 x\,\overline{h}^{\mu \nu}_{A}\,\Lambda_{0\,A}\,+\,
\,\eta_{\mu \nu}\,\int d^4 x\,\overline{h}^{\mu \nu}_{B}\,\Lambda_{0\,B}\,-\,
\nonumber \\
&-&
\frac{1}{2}\,\int d^4 x\,\int d^4 y\,\overline{h}^{\mu \nu}_{A}(x)\,\Le M^{1\,A A}_{\mu \nu\,\rho \sigma}\,+\,
M^{1\,B B}_{\mu \nu\,\rho \sigma} \Ra_{x y}\,
\,\overline{h}^{\rho \sigma}_{A}(y)\,-\,
\nonumber \\
&-&
\frac{1}{2}\,\int d^4 x\,\int d^4 y\,\overline{h}^{\mu \nu}_{B}(x)\,\Le M^{1\,A A}_{\mu \nu\,\rho \sigma}\,+\,
M^{1\,B B}_{\mu \nu\,\rho \sigma} \Ra_{x y}\,\,\overline{h}^{\rho \sigma}_{B}(y)\,-\,
\nonumber \\
&-&
\frac{1}{2}\,\int d^4 x\,\int d^4 y\,\overline{h}^{\mu \nu}_{A}(x)\,\Le M^{1\,A A}_{\mu \nu\,\rho \sigma}\,+\,
M^{1\,B B}_{\mu \nu\,\rho \sigma} \Ra_{x y}\,
\overline{h}^{\rho \sigma}_{B}(y)\,-\,
\nonumber \\
&-&
\frac{1}{2}\,\int d^4 x\,\int d^4 y\,\overline{h}^{\mu \nu}_{B}(x)\,\Le M^{1\,A A}_{\mu \nu\,\rho \sigma}\,+\,
M^{1\,B B}_{\mu \nu\,\rho \sigma} \Ra_{x y}\,
\overline{h}^{\rho \sigma}_{A}(y)\,-\,
\nonumber \\
&-&
\frac{1}{2}\,\int d^4 x\,\int d^4 y\,\overline{h}^{\mu \nu}_{A}(x)\, M^{1\,A B}_{\mu \nu\,\rho \sigma}\,\overline{h}^{\rho \sigma}_{B}(y)\,-\,
%\nonumber \\
%&-&
\frac{1}{2}\,\int d^4 x\,\int d^4 y\,\overline{h}^{\mu \nu}_{B}(x)\, M^{1\,B A}_{\mu \nu\,\rho \sigma}\,\overline{h}^{\rho \sigma}_{A}(y)\,
\eeqar
with
\beqar\label{CosC11}
&\,&
M_{1\,A A}^{\mu \nu\,\rho \sigma}(x,y)\,=\,\frac{1}{8}
\,\overline{\psi}_{A}(x)\,J_{x}^{\mu \nu}\,S_{F}(x,y)\,
J_{y}^{\rho \sigma}\,\psi_{A}(y)\,;
\\
\label{CosC12}\,
&\,&
M_{1\,B B}^{\mu \nu\,\rho \sigma}(x,y)\,=\,\frac{1}{8}
\,\overline{\psi}_{B}(x)\,J_{x}^{\mu \nu}\,S_{D}(x,y)\,
J_{y}^{\rho \sigma}\,\psi_{B}(y)\,;
\\
&\,&
M_{1\,A B}^{\mu \nu\,\rho \sigma}(x,y)\,=
\,\frac{1}{8}\,\overline{\psi}_{A}(x)\,J_{x}^{\mu \nu}\,S_{>}(x,y)\,
J_{y}^{\rho \sigma}\,\psi_{B}(y)\,;
\\
\label{CosC13}\,
&\,&
M_{1\,B A}^{\mu \nu\,\rho \sigma}(x,y)\,=
\,\frac{1}{8}\,\overline{\psi}_{B}(x)\,J_{x}^{\mu \nu}\,S_{<}(x,y)\,
J_{y}^{\rho \sigma}\,\psi_{A}(y)\,.
\eeqar
The  $T^{\mu \nu}_{A,B}$ and $\Lambda_{0\,A,B}$ in the expression are defined by the \eq{CosC1} effective action, we have for the A-manifold to leading order precision:
\beqar\label{CC9}
&\,&\,\left.\frac{\delta \Gamma(\overline{\psi},\psi,h)}{\delta \overline{h}^{\mu \nu}_{A}(z)}\right|_{\overline{h}_{A}\,=\,0}\,\,=\,
\nonumber \\
&=&\,
\frac{1}{4}\,
\frac{\delta }{\delta \overline{h}^{\mu \mu}_{A}(z)}\Le
\int d^4 x\,\overline{\psi}_{A}(x)\,\Le \overline{h}^{\mu \nu}_{A}(x)+\overline{h}^{\mu \nu}_{B}(x)\Ra
\Le J_{1\,\mu \nu}+\frac{1}{2}\,\eta_{\mu \nu}\, m\Ra\,\psi_{A}(x)\,-\,
\right. \nonumber \\
&-&
\left.
\int d^4 x\,\overline{\psi}_{B}(x)\,\Le \overline{h}^{\mu \nu}_{A}(x)+\overline{h}^{\mu \nu}_{B}(x)\Ra
\Le J_{1\,\mu \nu}+\frac{1}{2}\,\eta_{\mu \nu}\, m\Ra\,\psi_{B}(x)\Ra\,=
\nonumber \\
%&=&\,
%\frac{1}{4}\,\Le
%\overline{\psi}_{A}(z_{A})\,\Le J_{1\,\mu \nu}+\eta_{\mu \nu} J_{0}\Ra\,\psi_{A}(z_{A})\,-\,
%\overline{\psi}_{B}(x_{B}(z_{A}))\,\Le J_{1\,\mu \nu}+\eta_{\mu \nu} J_{0}\Ra\,\psi_{A}(x_{B}(z_{A}))\Ra\,=\,
%\nonumber \\
&=&\,
\frac{1}{4}\,\left[ \,\, \overline{\psi}_{A}(z)\, J_{1\,\mu \nu}\,\psi_{A}(z)\,-\,\overline{\psi}_{B}(z)\, J_{1\,\mu \nu}\psi_{B}(z)\,+\,
\right. \nonumber \\
&+&
\left.
\frac{1}{2}\,m\,\eta_{\mu \nu}\,\Le \overline{\psi}_{A}(z)\,\psi_{A}(z)\,-\,\overline{\psi}_{B}(z)\,\psi_{B}(z)\Ra\right]\,=\,
\nonumber \\
&=&\,
T_{\mu \nu\,A}\,+\,m_{p}^{2}\,\eta_{\mu \nu}\,\Lambda_{A}\,.
%\Le \rho_{A}(z)\,-\,\rho_{B}(x_{B}(z))\,\Ra\,
%=\,\frac{1}{8}\,T_{\mu \nu\,A}\,+\,\frac{m^{2}}{8\,}\,\eta_{\mu \nu}\,\Lambda_{A}
\eeqar
So, the cosmological constants we have here are
\beq\label{CC20}
\Lambda_{0A}\,=\,\Lambda_{0 B}\,=\,\frac{1}{8}\,\frac{m}{m^{2}_{p}}\,\Le \overline{\psi}_{A}(z)\,\psi_{A}(z)\,-\,\overline{\psi}_{B}(z)\,\psi_{B}(z)\Ra\,
%=\,\frac{1}{4\,m_{p}^{2}}\,\frac{1}{2}\,\Le \rho_{A}\,-\,\rho_{B} \Ra\,,
\eeq
or
\beq\label{CC20C1}
\Lambda_{0A}\,=\,\Lambda_{0 B}\,=\,\frac{1}{8\,m^{2}_{p}}\,\Le T^{\mu}_{A\,\mu}\,-\,T^{\mu}_{B\,\mu}\Ra\,;
%=\,\frac{1}{4\,m_{p}^{2}}\,\frac{1}{2}\,\Le \rho_{A}\,-\,\rho_{B} \Ra\,,
\eeq
the answer is different from the previous \eq{CC250015}-\eq{CC250016} expressions by absolute values and by sign of the $\Lambda_{B}$.
The cosmological constants in the expression depend on the difference of the traces of the energy-momentum tensors of the manifolds at the same moment of time but for the functions 
which can live in a  different time's arrows. Again, additionally to the trivial
\beq\label{CCC20}
T^{\mu}_{B\,\mu}\,=\,T^{\mu}_{A\,\mu}
\eeq
identity\footnote{This identity can not be absolutely correct due the non-zero value of $\Lambda_{A}$ of course.} which provides the zero value of the constant, we can consider a case when 
time argument is different for the functions:
%the time variable of the functions is
%. Therefore, we, assuming that the densities are changing with time, can write:
\beq\label{CC13}
t_{A,B}\,=\,\frac{T}{2}\,\pm\,t\,;\,\,\,\,\zeta\,=\,t/T\,
\eeq
obtaining
\beq\label{CC14}
\Lambda_{0A}\, =\,\frac{1}{8 m_{p}^{2}}\Le T^{\mu}_{A\,\mu}(\frac{T}{2}\,+\,t)\,-\,T^{\mu}_{B\,\mu}(\frac{T}{2}\,-\,t)\Ra\,\approx\,\frac{1}{4Tm_{p}^2}\,\Le\,
\frac{\D T^{\mu}_{0A\,\mu} }{\D \zeta}\,+\,\frac{\D T^{\mu}_{0B\,\mu} }{\D \zeta}\Ra\,t\,.
%&\approx&\,-\frac{1}{2Tm_{p}^{2}}\,\frac{\D \T^{\mu}_{\,\mu} }{\D \zeta}\,t\,,\,\,\,\,
\eeq
Here the $T/2$ we can consider roughly as a half of Universe time age,  the cosmological constant  is changing with time and rate of the  the change is defined by the change of the trace of the
energy-momentum tensor, which, in turn, is changing due a new matter creation and/or annihilation in our Universe. 
The model, in fact describes this effect. Considering the action structure, we see from \eq{CosC1} expression,
that there are terms which describe a creation of $\psi_{B}$ fermions through propagation of the $\overline{h}^{\mu \nu}_{A}$ gravitational wave. This field does not interact directly with an
ordinary matter and affects on the propagation of the gravitational fields similarly to the A-spinor fields. 
%In general, from the \eq{CosC1} effective action, it is clear that to this order of precision, the B-spinors with the negative masses affect on he propagation of the gravitational waves at the same way as A-spinors. 
Indeed, the Green's functions for this variant of the action can be obtained from \eq{CC25002} result by the following replacement: 
\beq\label{CC15}
M_{1\,A A},\,M_{1\,B B}\,\rightarrow\,M_{1\,A A}\,+\,M_{1\,B B}\,.
\eeq
Therefore, in comparison to \eq{CC2500131} answer, we obtain that in this particular case there is a modification of the propagator not only through interactions with the B manifold but also through the creation and/or
interaction of the gravitational wave with the B-spinors present in the A manifold. In this extend these spinors are dark matter, of course an additional change of the gravitational potential through 
the interaction with B manifold can be presented here as well.

  For the \eq{CosC8}-\eq{CosC9} weak fields we have the following effective action:
\beqar\label{CC21}
\Gamma\,&=&\,
\frac{m^{2}_{p}}{16}\,\int d^4 x\,\overline{h}^{\mu \nu}_{A}\,\Le G^{-1}_{AA} + G^{-1}_{BB}\Ra \,\overline{h}_{\mu \nu}^{A}\,+\,
\frac{m^{2}_{p}}{16}\,\int d^4 x\,\overline{h}^{\mu \nu}_{B}\,\Le G^{-1}_{BB} + G^{-1}_{AA}\Ra\,\overline{h}_{\mu \nu}^{B}\,+\,
%\frac{m^{2}_{p}}{4}\,\int d^4 x\,\overline{h}^{\mu \nu}_{A}\,G^{-1}_{BB}\,\overline{h}_{\mu \nu}^{A}\,+\,
\nonumber \\
&+&
%\frac{m^{2}_{p}}{4}\,\int d^4 x\,\overline{h}^{\mu \nu}_{B}\,G^{-1}_{AA}\,\overline{h}_{\mu \nu}^{B}\,
\frac{m^{2}_{p}}{8}\,\int d^4 x\,\overline{h}^{\mu \nu}_{A}\,\Le G^{-1}_{AA} + G^{-1}_{BB}\Ra\,\overline{h}_{\mu \nu}^{B}\,+\,
%\frac{m^{2}_{p}}{2}\,\int d^4 x\,\overline{h}^{\mu \nu}_{A}\,G^{-1}_{BB}\,\overline{h}_{\mu \nu}^{B}\,+\,
\int d^4 x\,\overline{h}^{\mu \nu}_{A}\,T_{\mu \nu\,A}\,+\,
\nonumber \\
&+&\,
\int d^4 x\,\overline{h}^{\mu \nu}_{B}\,T_{\mu \nu\,B}\,+\,
\eta_{\mu \nu}\,\int d^4 x\,\overline{h}^{\mu \nu}_{A}\,\Lambda_{0\,A}\,+\,
\eta_{\mu \nu}\,\int d^4 x\,\overline{h}^{\mu \nu}_{B}\,\Lambda_{0\,B}\,-\,
\nonumber \\
&-&
\frac{1}{2}\,\int d^4 x\,\int d^4 y\,\overline{h}^{\mu \nu}_{A}(x)\,\Le M^{1\,A A}_{\mu \nu\,\rho \sigma}\,+\,
M^{1\,B B}_{\mu \nu\,\rho \sigma} \Ra_{x y}\,
\,\overline{h}^{\rho \sigma}_{A}(y)\,-\,
\nonumber \\
&-&
\frac{1}{2}\,\int d^4 x\,\int d^4 y\,\overline{h}^{\mu \nu}_{B}(x)\,\Le M^{1\,B B}_{\mu \nu\,\rho \sigma}\,+\,
M^{1\,B B}_{\mu \nu\,\rho \sigma} \Ra_{x y}\,\,\overline{h}^{\rho \sigma}_{B}(y)\,-\,
\nonumber \\
&-&
\frac{1}{2}\,\int d^4 x\,\int d^4 y\,\overline{h}^{\mu \nu}_{A}(x)\,\Le M^{1\,A A}_{\mu \nu\,\rho \sigma}\,+\,
M^{1\,B B}_{\mu \nu\,\rho \sigma} \Ra_{x y}\,
\overline{h}^{\rho \sigma}_{B}(y)\,-\,
\nonumber \\
&-&
\frac{1}{2}\,\int d^4 x\,\int d^4 y\,\overline{h}^{\mu \nu}_{B}(x)\,\Le M^{1\,A A}_{\mu \nu\,\rho \sigma}\,+\,
M^{1\,B B}_{\mu \nu\,\rho \sigma} \Ra_{x y}\,
\overline{h}^{\rho \sigma}_{A}(y)\,-\,
\\
&-&
\frac{1}{2}\,\int d^4 x\,\int d^4 y\,\overline{h}^{\mu \nu}_{A}(x)\, M^{1\,A B}_{\mu \nu\,\rho \sigma}\,\overline{h}^{\rho \sigma}_{B}(y)\,-\,
%\nonumber \\
%&-&
\frac{1}{2}\,\int d^4 x\,\int d^4 y\,\overline{h}^{\mu \nu}_{B}(x)\, M^{1\,B A}_{\mu \nu\,\rho \sigma}\,\overline{h}^{\rho \sigma}_{A}(y)\,.
\eeqar
The difference between this action and the expression \eq{CosC10} in this order of precision is only in the propagators, 
the $G_{AA}$ and $G_{BB}$ propagators can be  different in general. For example, if they are Feynman and Dyson propagators
correspondingly, then in the expression we obtain the sum or difference of them, depending on the overall sign of the $G^{-1}_{BB}$ operator. We note also, that 
this issue is definitely important for the gravitational wave propagation and dark matter subjects discussion but is not will change drastically overall results obtained by the use of \eq{CosC10} action. 
Therefore, we do no discuss the \eq{CC21} action further preserving the task of it's detailed exploration for the future investigation.

\section{Main results}
\label{S10}

 The main result of the proposed approach is a construction of an effective action where a cancellation of the  one loop contributions of the matter spinor field to the cosmological constant occurs.
The result is based on the CPTM symmetry used for the construction of the the B-manifold actions, see \eq{Spin1805}-\eq{Spin1806} expressions. Consequently, for the full effective bare action,
the general energy-momentum vector of a quantum one-particle state is zero, see \eq{Sec4}-\eq{Sec5}, \eq{Sect3}-\eq{Sect7} and \eq{TS3}-\eq{TS7}. On this stage we have a four variants of the B-manifold Lagrangians with different forms of the spinors  propagators found. At the next step we construct an one loop effective action in an external gravitational field that allowed, requiring the cancellation of unwanted terms, to obtain a final variant of the B-manifold action, see \eq{FG204} and \eq{FG2042}. 

 In the proposed scheme we do not construct and consider an precise analog of the Schwinger-Keldysh mechanism of interactions between A and B spinor and gravitational fields, 
i.e. non-diagonal terms in the corresponding Green's functions matrix, see for example \eq{CC18}-\eq{CC25004} expressions. 
In general these contributions can be important when we discuss the one-loop contributions cancellation, see discussion further in the next Section. 
Without these contributions the structure of the action is simpler and 
cancellation of the ordinary one-loops contributions occurs if we assume the same gravitational field attached to the A,B spinors through corresponding vierbeins. If the fields are different, see
\eq{CC2500601}, then the cancellation has no place, the external gravity "legs" attached to spinor's loop  are different simply, see \eq{CC25001501}-\eq{CC25001503}. In the later case, the cosmological constant value is given by \eq{CC250011} expression, to the leading order it is proportional to the trace of the on-shell classical value of the corresponding energy-momentum tensor, but the problem of the vacuum loop contributions will remain.
In the former case, see \eq{CC25001503}, the cancellation is here and the cosmological constant to the leading order is defined by the difference of traces of the the A and B energy-momentum tensors, see \eq{CC25001504}.
The \eq{CC25001503} condition required for the weak gravitational fields in this case must be explained somehow therefore in the framework of the CPTM symmetry and/or Schwinger-Keldysh description of the interacting A,B systems. 

 The symmetry between the weak gravity fields of A and B manifolds, requested for the one-loops cancellation, can be introduced from the very beginning if we assume that any matter field is interacting with both fields
equivalently or that both metrics have a sum of the A,B weak gravitational fields as a weak field for each separate manifold (i.e. there is only one week field exists for the both Lagrangians), 
see \eq{CosC7} or \eq{CosC8}-\eq{CosC9} correspondingly. In the these cases the cosmological constant is proportional to the difference of the traces of the energy-momentum tensors, 
see \eq{CC20} and \eq{CC14}. 
There are also vertices $M_{1\,A A}$ and $M_{1\,BB}$ which allows an interaction between the A, B weak gravitational fields. These vertices, therefore, allows a creation of the B matter in the A manifold and vice verse creation of the A fields in the B manifold without direct interactions between the manifolds introduced. 
The model, through this mechanism of the A and B fields interactions through gravity only, describes some dark matter appearance in the A manifold therefore. Namely, 
there is the B spinor field which do not interact with A field directly but only through gravity and  
if we consider the dark matter manifestation as some corrections to the graviton propagator aka gravity potential, we see that the corrections appear inevitably. The form and structure 
the corrections depend on the form of interactions between A and B manifolds and/or on the interaction and creation of
of B spinor fields with and by the gravity in the A manifold. Anyway, these corrections are in the model simply by construction, see \eq{CC25001}-\eq{CC25004} and 
\eq{CC250013}-\eq{CC250014}. We do not really calculated these corrections in the present article, this task requires an additional exploration. Another important remark about the "dark" matter appearance defined here as B-spinor fields, is that if the \eq{CC25001506} condition is satisfied we obtain the correct sign for the cosmological constant in the approach. 
%Namely, indeed, to the first sight this condition is in accordance with the statement about a dark matter dominance.

\section{Discussion}
\label{S11}

 The present article is devoted to an exploration of a model intended to explain the smallness of the cosmological constant and, at the same time, the absence of unwanted contributions from the one-loop effective action of spinor field. The framework is based on an application of the CPTM discrete symmetry, introduced in \cite{Ser1,Ser2,Ser3}, to fermion fields. 
As an additional result, it is also demonstrated that dark matter appears in the approach in the form of a modified graviton propagator inevitably, the propagator's modification appears due to the second portion of the matter introduced in the form of B-spinors. Different possibilities of this construction are discussed through the article.

 Throughout the article we underlined that the approach is similar to the Schwinger-Keldysh approach in the non-equilibrium condensed matter physics. The similarities are clear of course. 
There is a request of the CPT symmetry in both approaches,  
doubling of degrees of freedom and two arrows of time. Therefore it is important to underline the differences between the frameworks.
The  Schwinger-Keldysh approach can be considered as a variant of the CPT invariant QFT intended describe non-equilibrium states of matter at nonzero temperature in general. In any case there is only one field which we define on the time contour defined in a complex time's plane. In the present approach the fields are different, there are A and B clones of the matter field related by the CPTM symmetry and the additional time direction arises due of the presence of an additional B manifold. The new time arrow
can be in parallel to the usual, "our", time direction or can be directed in an opposite direction, see \cite{Ser2}. Moreover, we fix the form of the final action by the introduction of the external gravitational field and by consequent request of the cancellation of the undesirable terms in the mutual effective action. It is not clear whether this can be done in the Schwinger-Keldysh approach with gravity included. 
Indeed, the \eq{Spin1806} Lagrangian is an initial point
of the consideration in the Schwinger-Keldysh approach whereas in the present framework we obtained this form of the Lagrangian only after mass sign inverse and negative vierbein introduction. 
%Beginning initially  from the \eq{Spin1806} Lagrangian we will need to preserve the positive vierbein sign thereafter in the case of the CPTM symmetry therefore. Is there the cancellation of the terms in %the effective action it is not clear and it definitely requires an additional investigation. This type of the cancellation, if happens, looks attractive because in this case the only one
%type of matter is required. 
There is an another problem will arise if we will follow the canonical Schwinger-Keldysh formalism and this a problem of the form of the gravity field. 
By construction there are two different gravity fields belong to the two different actions  in the Schwinger-Keldysh approach and then again there is no full one-loop cancellation can happen. 
%Moreover, the final conclusion is that we face the negative cosmological constant as the result in this formulation of the cosmological constant problem.

  However, an important lesson we learned from the Schwinger-Keldysh approach is that there are non-diagonal Green's functions constructed from the correlators of the fields belong to the different time's branches of the contour. In the Schwinger-Keldsysh approach these correlators describe a direct interaction between the fields with different complex time arguments  and in the present formalism
it can describe an interaction between the fields belong to different manifolds.
Correspondingly, there are a plenty of additional
spinor one-loop contributions into the cosmological constant arise when we account the loops with 
$S_{>}(x,y)$ and $S_{<}(y,x)$ non-diagonal propagators included. We do not consider these contributions in the article but formally
in the present framework we have the off-diagonal Green's functions as well. Important to note, that in the Schwinger-Keldysh formalism these functions in momentum representation 
are on-shell delta functions, see \cite{Kel1} for examples, and their contribution to the constant can be quite different from the ordinary matter loops.
The calculations of these type of contributions it is definitely very interesting task which we postpone for an additional publication. In the case when these 
non-diagonal interactions are absent, we stay with a variant of a dark matter model where the B-fields (dark matter) are created by the gravity and both kinds of the matter interact with the same gravity field.

  The problem of the cosmological constant, consequently, is resolved in the approach as following. Without the non-diagonal contributions, the one-loop contributions to the constant must be canceled independently on the number of the external gravitons attached\footnote{There is no cancellation, of course, if we talk about the diagrams with external legs attached to some given sources
of the gravity field, see discussion in \cite{Polchinsky} for example.}. Indeed, in this case there is only one internal momenta integration occurs and there are always a pair of the regularized diagrams which are continued to the Euclidean space differently with different signs in the front of them. The situation of course is more complicated with non-diagonal contribution introduced, then 
there are different contributions to the effective action constructed from the non-diagonal Green's functions, as mentioned above this case will be explored separately. 
%As mentioned above, we did not calculate this kind of diagrams in the article but this contribution is different from the known ordinary tadpole contributions to the energy-momentum tensor. Namely, it has %a structure different from the canelling one-loop contributions and 
%represents a direct interaction between the A and B manifolds. 
%These interactions are absent in the ordinary approaches to the
%calculations of the cosmological constant value. Of course, here we will have interactions between the spinor fields of both manifolds and, possibly, between the gravitational fields if they are different %for the manifolds. 
Assuming the cancellation of the large one-loop contributions, therefore, to the leading order the constant is proportional to the trace of the energy-momentum tensor if gravity fields are different, or to the difference of A and B tensors traces if we operate with the same
gravity field. In the later case we have two possibilities: it can be precisely zero if the values of the different spinor fields are the same at the same value of their arguments or
it can be non-zero value if the values are different, see discussion in \cite{Ser3}.  In the case of the first possibility, the value of the constant is suppressed even more in comparison to the represented result. In the second case we can obtain an equation of the \eq{CC14} type, there the cosmological constant is a rate of a change of the trace of the manifold's energy-momentum tensor during some large period of time which can be taken approximately as a lifetime of the Universe. Another interesting observation is the following one. The trace consists of two parts, classical and quantum on-shell contributions and
both of them can be small in general. Consequently, if we talk about the difference of the traces it could be that one parts of the trace is zero whereas the second provides the current small value of the constant. 

 In this article we did not discuss much a dark matter issue. Nevertheless, similarly to the modification of the propagators in the usual Schwinger-Keldysh model, here we obtain a similar result. 
Any additional matter introduced in the model inevitably leads to the additional terms in the graviton's propagator and modifies it.   
There are different scenario of the modifications depends on the matter properties. In the simplest variant, without non-diagonal Green's functions, the matter creation and/or absorbtion is due a gravity only whereas in the Schwinger-Keldysh like variant there are direct interactions between the A and B spinor and gravity fields. In is interesting to note, that in some extend the additional terms in the A-manifold effective action, which leads to the new graviton propagator,
are similar to the additional terms in the 
gravity action of \cite{Narlikar} model where a creation of the matter (C-field) in our branch of Universe was proposed. Also, the models perhaps can describe 
the MOND, \cite{Milgrom}, modification of the gravity at large distances, see for example \cite{Milgrom1} with a model of negative mass distribution
introduced for the MOND modification.
Of course, some detailed calculations are required in order to demonstrate whether and in which form the modified propagator will be able to mimic the main properties of the MOND model.

 One from the important properties of the framework is its fundamentally possible check for a falsification. Namely, each from the proposed and discussed or
mentioned possibilities for the cosmological constant appearance and value is different from an another. Therefore, any possible description of the experimental data based on the proposed model will depend on the particularly chosen scheme and can be verified in principal. Additionally to this type of a correctness check, there is an another interesting property we can to explore: model's renormalization properties will depend on it's particular construction and scheme. Therefore, in general, it is interesting to check the behavior of the
divergent diagrams, see for example \cite{FLoop}, in the case when we have a doubling of the degrees of freedom in any from the proposed ways. 

 An additional task which we can consider as a further application of the proposed ideas and which was not discussed here is the following one. It is very interesting to consider an application of the CPTM symmetry and/or Schwinger-Keldsyh like approach to the problem of the pure gravitational interactions between two manifolds without matter fields. Namely, the important question is if it possible to construct an extended effective action manifold with a similar cancellation of "bad" contributions at the level of the energy-momentum vector for a one-particle state and at the level of an one-loop effective action. This task could clarify the form and structure of the weak gravitational fields of A and B manifolds in the model that, in turn, will allow to complete the construction of the full one-loop effective action with both spinor and gravity fields included.
Of course, it is interesting and important as well to calculate the effective action of the model including fermion and graviton degrees of freedom to two-loops precision. At this precision a non-trivial mechanism of the possible cancellation of "bad" loop contributions with different Green's functions mixing could be revealed clearer. An another important direction of the model's developing it is an application of it's results to the experimental data description including a direct calculation of the cosmological constant value. Hopefully, these task will clarify yet non clear details of the approach  and will help to understand it's correctness.

%%%%%%%%%%%%%%%%%%%%%%%%%%%%%%%%%%%%%%% 

\newpage
\appendix
\section{CPT transform of solution of Dirac equation }
\label{appendix:a}
\renewcommand{\theequation}{A.\arabic{equation}}
\setcounter{equation}{0}

 We take the \eq{Spin1} quantum fields and perform the CPT transfrom for this field following, mainly, to definitions of \cite{Peskin}.
We have for the $\mathcal{P}$ (parity) transform:
\beqar\label{Spin2}
\hat{\mathcal{P}}\psi(x)& = & \psi_{P}(x)=\gamma^{0}\int\,\frac{d^3 p}{(2\pi)^{3/2}}\,\frac{1}{\sqrt{2\mathcal{E}{p}}}\,\sum_{s}\Le
a_{\bold{p}}^{s}\,u^{s}(p)\,e^{-\imath\,\Le p_{0} x^{0}+x^{i}p^{i}\Ra}\,+\,b_{\bold{p}}^{s\,\dag}\,v^{s}(p)\,e^{\imath\,\Le p_{0} x^{0}+x^{i}p^{i}\Ra}\,
\Ra =
\nonumber \\
&=&
\gamma^{0}\,\psi(t,-\bold{x})\,.
\eeqar
The next is $\mathcal{T}$ (time reversal) transform we account:
\beqar\label{Spin3}
\hat{\mathcal{T}}\psi_{P}(x)& = & \psi_{PT}(x) =
\nonumber \\
&=&
\gamma^{0}\,\gamma^{1}\,\gamma^{3}\,
\int\,\frac{d^3 p}{(2\pi)^{3/2}}\,\frac{1}{\sqrt{2\mathcal{E}{p}}}\,\sum_{s}\Le
a_{\bold{p}}^{-s}\,u^{-s}(p)\,e^{\imath\,\Le p_{0} x^{0}-x^{i}p^{i}\Ra}\,+\,b_{\bold{p}}^{-s\,\dag}\,v^{-s}(p)\,e^{-\imath\,\Le p_{0} x^{0}-x^{i}p^{i}\Ra}\,
\Ra =
\nonumber \\
&=&
\gamma^{0}\,\gamma^{1}\,\gamma^{3}\,\psi(-t,-\bold{x})\,.
\eeqar
An additional transform we account is the charge conjugation $\mathcal{C}$ which defined through
\beq\label{Spin4}
\mathcal{C}\,a_{\bold{p}}^{s} \,\mathcal{C}\,=\,b_{\bold{p}}^{s\,}\,,\,\,\,\mathcal{C}\,b_{\bold{p}}^{s} \,\mathcal{C}\,=\,a_{\bold{p}}^{s\,}
%\hat{\mathcal{C}}\psi_{PT}(x)\,=\,-\,\imath\,\gamma^{2}\,\psi_{PT}^{*}(x)
\eeq
that provides:
\beqar\label{Spin5}
&\,&\hat{\mathcal{C}}\psi_{PT}(x)=  \psi_{CPT}(x) =
\nonumber \\
&=&
\gamma^{0}\,\gamma^{1}\,\gamma^{3}\,
\int\,\frac{d^3 p}{(2\pi)^{3/2}}\,\frac{1}{\sqrt{2\mathcal{E}{p}}}\,\sum_{s}\Le
b_{\bold{p}}^{s}\,u^{s}(p)\,e^{\imath\,\Le p_{0} x^{0}-x^{i}p^{i}\Ra}\,+\,a_{\bold{p}}^{s\,\dag}\,v^{s}(p)\,e^{-\imath\,\Le p_{0} x^{0}-x^{i}p^{i}\Ra}\Ra\,.
\eeqar
Using the following identities
\beq\label{Spin13}
u^{s}(p)\,=\,-\,\imath\,\gamma^{2}\,v^{s *}(p)\,,\,\,\,v^{s}(p)\,=\,-\,\imath\,\gamma^{2}\,u^{s *}(p)\,
\eeq
we obtain instead \eq{Spin5}:
\beqar\label{Spin14}
\psi_{CPT}(x)& = & 
%-\imath\,\gamma^{0}\,\gamma^{1}\,\gamma^{3}\,\gamma^{2}\,\psi_{B}(x) = 
\gamma^{5}\,\psi^{*}(x) = \Le \overline{\psi} \gamma^{0} \gamma^{5}\Ra^{T}\,=\,
\nonumber \\
&=&
\gamma^{5}\,
\int\,\frac{d^3 p}{(2\pi)^{3/2}}\,\frac{1}{\sqrt{2\mathcal{E}{p}}}\,\sum_{s}\Le
b_{\bold{p}}^{s}\,v^{s*}(p)\,e^{\imath\,p x}\,+\,a_{\bold{p}}^{s\,\dag}\,u^{s*}(p)\,e^{-\imath\,p x}\Ra\,
\eeqar
and correspondingly finally we obtain for the classical solution \eq{Spin1} after the $\mathcal{CPT}$ transform:
\beq\label{Spin1301}
\psi(x)\,=\,
\int\,\frac{d^3 p}{(2\pi)^{3/2}}\,\frac{1}{\sqrt{2\mathcal{E}{p}}}\,\sum_{s}\Le
b_{\bold{p}}^{s\,\dag}\,v^{s}(p)\,e^{-\imath\,p x}\,+\,a_{\bold{p}}^{s}\,u^{s}(p)\,e^{\imath\,p x}\Ra\,.
\eeq
As it seems, the function can be obtained by the $x\,\rightarrow\,-x$ transformation of the initial \eq{Spin1}.

%%%%%%%%%%%%%%%%%%%%%%%%%%%%%%%%%%%%%%%%%%%%%%%%%%%%

\newpage
\section{ Propagators of fermion fields }
\label{appendix:b}
\renewcommand{\theequation}{B.\arabic{equation}}
\setcounter{equation}{0}

 We use the \eq{Sec8} definition of the Feynman propagator and write is as
\beqar\label{A1}
S_{F}(x-y)\, & = & \,-\imath\,\frac{\imath}{2\pi}\int\,d \om\,\frac{e^{-\imath\,\om\,(x^{0}-y^{0})}}{\om\,+\,\imath\,\varepsilon}\,
\sum_{s}\int\,\frac{d^{3}k}{(2\pi)^{3}\,2\,\om_{k}}\,u^{s}(k) \overline{u}^{s}(k)\,e^{-\imath\,\om_{k}\,(x^{0}-y^{0})+\imath\vec{k}(\vec{x}-\vec{y})}\,+\,
\nonumber \\
&+&
\imath\,\frac{\imath}{2\pi}\int\,d \om\,\frac{e^{-\imath\,\om\,(y^{0}-x^{0})}}{\om\,+\,\imath\,\varepsilon}\,
\sum_{s}\int\,\frac{d^{3}k}{(2\pi)^{3}\,2\,\om_{k}}\,v^{s}(k) \overline{v}^{s}(k)\,e^{-\imath\,\om_{k}\,(y^{0}-x^{0})+\imath\vec{k}(\vec{y}-\vec{x})}\,
\eeqar
with
\beq\label{A2}
\om_{k}\,=\,\sqrt{k^{2}\,+\,m^2}\,.
\eeq
After the variable change $\omega\,+\,\omega_{k}\,\rightarrow\,\tilde{\omega}$ in the both terms 
we arrive to
\beq\label{A3}
S_{F}(x-y)\,  =  \,\int\,\frac{d\tilde{\om}d^{3}k}{(2\pi)^{4}\,2\,\om_{k}}\,\Le \hat{k}\,+\,m \Ra\,\frac{e^{-\imath\,k\,(x-y)}}{\tilde{\om}\,-\,\om_{k}\,+\,\imath\varepsilon }\,-\,
\int\,\frac{d\tilde{\om}d^{3}k}{(2\pi)^{4}\,2\,\om_{k}}\,\Le \hat{k}\,-\,m \Ra\,\frac{e^{-\imath\,k\,(y-x)}}{\tilde{\om}\,-\,\om_{k}\,+\,\imath\varepsilon }\,
\eeq
and replacing
\beq\label{A4}
k\,\rightarrow\,-k\,
\eeq
in the second term we obtain finally:
\beqar\label{A5}
S_{F}(x-y)\, & = &  \,\Le \imath\hat{\D_{x}}\,+\,m \Ra\,\int\,\frac{d\tilde{\om}d^{3}k}{(2\pi)^{4}\,2\,\om_{k}}\,\frac{e^{-\imath\,k\,(x-y)}}{\tilde{\om}\,-\,\om_{k}\,+\,\imath\varepsilon }\,-\,
\Le \imath\hat{\D_{x}}\,+\,m \Ra\,\int\,\frac{d\tilde{\om}d^{3}k}{(2\pi)^{4}\,2\,\om_{k}}\,\frac{e^{-\imath\,k\,(x-y)}}{\tilde{\om}\,+\,\om_{k}\,-\,\imath\varepsilon }\,=\,
\nonumber \\
&=&
\Le \imath\hat{\D_{x}}\,+\,m \Ra\,\int\,\frac{d^{4}k}{(2\pi)^{4}}\,\frac{e^{-\imath\,k\,(x-y)}}{k^2\,-\,m^2\,+\,\imath\varepsilon }\,=\,\Le \imath\hat{\D_{x}}\,+\,m \Ra\,G_{F}
\eeqar
with $G_{F}$ as Feynman propagator of scalar field and with
\beq\label{A6}
\Le\,\imath\,\hat{\D}_{x}\,-\,m\,\Ra\,S_{F}(x,y)\,=\,\delta^{4}(x-y)\,
\eeq
satisfied.
Defining the Dyson propagator as
\beq\label{A7}
S_{D}(x-y)\, = \,
-\imath\,\Le \theta(x^{0}-y^{0})\,<0|\overline{\psi}(y) \psi(x)\,-\,
\theta(y^{0}-x^{0})\,<0|\psi(x) \overline{\psi}(y) |0>|0>\Ra\,
\eeq
we obtain after similar calculations
\beq\label{A8}
S_{D}(x-y)\,=\,-\,\Le \imath\hat{\D_{x}}\,+\,m \Ra\,\int\,\frac{d^{4}k}{(2\pi)^{4}}\,\frac{e^{-\imath\,k\,(x-y)}}{k^2\,-\,m^2\,-\,\imath\varepsilon }\,=\,\Le \imath\hat{\D_{x}}\,+\,m \Ra\,G_{D}
\eeq
with 
\beq\label{A9}
-\,\Le\,\imath\,\hat{\D}_{x}\,-\,m\,\Ra\,S_{D}(x,y)\,=\,\delta^{4}(x-y)\,
\eeq
satisfied.

%%%%%%%%%%%%%%%%%%%%%%%%%%%%%%%%%%%%%%%%%%%%%%%%%%%%%%
 
\newpage
\section{ One loop integrals}
\label{appendix:c}
\renewcommand{\theequation}{C.\arabic{equation}}
\setcounter{equation}{0}

\subsection*{First term}

 The first term contributes in \eq{OL12} has the following form:
\beq\label{CC01}
Int_{2a}=\int d^4 x\,s(x)\int d^4 y\,s(y)\Le \gamma^{\mu}\,\Le \D_{\mu x}\D_{\mu_{1} y}\,S_{F}(x,y)\Ra\,\gamma^{\mu_{1}}\,S_{F}(y,x) +
\gamma^{\mu}\,\Le \D_{\mu x}\,S_{F}(x,y)\Ra\,\gamma^{\mu_{1}}\,\Le \D_{\mu_{1} y}\,S_{F}(y,x)\Ra\Ra\,.
\eeq
After Fourier transform we have:
\beqar
&\,&
Int_{2a}= \int \frac{d^4 p_{i}}{(2\pi)^{4}} \tilde{s}(p_{i})\int d^4 p_{f} \tilde{s}(p_{f})\delta^{4}(p_{i}+p_{f})\int \frac{d^{4} p}{(2\pi)^{4}}
\left[ 
\gamma^{\mu}\frac{\gamma^{\rho} p_{\rho}+m}{p^2-m^2+\imath\varepsilon}\,
\gamma^{\mu_{1}}
\frac{ \gamma^{\rho_{1}}\Le p-p_{f}\Ra_{\rho_1}+m}{(p-p_{f})^2-m^2\,+\,\imath\varepsilon} p_{\mu}p_{\mu_{1}}\,-\,
\right.\nonumber \\
&-&
\left.
%\int\,\frac{d^{4} p}{(2\pi)^{4}}\,
\gamma^{\mu}\,\frac{\gamma^{\rho} p_{\rho}+m}{p^2\,-\,m^2\,+\,\imath\varepsilon}\,
\gamma^{\mu_{1}}\,
\frac{ \gamma^{\rho_{1}}\Le p-p_{f}\Ra_{\rho_1}+m}{(p-p_{f})^2\,-\,m^2\,+\,\imath\varepsilon}\,p_{\mu}\Le p-p_{f}\Ra_{\mu_{1}}\right]\,=\,
\nonumber \\
&=&\,
\int  \frac{d^4 p_{i}}{(2\pi)^{4}} \tilde{s}(p_{i})\int d^4 p_{f} \tilde{s}(p_{f})\delta^{4}(p_{i}+p_{f})\int \frac{d^{4} p}{(2\pi)^{4}}
\gamma^{\mu}\frac{\gamma^{\rho} p_{\rho}+m}{p^2-m^2+\imath\varepsilon}\,
\gamma^{\mu_{1}}
\frac{ \gamma^{\rho_{1}}\Le p-p_{f}\Ra_{\rho_1}+m}{(p-p_{f})^2-m^2\,+\,\imath\varepsilon} p_{\mu}p_{f\mu_{1}}\,
\label{CC02}
\eeqar
Introducing Feynman parameters and using a symmetry of the integral with respect to the $x\,\leftrightarrow \,(1-x)$ replace and also $\rho\,\leftrightarrow \,\rho_{1} $ symmetry in the expressions with gamma matrices trace, we obtain:
\beqar\label{CC03}
&\,& Int_{2a}\,=\,2\,\imath\,\int \frac{d^4 p_{i}}{(2\pi)^{4}} \tilde{s}(p_{i})\int d^4 p_{f} \tilde{s}(p_{f})\,p_{f}^{2}\,\delta^{4}(p_{i}+p_{f})\int_{0}^{1}\, x\,dx\,\int \frac{d^{4} l_{E}}{(2\pi)^{4}}\,
\frac{l_{E}^{2}-2 p_{f}^{2} x\Le 1-x \Ra +2 m^2  }{\Le l^{2}_{E}\,+\,\Delta\Ra^{2}}\,=\,
\nonumber \\
&=&
\,\imath\,\int \frac{d^4 p_{i}}{(2\pi)^{4}} \tilde{s}(p_{i})\int d^4 p_{f} \tilde{s}(p_{f})\,p_{f}^{2}\,\delta^{4}(p_{i}+p_{f})\int_{0}^{1}\,dx\,\int \frac{d^{4} l_{E}}{(2\pi)^{4}}\,
\frac{l_{E}^{2}-2 p_{f}^{2} x\Le 1-x \Ra +2 m^2  }{\Le l^{2}_{E}\,+\,\Delta\Ra^{2}}\,.
\eeqar
here $\Delta\,=\,m^{2}\,-\,x\,\Le 1-x\Ra\,p_{f}^{2}$, $l\,=\,p\,-\,x\,p_{f}$ and Wick rotation  $l_{0}\,\rightarrow\,\imath l_{0E}$ was performed. If the 
on-shell $p_{f}^{2}\,=\,0$ condition  is satisfied for the $s_{\mu \nu}(x)$ gravitational field, after a renormalization the integral is zero of course.

%pass in the Euclidean space for the $p_{i}$ and $p_{f}$ momenta performing
%$d^4p_{i,f}\,\rightarrow\,\imath\,d^4p_{i,f}$ and $p^{2}_{i,f}\,\rightarrow\,-\,p^{2}_{i,f}\,$ transforms in the integrals. Namely, we can integrate out one from the momenta by the use of delta function and 
%the remain momenta we can  rotate into Euclidean space; after that we can restore the delta function in the expression understanding the $p_{f,i}$ as momenta in Euclidean space. 
%In this case we will have for the \eq{CC03} answer:
%\beq\label{CC0301}
%Int_{2a}\,=\,-
%\int \frac{d^4 p_{i}}{(2\pi)^{4}} \tilde{s}(p_{i})\int d^4 p_{f} \tilde{s}(p_{f})\,p_{f}^{2}\,\delta^{4}(p_{i}+p_{f})\int_{0}^{1}\,dx\,\int \frac{d^{4} l_{E}}{(2\pi)^{4}}\,
%\frac{l_{E}^{2}-2 p_{f}^{2} x\Le 1-x \Ra +2 m^2  }{\Le l^{2}_{E}\,+\,\Delta\Ra^{2}}\,
%\eeq
%with integration on Euclidean $p_{i}$ and $p_{f}$ assumed.

\subsection*{Second term}

  Now we calculate the contribution of the following two (there is a symmetry with respect to $x$ and $y$ interchange in \eq{OL12})  terms  in \eq{OL12}:
\beqar
&\,& Int_{2b}=-\int d^4 x\,s(x)\int d^4 y\,s_{c_{1} \nu_{1}}(y)\,\eta^{\mu_{1} \nu_{1}}\, \gamma^{\mu}\,\Le \D_{\mu x}\D_{\mu_{1} y}\,S_{F}(x,y)\Ra\,\gamma^{c_{1}}\,S_{F}(y,x)\,-\,
\nonumber \\
&-&
\int d^4 x\,s(x)\int d^4 y\,s_{c_{1} \nu_{1}}(y)\,\eta^{\mu_{1} \nu_{1}}\, \gamma^{\mu}\,\Le \D_{\mu x}\,S_{F}(x,y)\Ra\,\gamma^{c_{1}}\,\Le \D_{\mu_{1} y}\,S_{F}(y,x)\Ra\,-\,
\nonumber \\
&-&
\int d^4 x\,s(x)\int d^4 y\,s_{c_{1} \nu_{1}}(y)\,\eta^{\mu_{1} \nu_{1}}\, \gamma^{\mu}\,\Le \D_{\mu_{1} y}\,S_{F}(x,y)\Ra\,\gamma^{c_{1}}\,\Le \D_{\mu x}\,S_{F}(y,x)\Ra\,-\,
\nonumber \\
&-&
\int d^4 x\,s(x)\int d^4 y\,s_{c_{1} \nu_{1}}(y)\,\eta^{\mu_{1} \nu_{1}}\, \gamma^{\mu}\,S_{F}(x,y) \,\gamma^{c_{1}}\,\Le \D_{\mu x}\D_{\mu_{1} y}\,S_{F}(x,y)\Ra\,.
\label{CC04}
\eeqar
Performing Fourier transform we obtain:
\beqar
Int_{2b}& = & \,-
\int  \frac{d^4 p_{i}}{(2\pi)^{4}} \tilde{s}(p_{i})\int d^4 p_{f} \tilde{s}_{c_{1} \nu_{1}}(p_{f})\delta^{4}(p_{i}+p_{f})\,\eta^{\mu_{1} \nu_{1}}\,
\nonumber \\
&\,&\,
\int \frac{d^{4} p}{(2\pi)^{4}}
\gamma^{\mu}\frac{\gamma^{\rho} p_{\rho}+m}{p^2-m^2+\imath\varepsilon}\,
\gamma^{c_{1}}
\frac{ \gamma^{\rho_{1}}\Le p-p_{f}\Ra_{\rho_1}+m}{(p-p_{f})^2-m^2\,+\,\imath\varepsilon} p_{f \mu}p_{f\mu_{1}}\,
\label{CC05}
\eeqar
that provides the final answer:
\beq\label{CC06}
Int_{2b} =  -\,2\,\imath\,\int  \frac{d^4 p_{i}}{(2\pi)^{4}} \tilde{s}(p_{i})\int d^4 p_{f} \tilde{s}_{\mu \nu}(p_{f})\,p^{\mu}_{f} p^{\nu}_{f} \,\delta^{4}(p_{i}+p_{f})\,
\int_{0}^{1} dx\,\int \frac{d^{4} l_{E}}{(2\pi)^{4}}\,
\frac{l_{E}^{2}-2p_{f}^{2} x\Le 1-x \Ra +2 m^2  }{\Le l^{2}_{E}\,+\,\Delta\Ra^{2}}\,.
\eeq
Because we do not really use the results for the one-loop action, the next simplification we can make is to simplify the integrals on $p_{i,f}$ variables obtaining in this particular case:
%For the Euclidean external momenta we obtain correspondingly:
\beq\label{CCC06}
Int_{2b} = - \frac{\imath}{2}\,\int  \frac{d^4 p_{i}}{(2\pi)^{4}} \tilde{s}(p_{i})\int d^4 p_{f} \tilde{s}(p_{f})\,p^{2}_{f} \,\delta^{4}(p_{i}+p_{f})\,
\int_{0}^{1} dx\,\int \frac{d^{4} l_{E}}{(2\pi)^{4}}\,
\frac{l_{E}^{2}-2p_{f}^{2} x\Le 1-x \Ra +2 m^2  }{\Le l^{2}_{E}\,+\,\Delta\Ra^{2}}\,.
\eeq
We note, that here for the case of polarization of outgoing (or incoming) classical gravitation field $s_{\mu \nu}\,p^{\mu}_{f}\,=\,0$ the integral is zero, see further also the next term contribution.

\subsection*{Third term}

 The third term (two terms together in \eq{OL12}) we calculate is the following one:
\beqar
&\,& Int_{2c}=-\frac{1}{2}\int d^4 x\,s(x)\int d^4 y\,s_{c_{1} a_{1}}(y)\,\gamma^{\mu}\,\Le \D_{\mu x}\D_{b_{1} y}\,S_{F}(x,y)\Ra\,\gamma^{c_{1}}\,[\gamma^{a_{1}}\,\gamma^{b_{1}}]\,
S_{F}(y,x)\,-\,
\nonumber \\
&-&
\frac{1}{2}\int d^4 x\,s(x)\int d^4 y\,s_{c_{1} a_{1}}(y)\,\gamma^{\mu}\,\Le \D_{\mu x}\,S_{F}(x,y)\Ra\,\gamma^{c_{1}}\,[\gamma^{a_{1}}\,\gamma^{b_{1}}]\,\Le \D_{b_{1} y}\,S_{F}(y,x)\Ra\,-\,
\nonumber \\
&-&
\frac{1}{2}\int d^4 x\,s(x)\int d^4 y\,s_{c_{1} a_{1}}(y)\,\gamma^{\mu}\,\Le \D_{b_{1} y}\,S_{F}(x,y)\Ra\,\gamma^{c_{1}}\,[\gamma^{a_{1}}\,\gamma^{b_{1}}]\,\Le \D_{\mu x}\,S_{F}(y,x)\Ra\,-\,
\nonumber \\
&-&
\frac{1}{2}\int d^4 x\,s(x)\int d^4 y\,s_{c_{1} a_{1}}(y)\,\gamma^{\mu}\,S_{F}(x,y) \,\gamma^{c_{1}}\,[\gamma^{a_{1}}\,\gamma^{b_{1}}]\,\Le \D_{\mu x}\D_{b_{1} y}\,S_{F}(x,y)\Ra\,.
\label{CC07}
\eeqar
Similarly to done before we have:
\beqar
Int_{2c}& = & \,-\frac{1}{2}\,
\int  \frac{d^4 p_{i}}{(2\pi)^{4}} \tilde{s}(p_{i})\int d^4 p_{f} \tilde{s}_{c_{1} a_{1}}(p_{f})\delta^{4}(p_{i}+p_{f})\,
\nonumber \\
&\,&\,
\int \frac{d^{4} p}{(2\pi)^{4}}
\gamma^{\mu}\frac{\gamma^{\rho} p_{\rho}+m}{p^2-m^2+\imath\varepsilon}\,
\gamma^{c_{1}}\,[\gamma^{a_{1}}\,\gamma^{b_{1}}]\,
\frac{ \gamma^{\rho_{1}}\Le p-p_{f}\Ra_{\rho_1}+m}{(p-p_{f})^2-m^2\,+\,\imath\varepsilon} p_{f \mu}p_{f b_{1}}\,.
\label{CC08}
\eeqar
Using the usual identities
\beq\label{CCC2}
\gamma^{\mu}\,\gamma^{\nu}\,\gamma^{\rho}\,=\,\eta^{\mu \nu}\gamma^{\rho}\, +\, \eta^{\nu \rho} \gamma^{\mu}\, -\,\eta^{\mu \rho}  \gamma^{\nu}\,-\,\imath\,\epsilon^{\mu \nu \rho \sigma}
\,\eta_{\sigma \sigma_{1}}  \gamma^{\sigma_{1}}\,\gamma^{5}\,
\eeq
we have
\beq\label{CC3}
\gamma^{\mu}\,[\gamma^{\nu}\,\gamma^{\rho}]\,=\,2\,\Le \eta^{\mu \nu}\gamma^{\rho}\, -\,\eta^{\mu \rho}  \gamma^{\nu}\,-\,\imath\,\epsilon^{\mu \nu \rho \sigma}
\,\eta_{\sigma \sigma_{1}}  \gamma^{\sigma_{1}}\,\gamma^{5}\Ra\,.
\eeq
and obtain
\beqar
Int_{2c}& = & \,-\,
\int  \frac{d^4 p_{i}}{(2\pi)^{4}} \tilde{s}(p_{i})\int d^4 p_{f} \tilde{s}_{c_{1} a_{1}}(p_{f})\delta^{4}(p_{i}+p_{f})\,
\nonumber \\
&\,&\,
\int \frac{d^{4} p}{(2\pi)^{4}}\,\left[ \eta^{c_{1} a_{1}}\,
\gamma^{\mu}\frac{\gamma^{\rho} p_{\rho}+m}{p^2-m^2+\imath\varepsilon}\,
\gamma^{b_{1}}\
\frac{ \gamma^{\rho_{1}}\Le p-p_{f}\Ra_{\rho_1}+m}{(p-p_{f})^2-m^2\,+\,\imath\varepsilon}\,-\,
\right. \nonumber \\
&-&\,
\left.
\eta^{c_{1} b_{1}}\,
\gamma^{\mu}\frac{\gamma^{\rho} p_{\rho}+m}{p^2-m^2+\imath\varepsilon}\,
\gamma^{a_{1}}\
\frac{ \gamma^{\rho_{1}}\Le p-p_{f}\Ra_{\rho_1}+m}{(p-p_{f})^2-m^2\,+\,\imath\varepsilon}\,-\,
\right. \nonumber \\
&-&\,
\left.
\imath\,\epsilon^{c_{1} a_{1} b_{1} d}\,\eta_{d d_{1}}
\gamma^{\mu}\frac{\gamma^{\rho} p_{\rho}+m}{p^2-m^2+\imath\varepsilon}\,
\gamma^{d_{1}}\,\gamma^{5}\,
\frac{ \gamma^{\rho_{1}}\Le p-p_{f}\Ra_{\rho_1}+m}{(p-p_{f})^2-m^2\,+\,\imath\varepsilon} \right]\,p_{f \mu}p_{f b_{1}}\,.
\label{CC09}
\eeqar
The second term in this expression is canceled with the previous one, the \eq{CC05} answer; the third term is zero due product of symmetric $s$ and antisymmetric tensors. So, we stay here therefore with only non-zero first term, which is
\beq\label{CC010}
Int_{2c}\,=\,-\,2\,\imath\,\int \frac{d^4 p_{i}}{(2\pi)^{4}} \tilde{s}(p_{i})\int d^4 p_{f} \tilde{s}(p_{f})\,p_{f}^{2}\,\delta^{4}(p_{i}+p_{f})\int_{0}^{1} dx\,\int \frac{d^{4} l_{E}}{(2\pi)^{4}}\,
\frac{l_{E}^{2}-2 p_{f}^{2} x\Le 1-x \Ra +2 m^2  }{\Le l^{2}_{E}\,+\,\Delta\Ra^{2}}\,.
\eeq
%and for Euclidean momenta:
%\beq\label{CC01001}
%Int_{2c}\,=\,2\,\imath\,\int \frac{d^4 p_{i}}{(2\pi)^{4}} \tilde{s}(p_{i})\int d^4 p_{f} \tilde{s}(p_{f})\,p_{f}^{2}\,\delta^{4}(p_{i}+p_{f})\int_{0}^{1} dx\,\int \frac{d^{4} l_{E}}{(2\pi)^{4}}\,
%\frac{l_{E}^{2}-2 p_{f}^{2} x\Le 1-x \Ra +2 m^2  }{\Le l^{2}_{E}\,+\,\Delta\Ra^{2}}\,,
%\eeq
%%the term has the same structure as the \eq{CC03} answer. 

\subsection*{Fourth term}

 The next term we have is the following one:
\beq\label{CC011}
Int_{2d}=\imath\,m\,\int d^4 x\,s(x)\int d^4 y\,s(y)\,\gamma^{\mu}\,\Le \Le \D_{\mu x}\,S_{F}(x,y)\Ra\,S_{F}(y,x) +
S_{F}(x,y)\,\Le \D_{\mu x}\,S_{F}(y,x)\Ra\Ra\,.
\eeq
Performing Fourier transform we obtain that this contribution is zero
\beqar
&\,& Int_{2d}= m\,
\int  \frac{d^4 p_{i}}{(2\pi)^{4}} \tilde{s}(p_{i})\int d^4 p_{f} \tilde{s}(p_{f})\delta^{4}(p_{i}+p_{f})\int \frac{d^{4} p}{(2\pi)^{4}}
\gamma^{\mu}\frac{\gamma^{\rho} p_{\rho}+m}{p^2-m^2+\imath\varepsilon}\,
\frac{ \gamma^{\rho_{1}}\Le p-p_{f}\Ra_{\rho_1}+m}{(p-p_{f})^2-m^2\,+\,\imath\varepsilon}\,p_{f\mu}\,=\,
\nonumber \\
&=&\,
m^2\,\int  \frac{d^4 p_{i}}{(2\pi)^{4}} \tilde{s}(p_{i})\int d^4 p_{f} \tilde{s}(p_{f})\delta^{4}(p_{i}+p_{f})\int \frac{d^{4} p}{(2\pi)^{4}}
\frac{\gamma^{\mu}\Le \gamma^{\rho} p_{\rho}+ \gamma^{\rho_{1}}\Le p-p_{f}\Ra_{\rho_1}\Ra}{\Le p^2-m^2+\imath\varepsilon\,\Ra\,\Le (p-p_{f})^2-m^2\,+\,\imath\varepsilon \Ra}\,p_{f\mu}\,=\,0\,.
\nonumber \\
&\,&
\label{CC012}
\eeqar
due the linear dependence of $l\,=\,p\,-\,x\,p_{f}$ in the numerator of the integral.

\subsection*{Fifth term}

 The fifth term is the new one, we have:
\beqar\label{CC013}
Int_{2e}& = &\int d^4 x\,s_{c \nu}(x)\int d^4 y\,s_{c_{1} \nu_{1}}(y)\eta^{\mu \nu}\eta^{\mu_{1} \nu_{1}} \,
\nonumber \\
&\,&
\Le \gamma^{c}\,\Le \D_{\mu x}\D_{\mu_{1} y}\,S_{F}(x,y)\Ra\,\gamma^{c_{1}}\,S_{F}(y,x) +
\gamma^{c}\,\Le \D_{\mu x}\,S_{F}(x,y)\Ra\,\gamma^{c_{1}}\,\Le \D_{\mu_{1} y}\,S_{F}(y,x)\Ra\Ra\,.
\eeqar
After the Fourier transform we have:
\beqar\label{CC014}
Int_{2e} & = & 
\int  \frac{d^4 p_{i}}{(2\pi)^{4}} \tilde{s}_{c \nu}(p_{i})\int d^4 p_{f} \tilde{s}_{c_{1}\nu_{1}}(p_{f})\delta^{4}(p_{i}+p_{f})\eta^{\mu \nu}\eta^{\mu_{1} \nu_{1}} \,
\nonumber \\
&\,&\,
\int \frac{d^{4} p}{(2\pi)^{4}}
\gamma^{c}\frac{\gamma^{\rho} p_{\rho}+m}{p^2-m^2+\imath\varepsilon}\,
\gamma^{c_{1}}
\frac{ \gamma^{\rho_{1}}\Le p-p_{f}\Ra_{\rho_1}+m}{(p-p_{f})^2-m^2\,+\,\imath\varepsilon} p_{\mu}p_{f\mu_{1}}\,.
\eeqar
The calculation gives for this term:
\beqar\label{CC015}
&\,& Int_{2e} = \,\imath
\int \frac{d^4 p_{i}}{(2\pi)^{4}}\tilde{s}_{c \nu}(p_{i})\,\int d^4 p_{f}\, \tilde{s}_{\nu_{1}}^{c}(p_{f})p_{f}^{\nu}\, p_{f}^{\nu_{1}}\delta^{4}(p_{i}+p_{f})\,
\int_{0}^{1}\,dx\,\int \frac{d^{4} l_{E}}{(2\pi)^{4}}
\frac{l_{E}^{2}+2 p_{f}^{2} x\Le 1-x \Ra +2 m^2  }{\Le l^{2}_{E}\,+\,\Delta\Ra^{2}}-
\nonumber \\
&-&\,
4\,\imath\,
\int  \frac{d^4 p_{i}}{(2\pi)^{4}}\,\tilde{s}_{c \nu}(p_{i})p_{f}^{c}\,p_{f}^{\nu}\,\int d^4 p_{f} \tilde{s}_{c_{1}\nu_{1}}(p_{f})\,p_{f}^{c_{1}}\, p_{f}^{\nu^{1}}\,\delta^{4}(p_{i}+p_{f})\,
\int_{0}^{1}\,dx\,\int \frac{d^{4} l_{E}}{(2\pi)^{4}}\,
\frac{x\Le 1-x \Ra}{\Le l^{2}_{E}\,+\,\Delta\Ra^{2}}\,.
\eeqar
or
\beqar\label{CC01501}
&\,& Int_{2e} = \,\frac{\imath}{4}
\int \frac{d^4 p_{i}}{(2\pi)^{4}}\tilde{s}_{c \nu}(p_{i})\,\int d^4 p_{f}\, \tilde{s}^{c \nu}(p_{f})p_{f}^{2}\,\delta^{4}(p_{i}+p_{f})\,
\int_{0}^{1}\,dx\,\int \frac{d^{4} l_{E}}{(2\pi)^{4}}
\frac{l_{E}^{2}+2 p_{f}^{2} x\Le 1-x \Ra +2 m^2  }{\Le l^{2}_{E}\,+\,\Delta\Ra^{2}}-
\nonumber \\
&-&\,
\frac{\imath}{4}\,
\int  \frac{d^4 p_{i}}{(2\pi)^{4}}\,\tilde{s}(p_{i})\,\int d^4 p_{f} \tilde{s}(p_{f})\,p_{f}^{4}\, \delta^{4}(p_{i}+p_{f})\,
\int_{0}^{1}\,dx\,\int \frac{d^{4} l_{E}}{(2\pi)^{4}}\,
\frac{x\Le 1-x \Ra}{\Le l^{2}_{E}\,+\,\Delta\Ra^{2}}\,.
\eeqar

\subsection*{Sixth term}

 For the sixth term we write:
\beqar\label{CC016}
&\,& Int_{2f}= \frac{1}{2}\,\int d^4 x\,s_{c \nu}(x)\int d^4 y\,s_{c_{1} a_{1}}(y)\eta^{\mu \nu}\,
\\
&\,&
\Le \gamma^{c}\,\Le \D_{\mu x}\D_{b_{1} y}\,S_{F}(x,y)\Ra\,\gamma^{c_{1}}\,[\gamma^{a_{1}}\,\gamma^{b_{1}}] \,S_{F}(y,x) +
\gamma^{c}\,\Le \D_{\mu x}\,S_{F}(x,y)\Ra\,\gamma^{c_{1}}\,[\gamma^{a_{1}}\,\gamma^{b_{1}}]\,\Le \D_{b_{1} y}\,S_{F}(y,x)\Ra\right.
\nonumber \\
&+&
\left.
\gamma^{c}\,\Le \D_{b_{1} y}\,S_{F}(x,y)\Ra\,\gamma^{c_{1}}\,[\gamma^{a_{1}}\,\gamma^{b_{1}}] \,\Le \D_{\mu x}S_{F}(y,x)\Ra +
\gamma^{c}\,S_{F}(x,y)\,\gamma^{c_{1}}\,[\gamma^{a_{1}}\,\gamma^{b_{1}}]\,\Le \D_{\mu x}\D_{b_{1} y}\,S_{F}(y,x)\Ra\,
\Ra\,\nonumber \,.
\eeqar
After the Fourier transform we obtain:
\beqar\label{CC017}
Int_{2f} & = & 
\frac{1}{2}\int  \frac{d^4 p_{i}}{(2\pi)^{4}} \tilde{s}_{c \nu}(p_{i})\int d^4 p_{f} \tilde{s}_{c_{1}a_{1}}(p_{f})\delta^{4}(p_{i}+p_{f})\eta^{\mu \nu}\,
\nonumber \\
&\,&\,
\int \frac{d^{4} p}{(2\pi)^{4}}
\gamma^{c}\frac{\gamma^{\rho} p_{\rho}+m}{p^2-m^2+\imath\varepsilon}\,
\gamma^{c_{1}}\,[\gamma^{a_{1}}\,\gamma^{b_{1}}]\,
\frac{ \gamma^{\rho_{1}}\Le p-p_{f}\Ra_{\rho_1}+m}{(p-p_{f})^2-m^2\,+\,\imath\varepsilon} p_{f\mu}p_{f b_{1}}\,.
\eeqar
After some calculation we get:
\beqar\label{CC018}
&\,& Int_{2f} = 2\imath
\int \frac{d^4 p_{i}}{(2\pi)^{4}}\,\int d^4 p_{f}\, \tilde{s}(p_{i})\,\tilde{s}_{c \nu}(p_{f})\, p_{f}^{c}\,p_{f}^{\nu}\delta^{4}(p_{i}+p_{f})\,
\int_{0}^{1}\,dx\,\int \frac{d^{4} l_{E}}{(2\pi)^{4}}
\frac{l_{E}^{2}-2 p_{f}^{2} x\Le 1-x \Ra +2 m^2  }{\Le l^{2}_{E}\,+\,\Delta\Ra^{2}}\,-
\nonumber \\
&-&\,
2\,\imath\,
\int  \frac{d^4 p_{i}}{(2\pi)^{4}}\,\tilde{s}_{c \nu}(p_{i})p_{f}^{\nu}\,\int d^4 p_{f} \tilde{s}_{c_{1}}^{c}(p_{f})p_{f}^{c_{1}}\,\delta^{4}(p_{i}+p_{f})\,
\int_{0}^{1}\,dx\,\int \frac{d^{4} l_{E}}{(2\pi)^{4}}\,
\frac{l_{E}^{2} + 2 p_{f}^{2} x\Le 1-x \Ra + 2m^2  }{\Le l^{2}_{E}\,+\,\Delta\Ra^{2}}\,+\,
\nonumber \\
&+&\,
8\,\imath\,
\int  \frac{d^4 p_{i}}{(2\pi)^{4}}\,\tilde{s}_{c \nu}(p_{i})p_{f}^{c}\,p_{f}^{\nu}\,\int d^4 p_{f} \tilde{s}_{c_{1}\nu_{1}}(p_{f})\,p_{f}^{c_{1}}\, p_{f}^{\nu_{1}}\,\delta^{4}(p_{i}+p_{f})\,
\int_{0}^{1}\,dx\,\int \frac{d^{4} l_{E}}{(2\pi)^{4}}\,
\frac{x\Le 1-x \Ra}{\Le l^{2}_{E}\,+\,\Delta\Ra^{2}}\,
\eeqar
or
%For the Euclidean external momenta we have:
\beqar\label{CC01801}
&\,& Int_{2f} = \frac{\imath}{2}
\int \frac{d^4 p_{i}}{(2\pi)^{4}}\,\tilde{s}(p_{i})\,\int d^4 p_{f}\, \tilde{s}(p_{f})\, p_{f}^{2}\,\delta^{4}(p_{i}+p_{f})\,
\int_{0}^{1}\,dx\,\int \frac{d^{4} l_{E}}{(2\pi)^{4}}
\frac{l_{E}^{2}-2 p_{f}^{2} x\Le 1-x \Ra +2 m^2  }{\Le l^{2}_{E}\,+\,\Delta\Ra^{2}}\,-
\nonumber \\
&-&\,
\frac{\imath}{2}\,
\int  \frac{d^4 p_{i}}{(2\pi)^{4}}\,\tilde{s}_{c \nu}(p_{i})\,\int d^4 p_{f} \tilde{s}^{c \nu}(p_{f})p_{f}^{2}\,\delta^{4}(p_{i}+p_{f})\,
\int_{0}^{1}\,dx\,\int \frac{d^{4} l_{E}}{(2\pi)^{4}}\,
\frac{l_{E}^{2} + 2 p_{f}^{2} x\Le 1-x \Ra + 2m^2  }{\Le l^{2}_{E}\,+\,\Delta\Ra^{2}}\,+\,
\nonumber \\
&+&\,
\frac{\imath}{2}\,
\int  \frac{d^4 p_{i}}{(2\pi)^{4}}\,\tilde{s}(p_{i})\,\int d^4 p_{f} \tilde{s}(p_{f})\,p_{f}^{4}\, \delta^{4}(p_{i}+p_{f})\,
\int_{0}^{1}\,dx\,\int \frac{d^{4} l_{E}}{(2\pi)^{4}}\,
\frac{x\Le 1-x \Ra}{\Le l^{2}_{E}\,+\,\Delta\Ra^{2}}\,.
\eeqar

\subsection*{Seventh term}

 The next term we consider is the following one:
\beq\label{CC019}
Int_{2g}=-\imath\,m\,\int d^4 x\,s(x)\int d^4 y\,s_{c \nu}(y)\,\eta^{\mu \nu }\,\gamma^{c}\,\Le \Le \D_{\mu x}\,S_{F}(x,y)\Ra\,S_{F}(y,x) +
S_{F}(x,y)\,\Le \D_{\mu x}\,S_{F}(y,x)\Ra\Ra\,=\,0\,,
\eeq
similarly to fourth term.

\subsection*{Eighth term}

 The eighth term is
\beqar\label{CC020}
&\,& Int_{2h}= \frac{1}{4}\,\int d^4 x\,s_{c a}(x)\int d^4 y\,s_{c_{1} a_{1}}(y)\,\\
&\,&
\Le \gamma^{c}\,[\gamma^{a}\,\gamma^{b}]\,\Le \D_{b x}\D_{b_{1} y}\,S_{F}(x,y)\Ra\,\gamma^{c_{1}}\,[\gamma^{a_{1}}\,\gamma^{b_{1}}] \,S_{F}(y,x) +
\gamma^{c}\,[\gamma^{a}\,\gamma^{b}]\,\Le \D_{b x}S_{F}(x,y)\Ra\,\gamma^{c_{1}}\,[\gamma^{a_{1}}\,\gamma^{b_{1}}] \,\Le \D_{b_{1} y}\,S_{F}(y,x)\Ra\Ra\,.
\nonumber
\eeqar
After the Fourier transform we have:
\beqar\label{CC014C1}
Int_{2h} & = & 
\frac{1}{4}\,\int  \frac{d^4 p_{i}}{(2\pi)^{4}} \tilde{s}_{c a}(p_{i})\int d^4 p_{f} \tilde{s}_{c_{1}a_{1}}(p_{f})\delta^{4}(p_{i}+p_{f})\,
\nonumber \\
&\,&\,
\int \frac{d^{4} p}{(2\pi)^{4}}
\gamma^{c}\,[\gamma^{a}\,\gamma^{b}]\,\frac{\gamma^{\rho} p_{\rho}+m}{p^2-m^2+\imath\varepsilon}\,
\gamma^{c_{1}}\,[\gamma^{a_{1}}\,\gamma^{b_{1}}]
\frac{ \gamma^{\rho_{1}}\Le p-p_{f}\Ra_{\rho_1}+m}{(p-p_{f})^2-m^2\,+\,\imath\varepsilon} p_{b}p_{f b_{1}}\,.
\eeqar
With the help of
\beq\label{CC2}
\gamma^{\mu}\,\gamma^{\nu}\,\gamma^{\rho}\,=\,\eta^{\mu \nu}\gamma^{\rho}\, +\, \eta^{\nu \rho} \gamma^{\mu}\, -\,\eta^{\mu \rho}  \gamma^{\nu}\,-\,\imath\,\epsilon^{\mu \nu \rho \sigma}
\,\eta_{\sigma \sigma_{1}}  \gamma^{\sigma_{1}}\,\gamma^{5}\,
\eeq
and
\beq\label{CCC3}
\gamma^{\mu}\,[\gamma^{\nu}\,\gamma^{\rho}]\,=\,2\,\Le \eta^{\mu \nu}\gamma^{\rho}\, -\,\eta^{\mu \rho}  \gamma^{\nu}\,-\,\imath\,\epsilon^{\mu \nu \rho \sigma}
\,\eta_{\sigma \sigma_{1}}  \gamma^{\sigma_{1}}\,\gamma^{5}\Ra\,
\eeq
we have for the trace we need:
\beqar\label{CC4}
Tr_{1}\,&=&\,Tr\left[ \gamma^{c}\,[\gamma^{a}\,\gamma^{b}]\,\gamma^{\rho}\,\gamma^{c_{1}}\,[\gamma^{a_{1}}\,\gamma^{b_{1}}]\,\gamma^{\rho_{1}}  \right]\,=\,Tr\left[
\Le \eta^{c a}\gamma^{b}\, -\,\eta^{c b}  \gamma^{a}\,-\,\imath\,\epsilon^{c a b \sigma}
\,\eta_{\sigma \sigma_{1}}  \gamma^{\sigma_{1}}\,\gamma^{5}  \Ra\,\gamma^{\rho}\,
\right. 
\nonumber \\
&\,&
\left.
\Le \eta^{c_{1} a_{1}}\gamma^{b_{1}}\, -\,\eta^{c_{1} b_{1}}  \gamma^{a_{1}}\,-\,\imath\,\epsilon^{c_{1} a_{1} b_{1} \gamma}
\,\eta_{\gamma \gamma_{1}}  \gamma^{\gamma_{1}}\,\gamma^{5}  \Ra\,
\gamma^{\rho_{1}}\right]\,.
\eeqar
For the symmetrical $s_{\mu \nu}$ we stay therefore with
\beq\label{CC41}
Tr_{1}\,=\,Tr\left[
\Le \eta^{c a}\gamma^{b}\, -\,\eta^{c b}  \gamma^{a}\,\Ra\,\gamma^{\rho}\,
\Le \eta^{c_{1} a_{1}}\gamma^{b_{1}}\, -\,\eta^{c_{1} b_{1}}  \gamma^{a_{1}}\,\Ra\,\gamma^{\rho_{1}}\right]\,.
\eeq
Now, after some algebra, we obtain finally:
\beqar\label{CC5}
&\,& Int_{2h} \,=\,\frac{\imath}{4}\,
\int \frac{d^4 p_{i}}{(2\pi)^{4}}\,\int d^4 p_{f}\, \tilde{s}(p_{i})\,\tilde{s}(p_{f})\, p_{f}^{2}\,\delta^{4}(p_{i}+p_{f})\,
\int_{0}^{1}\,dx\,\int \frac{d^{4} l_{E}}{(2\pi)^{4}}
\frac{l_{E}^{2}-2 p_{f}^{2} x\Le 1-x \Ra +2 m^2  }{\Le l^{2}_{E}\,+\,\Delta\Ra^{2}}\,-\,
\nonumber \\
&-&
\frac{\imath}{2}\,
\int \frac{d^4 p_{i}}{(2\pi)^{4}}\,\int d^4 p_{f}\, \tilde{s}(p_{i})\,\tilde{s}_{c \nu}(p_{f})\, p_{f}^{c}\,p_{f}^{\nu}\delta^{4}(p_{i}+p_{f})\,
\int_{0}^{1}\,dx\,\int \frac{d^{4} l_{E}}{(2\pi)^{4}}
\frac{l_{E}^{2}-2 p_{f}^{2} x\Le 1-x \Ra +2 m^2  }{\Le l^{2}_{E}\,+\,\Delta\Ra^{2}}\,+
\nonumber \\
&+&\,
\,\frac{\imath}{4}\,
\int  \frac{d^4 p_{i}}{(2\pi)^{4}}\,\tilde{s}_{c \nu}(p_{i})p_{f}^{\nu}\,\int d^4 p_{f} \tilde{s}_{\nu_{1}}^{c}(p_{f})p_{f}^{\nu_{1}}\,\delta^{4}(p_{i}+p_{f})\,
\int_{0}^{1}\,dx\,\int \frac{d^{4} l_{E}}{(2\pi)^{4}}\,
\frac{l_{E}^{2} + 2  p_{f}^{2} x\Le 1-x \Ra +2 m^2  }{\Le l^{2}_{E}\,+\,\Delta\Ra^{2}}\,-\,
\nonumber \\
&-&\,
\imath\,
\int  \frac{d^4 p_{i}}{(2\pi)^{4}}\,\tilde{s}_{a c }(p_{i})p_{f}^{a}\,p_{f}^{c}\,\int d^4 p_{f} \tilde{s}_{a_{1} c_{1}}(p_{f})\,p_{f}^{a_{1}}\, p_{f}^{c_{1}}\,\delta^{4}(p_{i}+p_{f})\,
\int_{0}^{1}\,dx\,\int \frac{d^{4} l_{E}}{(2\pi)^{4}}\,
\frac{x\Le 1-x \Ra}{\Le l^{2}_{E}\,+\,\Delta\Ra^{2}}\,.
\eeqar
and
\beqar\label{CC501}
&\,& Int_{2h} \,=\,\frac{\imath}{4}\,
\int \frac{d^4 p_{i}}{(2\pi)^{4}}\,\int d^4 p_{f}\, \tilde{s}(p_{i})\,\tilde{s}(p_{f})\, p_{f}^{2}\,\delta^{4}(p_{i}+p_{f})\,
\int_{0}^{1}\,dx\,\int \frac{d^{4} l_{E}}{(2\pi)^{4}}
\frac{l_{E}^{2}-2 p_{f}^{2} x\Le 1-x \Ra +2 m^2  }{\Le l^{2}_{E}\,+\,\Delta\Ra^{2}}\,-\,
\nonumber \\
&-&
\frac{\imath}{8}\,
\int \frac{d^4 p_{i}}{(2\pi)^{4}}\,\tilde{s}(p_{i})\,\int d^4 p_{f}\, \tilde{s}(p_{f})\, p_{f}^{2}\,\delta^{4}(p_{i}+p_{f})\,
\int_{0}^{1}\,dx\,\int \frac{d^{4} l_{E}}{(2\pi)^{4}}
\frac{l_{E}^{2}-2 p_{f}^{2} x\Le 1-x \Ra +2 m^2  }{\Le l^{2}_{E}\,+\,\Delta\Ra^{2}}\,+
\nonumber \\
&+&\,
\,\frac{\imath}{16}\,
\int  \frac{d^4 p_{i}}{(2\pi)^{4}}\,\tilde{s}_{c \nu}(p_{i})\,\int d^4 p_{f} \tilde{s}^{c \nu}(p_{f})p_{f}^{2}\,\delta^{4}(p_{i}+p_{f})\,
\int_{0}^{1}\,dx\,\int \frac{d^{4} l_{E}}{(2\pi)^{4}}\,
\frac{l_{E}^{2} + 2  p_{f}^{2} x\Le 1-x \Ra +2 m^2  }{\Le l^{2}_{E}\,+\,\Delta\Ra^{2}}\,-\,
\nonumber \\
&-&\,
\frac{\imath}{16}\,
\int  \frac{d^4 p_{i}}{(2\pi)^{4}}\,\tilde{s}(p_{i})\,\int d^4 p_{f} \tilde{s}(p_{f})\,p_{f}^{4}\, \delta^{4}(p_{i}+p_{f})\,
\int_{0}^{1}\,dx\,\int \frac{d^{4} l_{E}}{(2\pi)^{4}}\,
\frac{x\Le 1-x \Ra}{\Le l^{2}_{E}\,+\,\Delta\Ra^{2}}\,.
\eeqar

\subsection*{Ninth term}

 Similarly to the previous similar terms, the term
\beq\label{CC6}
Int_{2I}=-\frac{\imath m}{2}\,\int d^4 x\,s(x)\int d^4 y\,s_{c a}(y)\,\gamma^{c}\,[\gamma^{a}\,\gamma^{b}]\,\Le \Le \D_{b x}\,S_{F}(x,y)\Ra\,S_{F}(y,x) +
S_{F}(x,y)\,\Le \D_{b x}\,S_{F}(y,x)\Ra\Ra\,=\,0\,,
\eeq
is zero as well.

\subsection*{Tenth term}
 
The last term we need is
\beq\label{CC7}
Int_{2j}=-\frac{ m^2}{2}\int d^4 x\,s(x)\int d^4 y\,s(y)\,S_{F}(x,y)\,S_{F}(y,x)\,.
\eeq
Fourier transform provides:
\beq\label{CC701}
Int_{2j}=-\frac{ m^2}{2} \int \frac{d^4 p_{i}}{(2\pi)^{4}} \tilde{s}(p_{i})\int d^4 p_{f} \tilde{s}(p_{f})\delta^{4}(p_{i}+p_{f})\int \frac{d^{4} p}{(2\pi)^{4}}
\frac{\gamma^{\rho} p_{\rho}+m}{p^2-m^2+\imath\varepsilon}\,
\frac{ \gamma^{\rho_{1}}\Le p-p_{f}\Ra_{\rho_1}+m}{(p-p_{f})^2-m^2\,+\,\imath\varepsilon} 
\eeq
and after a calculation we obtain the followinganswer
\beq\label{CC702}
Int_{2j}= 2\,\imath\, m^2\,\int \frac{d^4 p_{i}}{(2\pi)^{4}} \tilde{s}(p_{i})\int d^4 p_{f} \tilde{s}(p_{f})\delta^{4}(p_{i}+p_{f})\,
\int_{0}^{1}\,dx\,\int \frac{d^{4} l_{E}}{(2\pi)^{4}}\,
\frac{l_{E}^{2} + p_{f}^{2} x\Le 1-x \Ra - m^2  }{\Le l^{2}_{E}\,+\,\Delta\Ra^{2}}\,.
\eeq
%and correspondingly 
%for the Euclidean momenta:
%\beq\label{CC703}
%Int_{2j}= -2\,\imath\, m^2\,\int \frac{d^4 p_{i}}{(2\pi)^{4}} \tilde{s}(p_{i})\int d^4 p_{f} \tilde{s}(p_{f})\delta^{4}(p_{i}+p_{f})\,
%\int_{0}^{1}\,dx\,\int \frac{d^{4} l_{E}}{(2\pi)^{4}}\,
%\frac{l_{E}^{2} + p_{f}^{2} x\Le 1-x \Ra - m^2  }{\Le l^{2}_{E}\,+\,\Delta\Ra^{2}}\,.
%\eeq

%%%%%%%%%%%%%%%%%%%%%%%%%%%%%%%%%%%%%%%%%%%%%%%

\end{document}